\documentclass[10pt]{article}

\usepackage[english]{babel}
\usepackage{latexsym,amsmath,amssymb,amsbsy,amstext,amscd,amsfonts}
\usepackage{graphics,graphicx}
\usepackage{color}
\usepackage{tabularx}
\usepackage{multirow}
\usepackage{booktabs}
\usepackage{epstopdf}
\usepackage{natbib}
\usepackage{ulem}
\newcommand{\bPhi}{\mbox{\boldmath $\Phi$}}

\date{\today}

\title{Energy Release Rate in hydraulic fracture:\\ can we neglect an impact of the hydraulically induced shear stress?}

\author{Michal Wrobel$^{(1)}$, Gennady Mishuris$^{(1)}$ Andrea Piccolroaz$^{(2)}$,
\\
{\it $^{(1)}$\!Department of Mathematics,
Aberystwyth University, }
\\ {\it Ceredigion SY23 3BZ, Wales U.K.,}
\\{\it $^{(2)}$\! Dipartimento di Ingegneria Civile, Ambientale e Meccanica,}\\
{\it Università di Trento, }
 {\it via Mesiano, 77 I-38123 Trento (Italia)}}

\begin{document}

\maketitle

\begin{abstract}
 A novel hydraulic fracture (HF) formulation is introduced which accounts for the hydraulically induced shear stress at the crack faces. It utilizes a general form of the elasticity operator alongside a revised fracture propagation condition based on the critical value of the energy release rate. It is shown that the revised formulation describes the underlying physics of HF in a more accurate way and is in agreement with the asymptotic behaviour of the linear elastic fracture mechanics.  A number of numerical simulations by means of the universal HF algorithm previously developed in \citet{wr_mish_2015} are performed in order to:  i) compare the modified HF formulation with its classic counterpart and ii) investigate the peculiarities of the former. Computational advantages of the revised HF model are demonstrated. Asymptotic estimations of the main solution elements are provided for the cases of small and large toughness. The modified formulation opens new ways to analyse the physical phenomenon of HF and also improves the reliability and efficiency of its numerical simulations.

\end{abstract}

\section{Introduction}

The term of hydraulic fracture refers to a propagation of a fluid driven crack. This can be encountered in many natural processes such as expansion of the magmatic intrusion in the earth crust \citep{rubin_1995} or subglacial drainage of water \citep{tsai_rice_2010}. This physical mechanism is also used in technology, where it can act as an accompanying effect to main processes or be an intentionally created treatment. Hydraulic fracture can have a detrimental impact, in which case  efforts are taken to counteract it (e.g. CO$_2$ sequestration, underground waste storage),  but it can also introduce a desired effect, and then its stimulation is needed (fracking technologies). All these considerations create a demand for the proper understanding and prediction of the underlying physical process.

The multiphysics nature of hydraulic fracture requires careful analysis of the interactions between component physical fields. As a result, the mathematical description and solution of the problem pose a formidable task. The main difficulties stem from issues including: i) strong non-linearities resulting from interaction between the solid and fluid phases, ii) singularities in the physical fields, iii) moving boundaries, iv) degeneration of the governing equations at the singular points of the domain, v) leak-off to the rock formation, vi) pronounced multiscaling effects, vii) complex geometry.

The complexity of the problem motivates various simplifications in theoretical analyses. The first mathematical models of HF were proposed in the 1950s (see for example \citet{Harrison} and \citet{Hubbert}). Together with some later works, these simplified approaches can be summarized in the following three main classic models.
The so-called PKN model was considered by \citet{Perkins_Kern} and further developed by \citet{Nordgren}.
The KGD (plane strain) model was developed independently by \citet{Khristianovic} and \citet{Geertsma}. Finally, the penny-shaped (radial) model was introduced by \citet{Sneddon_Elliot} for constant fluid pressure and was extended for the general case by \citet{Spence&Sharp}.

Although these simplified models have been superseded in practical use by more advanced versions, they are still used in the analyses of the basic features of solutions and the verification of computational algorithms. For example, as the KGD model reflects the tip behaviour of a planar 3D fracture, it has been used to analyse the near-tip region, in which the mulitscaling character of the problem becomes important \citep{Bunger_2005,MKP,Gar_Det_Ad}. It has been shown that the coupling between non-linear, non-local and history dependent physical fields is reflected in the complex solution structure that describes coupled processes at various scales near the crack tip. As a result, at certain spatial and temporal scales some of the competing processes become predominant at the expense of others which become irrelevant. It is now understood that the fracture evolves between different propagation regimes in the parametric space encompassed by four basic modes \citep{Gar_Det_Ad}: (i) leak-off dominated, (ii) storage dominated, (iii) toughness dominated, (iv) viscosity dominated. Each of these limiting regimes is associated with qualitatively different tip asymptotic behaviour.

The mechanics of fluid flow inside the fracture is of primary importance. In the general case of rough walled three-dimensional fracture, accurate computation of the fluid flow would require solving the full 3D Navier-Stokes equations, where the computational cost would be rather prohibitive. For this reason, a number of simplifications have been introduced into the models designed to date. One of the important concepts adopted here is the so called effective (or equivalent) hydraulic aperture. It defines the opening of a virtual crack  with parallel smooth walls, for which the given gradient pressure would produce the same fluid flow rate as in the real rough-walled fracture \citep{Brown_1987,Zimmerman_1991}. This concept allows application of the lubrication theory approximation to the fluid flow \citep{Brown_1987,Zimmerman_1991,Lavrov_2013}, so that the fluid velocity is averaged over the fracture width which is assumed to be significantly smaller than the other dimensions of the crack.

Another consequence of applying lubrication theory is the common a priori assumption that the hydraulically induced shear stresses on the fracture walls are small compared to the fluid pressure and thus can be neglected. Although explicitly stated only in a few papers \citep{Spence&Sharp,tsai_rice_2012}, this expectation has been used commonly throughout the years \citep{Lenoach_1995,Adachi_Detournay,Bunger_2005,garagash_large_toughenss,Gar_Det_Ad}. On the other hand, when considering the near-tip asymptotics of the solution, the singularity of the shear stress is always stronger than that of the fluid pressure. Thus, the assumption of the dominance of the fluid pressure over the shear stress is questionable, at least in the near-tip zone. Surprisingly, even in those papers where the numerical solutions are delivered \citep{Spence&Sharp,Adachi_Detournay,garagash_large_toughenss}, no a posteriori estimation of the ratio fluid pressure/shear stress has been given. Moreover, the numerical data provided in the said papers suggest that the mentioned underlying assumption may not be justified. Thus, an analysis of the influence of the tangential stresses on the fracture walls during hydraulic fracture propagation is needed.

We address the problem of hydraulically induced tangential traction in hydrofracturing by developing a revised HF model. A new form of the elasticity operator which accounts for this additional component of the loading is introduced. As a result, the tip asymptote of the toughness dominated regime holds regardless of the material toughness and, thus, the classic asymptotic behaviour of the viscosity dominated regime is no longer permissible. We introduce a new form of the fracture propagation criterion, based on the critical value of the energy release rate (ERR),  that takes into account the hydraulically induced singular shear stress. We prove that the standard assumption of the fluid pressure dominance over the tangential traction is not justified except in the case of large material toughness. We show however, that the recalled assumption is not a valid criterion to assess the
overall influence of the hydraulically induced shear stress on the underlying physics. On the other hand, we demonstrate that this load component plays a crucial role in the HF process and thus cannot be neglected, which surprisingly leads to significant  computational advantages.

The structure of the paper is given by the following. In Section 2 we present the basic equations describing the problem of a planar 2D hydraulic fracture. Preliminary information on the hydraulically induced tangential traction at the crack faces is presented. Justification for the introduction of a modified HF formulation which accounts for this additional feature is given. Section 3 contains a reduction of the problem to a 1D plane-strain  model of hydraulic fracture, where a new form of elasticity operator, employing the tangential traction, is proposed. A novel crack propagation condition, based on the critical value of the energy release rate and which accounts for for the shear stress component, is derived. A complete formulation of the revised HF model is given. In Section 4, a time independent self-similar version of the problem is introduced. The self-similar equivalent of the modified HF formulation is used in numerical simulations in Section 5 to: i) compare it with the classical KGD model and its amended version, ii) investigate the basic features of the revised model as they relate to the introduction of the tangential traction. The asymptotic estimations of  some important HF parameters for small and large toughness regimes of crack propagation are delivered. A discussion and some final conclusions can be found in Section 6.

\section{Introduction and preliminary results}

Let us consider the governing equations for the 2D planar hydraulic fracture. The continuity equation assumes the form  (see e.g. \cite{Adachi-et-Al-2007}):
\begin{equation}
\label{continuos_0} \frac{\partial w}{\partial t}+\nabla \cdot {\bf q}+q_l=0,\quad t > 0,\quad (x,y)\in \Omega(t),
\end{equation}
where $w(t,x,y)$ stands for the crack aperture, ${\bf q}(t,x,y)$ is the fluid flow rate, and $q_l(t,x,y)$ is the function that describes the fluid loss to the rock formation (the leak-off function). We assume $q_l(t,x,y)$ to be known.

Fluid flow inside the fracture is described by the Poiseuille equation which, in the case of Newtonian fluid, is:
\begin{equation}\label{Poiseulle_0}
{\bf q}=-\frac{1}{M}w^3\nabla  p,\quad t > 0,\quad (x,y)\in \Omega(t).
\end{equation}
Here $M=12\mu$, where $\mu$ is the dynamic viscosity (e.g. \cite{Adachi-et-Al-2007}), and $p(t,x,y)$ is the net fluid pressure, i.e. the difference between the actual pressure, $p_{f}$, and the confining stress, $\sigma_0$ ($p=p_{f}-\sigma_0$).
When introducing the average velocity of fluid flow over the cross section  (the particle velocity) \citep{Linkov_Brisbane},
\begin{equation}\label{Velocity_0}
{\bf V}=-\frac{1}{M}w^2\nabla  p ,\quad t > 0,\quad (x,y)\in \Omega(t),
\end{equation}
equation (\ref{Poiseulle_0}) can be written as:
\begin{equation}\label{Velocity_1}
{\bf q}={\bf V}w,\quad t > 0,\quad (x,y)\in \Omega(t).
\end{equation}
We assume that the particle velocity is bounded such that:
\begin{equation}\label{Bound_0}
|{\bf V}|<\infty,\quad t > 0,\quad (x,y)\in \Omega(t),
\end{equation}
which is in line with the lubrication theory approximation accepted in most of the hydraulic fractures models (along with the omission of the inertial
terms).

A zero-opening boundary condition at the crack front is imposed:
\begin{equation}
\label{w_tip}
w\big|_{\partial \Omega}=0.
\end{equation}
Boundary conditions for the fluid flux across the fracture, $\bf q$, are specified as follows:
\begin{itemize}
\item{along the crack front}
\begin{equation}
\label{q_tip}
{\bf{q}}\big|_{\partial \Omega}=0.
\end{equation}
\item{at the fracture inlet where the influx value is given}
\begin{equation}
\label{q_or}
\oint_{S_0}{\bf{q} \cdot {\bf n}}\,ds=Q_0(t),
\end{equation}
where $S_0$ is an arbitrary small contour containing the origin $O$.
\end{itemize}

Note that conditions \eqref{Bound_0} and \eqref{w_tip} imply \eqref{q_tip}, while the condition \eqref{Bound_0} does not necessarily follow
from \eqref{w_tip} and \eqref{q_tip}.

The fluid flow equations are to be supplemented by relations for the interaction between the solid and fluid phases.
The normal and tangential stresses on crack faces induced by the fluid pressure are given according to lubrication theory
(see for example \cite{tsai_rice_2010}) in the following way:
\begin{equation}
\label{tau_0}
\sigma_n=-p, \quad {\boldsymbol{\tau}}=-\frac{w}{2}\nabla  p,\quad (x,y)\in \Omega(t).
\end{equation}

The classic approach for describing rock deformation under applied hydraulic loading neglects the shear stresses  ${\boldsymbol{\tau}}$
and employs the planar elasticity equation in its hyper-singular form combined with the Boundary Element Method (BEM)
(see for example \citep{Adachi-et-Al-2007}):
\begin{equation}
\label{elastic_0}
\int_{\Omega(t)}C(x,y;\xi,\eta)w(t,\xi,\eta)d\xi d\eta=p,\quad (x,y)\in \Omega(t).
\end{equation}
In the case of an arbitrary crack geometry, the respective system of hyper-singular integral equations can be found in the work of \cite{Linkov_2002}.

Equations (\ref{continuos_0}), (\ref{Poiseulle_0}) and (\ref{elastic_0}), together with boundary conditions \eqref{w_tip} -- \eqref{q_tip} and respective initial conditions (initial crack opening and length) constitute the system of governing equations describing the problem of hydraulic fracture in its classic formulation.

The above system has to be complemented by a fracture propagation criterion which is usually the one of the linear elastic fracture mechanics based on the critical value of the stress intensity factor:
\begin{equation}
\label{K_IC}
K_I=K_{IC},
\end{equation}
where $K_{IC}$ is the material toughness, while $K_I$ denotes the stress intensity factor provided that the following crack opening tip asymptotics holds:
\begin{equation}
\label{Ass_K_IC}
w(t,r) \sim \gamma
K_I(t)
\sqrt{r},\quad r\to0.
\end{equation}
Here $r$ is the radial distance from the crack front and
\begin{equation}
\label{gamma}
\gamma=\frac{8}{\sqrt{2\pi} }\frac{1-\nu^2}{E},
\end{equation}
where $E$ and $\nu$ are the Young modulus and Poisson's ratio respectively.

In the case of the so called viscosity dominated regime of fracture propagation, that is when $K_{IC}=0$, the pertinent tip asymptotics yields \citep{Gar_Det_Ad}:
\begin{equation}
\label{fluid_as}
w(t,r) \sim
w_0(t)r^{2/3},\quad r\to0.
\end{equation}

This {\it classic} formulation has been widely used for mathematical description of the HF problem. A number of its simplified variants have been introduced to define the basic models such as: PKN \citep{Nordgren}, KGD and radial \citep{Spence&Sharp}, pseudo 3D \citep{Warpinski} and others. They enable investigation of the basic physical features of the HF phenomenon. In particular, important results concerning the solution tip asymptotics and its multiscale character have been delivered in \cite{Gar_Det_Ad}, by analysis of the plane strain model (KGD). The relevance of the tip asymptotics for proper understanding and modelling of the hydraulic fractures is clear and has been emphasized in numerous papers \citep{Bunger_2005,garagash_large_toughenss,MKP,Lecampion_Brisbane,wr_mish_2015,Linkov_2015,Perkowska_2016}.

\subsection{Hydraulically induced tangential traction}
\label{tan_str}

One of the basic assumptions of the {\it classic} HF models is that the hydraulically induced shear traction acting along the fracture surfaces
is negligible as compared to the fluid pressure:
\begin{equation}
\label{tau_p}
|{\boldsymbol{\tau}}|\ll |{p}|,
\end{equation}
and thus, related effects can be neglected. This was first stated in \cite{Spence&Sharp}, while a rough estimation of the ratio $\tau/p$ for the crack inlet was given in \cite{tsai_rice_2012}.
We do not discuss possible reasons why this simplification has been commonly accepted.
The primary goal of this paper is to provide a thorough analysis of the HF problem where the tangential stresses are not neglected.

We first justify our interest in revisiting the commonly accepted omission of $\tau$ and try to answer the following question:
\begin{itemize}
\item \textbf{Can estimate \eqref{tau_p} be satisfied in the framework of the classic formulation?}
\end{itemize}
In other words, we would like to estimate the value of the ratio (compare with (\ref{Velocity_0})):
\begin{equation}
\label{ratio_0}
\delta(t,{\bf x})=\frac{|{\boldsymbol{\tau}}|}{p}\equiv \frac{w}{2p}|\nabla p|=\frac{M}{2}\frac{|{\bf V}|}{wp},\quad t > 0,\quad (x,y)\in \Omega(t),
\end{equation}
that depends on spatial variables and time. When considering the tip asymptotic behaviour of respective functions in two basic modes of crack propagation, the so-called toughness dominated and viscosity dominated regimes (see e.g. \cite{Gar_Det_Ad,wr_mish_2015}),  we obtain:
\begin{equation}
\label{ratio_0a}
\delta_t=O\big(r^{-1/2}\log^{-1}r\big), \quad  \delta_v=O\big(r^{-1/3}\big), \quad r\to 0,
\end{equation}
where $r$ is the distance to the crack front ($\delta_t$ refers to the toughness dominated regime, $\delta_v$ pertains to the viscosity dominated mode).
The first straightforward conclusion from \eqref{ratio_0a} is rather worrying:
\begin{itemize}
\item[$\clubsuit$]\textbf{The basic assumption of the classic HF theory is violated, at least near the crack front.}
\end{itemize}

It is even more alarming when one recalls that in the classic HF theory the near-tip behaviour of the solution has been recognized to control the global response of the fracture \citep{Gar_Det_Ad} and its relevance in the accurate and efficient numerical simulation of the problem has been identified \citep{Lecampion_Brisbane,wr_mish_2015}.

When applying asymptotic results for the classic KGD model (in both toughness and viscosity dominated regimes) to the definition of the tangential stress \eqref{tau_0}$_2$, we have the following asymptotic estimates for $\tau$:
\begin{equation}
\label{tau_00}
|{\boldsymbol{\tau}_t}|=O\big(r^{-1/2}\big), \quad  |{\boldsymbol{\tau}_v}|=O\big(r^{-2/3}\big), \quad r\to 0.
\end{equation}
Obviously, in both cases the shear stress singularity is much stronger than that of the fluid pressure (see e.g. \cite{wr_mish_2015}).

This, in turn, raises questions, as to what degree the omission of hydraulically induced shear stresses in the classic HF formulation affects:
\begin{itemize}
\item[$\diamond$]\textbf{asymptotic behaviour of the solution},
\item[$\diamond$]\textbf{elastic response of the solid material},
\item[$\diamond$]\textbf{fracture propagation criteria}.
\end{itemize}

We respond to the above questions by introducing a revised HF formulation that accounts for the tangential traction at the crack faces induced by the fluid flow.
The analytical and numerical results provided below demonstrate both the qualitative and the quantitative consequences of accounting for $\tau$. The conclusions drawn in the final section suggest that the impact of the shear stress is {\it  rather significant}.

\section{The revised HF formulation}

\subsection{Governing equations for 1D KGD model of hydraulic fracture}

We focus on the 1D plane strain model (KGD) which represents the tip behaviour of a planar 3D fracture.

\begin{figure}[h!]
%M/N=1/300
    \center
    %\hspace{-2mm}
    \includegraphics [scale=0.70]{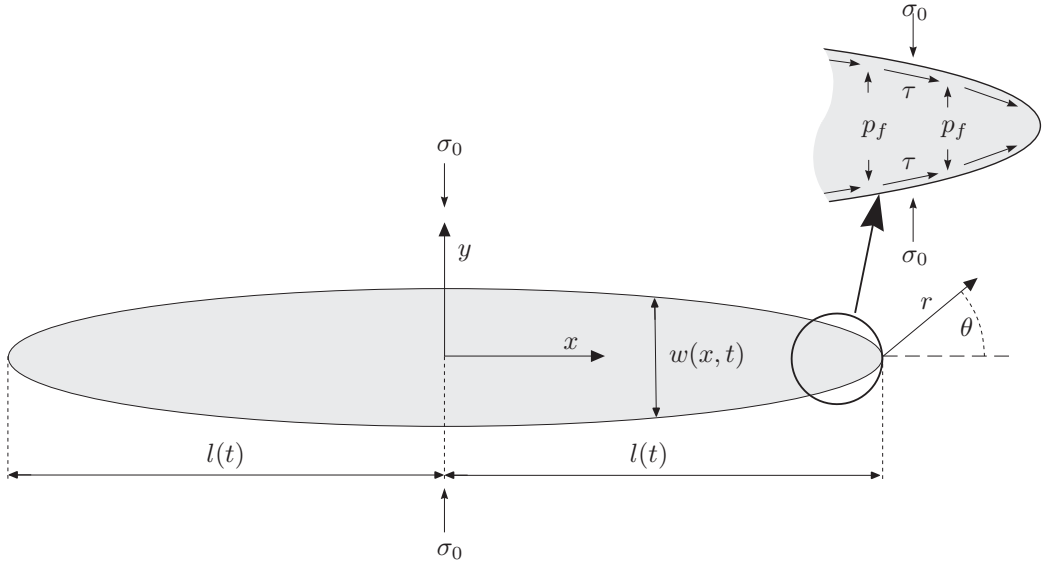}
    \put(-180,70){$x$}
    \put(-220,105){$y$}
    \put(-140,65){$w(x,t)$}
    \put(-155,27){$l(t)$}
    \put(-315,27){$l(t)$}
    \put(-30,75){$\theta$}
    \put(-45,85){$r$}
    \put(-67,152){$p_f$}
    \put(-37,152){$p_f$}
    \put(-52,165){$\tau$}
    \put(-52,138){$\tau$}
    \put(-228,145){$\sigma_0$}
    \put(-228,-8){$\sigma_0$}
    \put(-52,196){$\sigma_0$}
    \put(-52,103){$\sigma_0$}
    \caption{Sketch of a plane-strain fluid driven fracture.}
\label{ERR}
\end{figure}

Let us consider a rectilinear crack  of length $2l$ ($-l<x<l$),  completely filled with Newtonian fluid injected at the midpoint ($x=0$) at  given rate $q_0(t)$ (Fig.\ref{ERR}). As a result, the crack front ($x=\pm l$) moves, and the crack half-length, $l=l(t)$, is a function of time.
As usual, given the symmetry of the problem, we analyse only one of the symmetrical parts of the crack $x\in [0,l(t)]$.

The continuity equation has the form:
\begin{equation}
\label{continuos_1} \frac{\partial w}{\partial t}+\frac{\partial
q}{\partial x}+q_l=0,\quad t > 0,\quad 0< x < l(t).
\end{equation}
The fluid flow inside the fracture is described by the Poiseuille equation which, in the case of Newtonian fluids, is given by:
\begin{equation}\label{Poiseulle_1}
q=-\frac{1}{M}w^3\frac{\partial p}{\partial x},\quad t > 0,\quad 0< x < l(t),
\end{equation}
where the respective variables are now scalars. The function $q_l=q_l(t,x)$, in the right-hand
side of the continuity equation \eqref{continuos_1},  is the volumetric rate of fluid loss to  the rock
formation in the direction perpendicular to the crack surfaces per unit
length of the fracture.  For simplicity, we assume it to be given and non-singular (see discussion in \cite{wr_mish_2015}).
A more comprehensive analysis would involve a nonlocal formulation, where the mass transfer
in the entire external domain would be taken into account (e.g. \cite{Kovalyshen_PhD}).

The hydraulically induced tangential stress on the crack faces is computed in this case as (compare (\ref{tau_0})$_2$):
\begin{equation}
\label{tau}
\tau=-\frac{w}{2}\frac{\partial p}{\partial x}.
\end{equation}

The elasticity relation defining the deformation of rock under both normal and shear stresses can be derived from the general BEM formulation \citep{Linkov_2002} or adopted from \cite{Picc_2013}
\begin{equation}
\label{elasticity_1}
p(x) =
-\frac{k_1}{2} \int_{-l(t)}^{l(t)} \tau(s) \frac{ds}{x - s} +
\frac{k_2}{2} \int_{-l(t)}^{l(t)} \frac{\partial w}{\partial s} \frac{ds}{x - s}, \quad -l(t)<x<l(t).
\end{equation}

Here the constants $k_{1(2)}$ are computed as:
\begin{equation}
\label{k_1_2}
k_1=\frac{1-2\nu}{\pi(1-\nu)}, \quad k_2=\frac{1}{2\pi}\frac{E}{1-\nu^2}.
\end{equation}

Taking into account the symmetry of the problem,  equation \eqref{elasticity_1} can be conveniently rewritten in the form
\begin{equation}
\label{elasticity_1a}
p(x)=  \int_{0}^{l(t)} \left[k_2\frac{\partial w}{\partial s}-k_1 \tau(s) \right] \frac{s\, ds}{x^2 - s^2}, \quad 0\le x<l(t),
\end{equation}
which, in turn, transforms to the standard KGD operator (see, for example, \cite{Spence&Sharp}) when $\tau=0$. As can be seen from \eqref{k_1_2} and \eqref{elasticity_1a}, the influence of the shear stress on the elastic response of the solid material decreases as Poisson's ratio approaches 0.5 (for the incompressible material one has $k_1=0$ which reduces \eqref{elasticity_1a} to the classic KGD form).

The considered problem is equipped with two tip boundary conditions:
\begin{equation}
\label{w_pp_cond}
w(t,l(t))=0,\quad q(t,l(t))=0,
\end{equation}
and the influx boundary condition at the crack inlet:
\begin{equation}
\label{q_0}
q(t,0)=q_0(t), \quad t > 0.
\end{equation}

Similarly to \cite{wr_mish_2015}, we employ a new dependent variable, the particle velocity function $V$, which describes the average speed of fluid flow over the fracture cross section:
\begin{equation}
\label{V_eq}
V=\frac{q}{w}=-\frac{1}{M}w^2\frac{\partial p}{\partial x}.
\end{equation}
The advantages of employing this variable have been shown in \cite{wr_mish_2015} and \cite{Perkowska_2016}.

The mechanism of fracture front tracing will be based on the Stefan type condition formulated as:
\begin{equation}
\label{SE}
l'(t)=-\frac{1}{M}w^2\frac{\partial p}{\partial x}\big|_{x=l(t)}>0.
\end{equation}
Condition \eqref{SE} was derived on the assumption that the crack tip coincides with the fluid front and that the leak-off function is bounded at the crack tip. The convenience and efficiency of such an approach, as well as the technical details of its implementation,  have been demonstrated in \cite{M_W_L,Kusmierczyk,solver_calkowy,wr_mish_2015,Perkowska_2016}.

The above system of equations, together with the respective initial conditions, constitutes the revised HF formulation which will be analysed in this paper. At this stage, the only difference between this formulation and the classic plane strain (KGD) model is the presence of the tangential traction on the fracture walls (see elasticity relation \eqref{elasticity_1}). The introduction of $\tau$, however, has consequences also for the fracture propagation condition, which shall be discussed later.

In the classic KGD model, one defines two basic modes of crack propagation related to the solid material toughness: the viscosity dominated regime ($K_{Ic}=0$) and the toughness dominated regime ($K_{Ic}>0$). Each  exhibits qualitatively different asymptotic behaviour \citep{Gar_Det_Ad}. One can easily check that in the case of the revised HF formulation, due to the introduction of $\tau$,
only the asymptotics of the toughness dominated regime is permissible irrespectively of the value of $K_{Ic}$.
In this way, the following tip asymptotes for the crack opening and net fluid pressure are valid for the revised formulation:
\begin{equation}
\label{w_asymp}
w(x,t)=w_0(t)\sqrt{l(t)-x}+w_1(t)(l(t)-x)+w_2(t)(l(t)-x)^{3/2}\ln(l(t)-x)+...,\quad x\to l(t),
\end{equation}
\begin{equation}
\label{p_asymp}
p(x,t)=p_0(t)\ln(l(t)-x)+p_1(t)+p_2(t)\sqrt{l(t)-x}+p_3(t)(l(t)-x)\ln(l(t)-x)+...,\quad x\to l(t).
\end{equation}

Interestingly, and importantly for further analysis, the tip asymptote of the shear stress when derived from (\ref{tau}) yields:
\begin{equation}
\label{tau_asymp_1}
\tau(x,t)=\frac{\tau_0(t)}{\sqrt{l(t)-x}}+\tau_1(t)+\tau_2(t)\sqrt{l(t)-x}\ln(l(t)-x)+\tau_3\sqrt{l(t)-x}+...,\quad x\to l(t).
\end{equation}
Note that qualitative asymptotic behaviour of $\tau$ is the same as that of $w'_x$ and thus both integrands in \eqref{elasticity_1} exhibit the same asymptotics. The multipliers of the leading asymptotic terms are interrelated:
\begin{equation}
\label{tau_10}
\tau_0(t)=\frac{1}{2}p_0(t) w_0(t).
\end{equation}

Additionally, when taking into account \eqref{w_asymp} -- \eqref{p_asymp}, the equation \eqref{SE} can be transformed to:
\begin{equation}
\label{SE_1}
l'(t)=\frac{1}{M}p_0(t)w_0^2(t).
\end{equation}

The fracture propagation criterion \eqref{K_IC} and estimation \eqref{Ass_K_IC} no longer hold true in the revised HF formulation. Indeed, the stress intensity factor, $K_I$, is now  not only the nonlocal parameter, but also depends on the local singular loading produced by the hydraulically induced shear stress at the crack tip (compare \eqref{tau_asymp_1}). Thus, in the case under consideration, the condition (\ref{K_IC}) should
be replaced by a more general one based on the critical value of the energy release rate \citep{Rice_1968}.

\subsection{Energy Release Rate in hydrofracturing accounting for the tangential traction}

To compute the energy release rate, accurate analysis of the asymptotic
behaviour of the solution near the crack tip is crucial. Let us consider a near-tip zone of the fracture shown schematically in Fig.\ref{ERR}.

Note that the solution of the problem is still symmetrical with respect to the axis $y=0$
even though the shear stress does not vanish along the crack surfaces. Let us accept the following vectorial representation of the displacement field in a vicinity of the crack tip:
\begin{align}
\label{ass_U}
{\bf u}(r,\theta) &= u_r(r,\theta) {\bf e}_r + u_\theta(r,\theta) {\bf e}_\theta =
 r^{1/2}\Big[C_1 \bPhi_1(\theta) + \tau_0 \bPhi_{\tau_0}(\theta)\Big] + \nonumber \\
& r \log r \Big[p_0 \tilde\bPhi_{p_0}(\theta) + \tau_1 \tilde\bPhi_{\tau_1}(\theta)\Big]
+ r \Big[C_2 \bPhi_2(\theta) + p_0 \bPhi_{p_0}(\theta) + p_1 \bPhi_{p_1}(\theta) + \tau_1 \bPhi_{\tau_1}(\theta)\Big] + \\
& r^{3/2} \log r \Big[p_2 \tilde\bPhi_{p_2}(\theta) + \tau_2 \tilde\bPhi_{\tau_2}(\theta)\Big]
+ r^{3/2} \Big[C_3 \bPhi_3(\theta) + p_2 \bPhi_{p_2}(\theta) + \tau_2 \bPhi_{\tau_2}(\theta)\Big]
+o(r^{3/2}), \quad r \to 0, \nonumber
\end{align}
where $C_1 = K_I/\sqrt{2\pi}$. $C_2$, $C_3$ are constants related to the loading away from the crack tip.  The polar coordinates, $r$ and $\theta$,
are defined in the standard way (see Fig.\ref{ERR}). The vector-function, ${\bf u}(r,\theta)$, satisfies the Lam\'e equations, while the constants involved in the representation may be interrelated. Particular forms of the functions $\bPhi_j$ and $\tilde \bPhi_j$,  as well as a justification for their use, are given in the Appendix.
Representation \eqref{ass_U} was constructed as a superposition of two displacement fields, one related to classic fracture mechanics (where the traction vanishes at the crack surfaces)
and another resulting from hydraulic loading (compare \eqref{p_asymp} --  \eqref{tau_asymp_1}).

The corresponding asymptote of the crack opening, $w$, computed as:
\begin{equation}
\label{w_fi}
w(r)= u_{\theta}(r,-\pi)-u_{\theta}(r,\pi)
\end{equation}
yields:
\begin{multline}
\label{w_r}
w(r) =
\frac{1}{E^*}
\Big\{
\sqrt{r} \left[8 C_1 + 4 \left(1+\nu^*\right) \tau_0\right] +
r \left[-2\left(1-\nu^*\right) \tau_1 + 4 \pi p_0\right] + \\
r^{3/2} \left[-8 C_3 - \frac{8}{9} \left(1+\nu^*\right) \tau_2\right] +
r^{3/2} \log r \left[-\frac{4}{3} \left(1-\nu^*\right) \tau_2 - \frac{8 p_2}{3 \pi}\right]
\Big\}, \quad r \to 0,
\end{multline}
where
$
r=l(t)-x,$ and

\begin{equation}
E^*=\frac{E}{1-\nu^2}, \quad \nu^*=\frac{\nu}{1-\nu}.
\end{equation}
Note also that $u_{r}(r,-\pi)-u_{r}(r,\pi)=0$.

Thus, when comparing (\ref{w_r}) with (\ref{w_asymp}), and computing the respective stress tensor components, we conclude:
\begin{equation}
\label{w_0_K_I}
w_0(t)=\gamma \big(K_I(t)+K_f(t)\big),\quad K_f=B^{-1}\tau_0,\quad  B=\sqrt{\frac{2}{\pi}}\frac{1}{1+\nu^*}.
\end{equation}
Here the first term (related to the terms multiplied by $C_1$ in \eqref{ass_U} and \eqref{w_r}) corresponds to the standard Mode I pertaining to the nonlocal parameter - stress intensity factor $K_I$. It  produces zero tangential traction along the crack surfaces. The second (special) term contains a new parameter, $K_f$, henceforth called the shear stress intensity factor. It refers to the solution component which yields zero normal stress (fluid pressure) and non-zero, singular at the crack tip, tangential traction on the fracture faces. However, as can be deduced from \eqref{tau_10}, this term depends also on the multiplier of the logarithmic term of the fluid pressure.

\begin{figure}[h!]
%M/N=1/300
    \center
    %\hspace{-2mm}
    \includegraphics [scale=0.60]{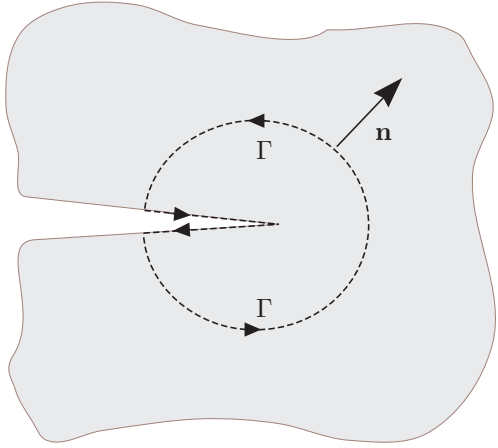}
    \put(-90,108){$\Gamma$}
    \put(-90,48){$\Gamma$}
    \put(-45,115){$\bf n$}
    \caption{Fracture tip zone - J-integral calculation.}
\label{J_int}
\end{figure}

Using expression \eqref{ass_U} one can compute the energy release rate (ERR), based on the standard formula \citep{Rice_1968}:
\begin{equation}
{\cal E} = \int_{\Gamma} \Big\{ \frac{1}{2} ({\boldsymbol \sigma} \cdot {\boldsymbol \varepsilon}) n_x - {\boldsymbol t}_n \cdot \frac{\partial {\boldsymbol u}}{\partial x} \Big\} ds,
\end{equation}
where $\Gamma$ is the limiting contour, shown in Fig.\ref{J_int}, ${\boldsymbol n}$ is the outward normal to the contour $\Gamma$,
${\bf t}_n = {\boldsymbol \sigma}{\bf n}$ is the traction vector along $\Gamma$.

As a result, we arrive at the following expression:
\begin{equation}
\label{ERR_def}
{\cal E}=\frac{1}{E^*}\left( K_I^2+ a K_IK_f\right),
\end{equation}
where
\begin{equation}
\label{a_def}
a=\frac{4}{1+\nu^*}=4(1-\nu).
\end{equation}
The general fracture propagation criterion based on the ERR yields \citep{Rice_1968}:
\begin{equation}
\label{ERR_def_1}
{\cal E}={\cal E}_c\equiv\frac{1}{E^*} K_{Ic}^2.
\end{equation}

%%%%%%%%%%%%%%%%%%%%%%%%%%%%%%%%%%%%%%%%%%%%%%%%%%%%%%%%%%%%%%%%%
%\begin{figure}[!htcb]
%\centering
%\includegraphics[width=80mm]{fig_Jint.eps}
%\caption{Contour path for the calculation of the energy release rate.}
%\label{fig_Jint}
%\end{figure}
%%%%%%%%%%%%%%%%%%%%%%%%%%%%%%%%%%%%%%%%%%%%%%%%%%%%%%%%%%%%%%%%%

Only the leading asymptotic  term in (\ref{w_r}) contributes to the
ERR, but the next terms allow us to establish important relationships between the multipliers in the expansions of the stresses and the displacements,
which is useful in the numerical realization. Interestingly, in the case of zero material toughness (${\cal E}_c=0$), only the nonlocal component of ERR, the stress intensity factor $K_I$,
vanishes, while that related to the local behaviour near the crack tip (shear stress intensity factor) takes non zero value $K_f>0$.

We note that  $K_f$ and $K_I$ are not independent. Indeed, collecting together (\ref{tau_10}) and (\ref{w_0_K_I}),
one has
\begin{equation}
\label{K_f_K_I}
K_f=\varpi K_I,\quad \varpi=\frac{p_0\gamma}{2B-p_0\gamma}=\frac{p_0}{G-p_0}>0,
\end{equation}
where $G$ is the shear modulus ($\varpi$ is dimensionless). This relationship allows us to finally formulate the toughness fracture criterion in the form:
\begin{equation}
\label{criterion_1}
K_I=\frac{K_{IC}}{\sqrt{1+a\varpi}},\quad K_f=\frac{K_{IC}\varpi}{\sqrt{1+a\varpi}},
\end{equation}
or equivalently,
\begin{equation}
\label{criterion_2}
w_0=\frac{\gamma(1+\varpi)}{\sqrt{1+a\varpi}}K_{IC}.
\end{equation}
 It is clear that the following condition should always be  satisfied:
\begin{equation}
\label{super_criterion}
0<p_0<G.
\end{equation}
If $p_0\ll G$ then $\varpi\ll 1$ and, as a result, criteria (\ref{criterion_1}) or (\ref{criterion_2}) are approximately equal to the classic criterion (\ref{K_IC}).
The other end of the interval (\ref{super_criterion}), $p_0\to G$, yields that  $\varpi\to\infty$ and the multipliers of $K_{IC}$ in both formulae (\ref{criterion_1})$_2$ and (\ref{criterion_2})
go to infinity. Thus, the new fracture criterion which takes into account the shear stress will produce pronouncedly different results from those obtained on the basis of the classic condition \eqref{K_IC}.
We observe that the non-local parameter $p_0$ defining the pressure drop near the crack tip depends on the given material toughness:
\[
p_0=p_0(K_{IC}),\quad \varpi=\varpi(K_{IC}).
\]

It will be shown later on that in the revised HF formulation the so-called small toughness regime ($K_{Ic}\to 0$) has its natural limit when $K_I=0$ but $K_f>0$.

In summary, we have delivered a revised fracture propagation criterion for HF that accounts for the specific peculiarity of the solution near the crack tip.
Moreover, we expect a solution of the revised problem to have the following properties:
\begin{equation}
\begin{array}{c}
p_0(K_{IC})\to0 \quad \mbox{as}\quad K_{IC}\to\infty,\\[2mm]
p_0(K_{IC})\to G \quad \mbox{as}\quad K_{IC}\to0.
\end{array}
\end{equation}

The fracture criterion in the form \eqref{criterion_1}--\eqref{criterion_2}, derived in this section, complements the revised HF formulation introduced previously.

\subsection{Inversion of the elasticity equation (\ref{elasticity_1a})}

Before we analyse the revised HF formulation, let us define an inverse of the elasticity operator \eqref{elasticity_1a}. In \cite{wr_mish_2015} and \cite{Perkowska_2016},  it was shown that employing the inverse form rather than its original is very conducive to numerical implementation.

Equation (\ref{elasticity_1a}) can be rewritten in the form:
\begin{equation}
\label{elasticity_2a}
p(x)=\int_0^{l(t)}g'(s)\frac{s}{x^2-s^2}ds, \quad 0<x<l(t),
\end{equation}
where we have introduced a new notation:
\begin{equation}
\label{tau_g}
g(x)=\int_x^{l(t)}\left (k_1\tau(s)-k_2\frac{\partial w}{\partial x}(s)  \right)ds=\int_x^{l(t)}k_1\tau(s)ds+k_2w(x).
\end{equation}
We note that equations \eqref{elasticity_1} and \eqref{elasticity_1a} are equivalent, provided that $g'(0)=0$,
or in other words:
\begin{equation}
\label{tau_dw}
k_1\tau\big |_{x=0}=k_2\frac{\partial w}{\partial{x}}\big |_{x=0}.
\end{equation}
When $\tau=0$ or $k_1=0$, condition \eqref{tau_dw} reduces to $w'_x(0)=0$ which is a standard one accepted for the classic KGD model (see e.g. \cite{wr_mish_2015}).

The integral operator on the right hand side of \eqref{elasticity_2a} can be inverted (see e.g. \cite{Adachi_Detournay}) to give:
\begin{equation}
\label{tau_w_S_new}
k_2w(x)+k_1\int_x^{l(t)}\hspace{-2mm}\tau ds=\frac{2}{\pi^2}\int_0^{l(t)} p(t,s) \ln \left|\frac{\sqrt{l^2(t)-x^2}+\sqrt{l^2(t)-s^2}}{\sqrt{l^2(t)-x^2}-\sqrt{l^2(t)-s^2}}\right|ds.
\end{equation}

From \eqref{tau_w_S_new} and the asymptotics \eqref{w_asymp} -- \eqref{p_asymp}, it also follows that:
\begin{equation}
\label{K1_equiv}
k_2w_0+k_1w_0p_0=\frac{8}{\sqrt{2}\pi^2}\sqrt{l(t)}\int_0^{l(t)}p(s,t)\frac{ds}{\sqrt{l^2(t)-s^2}}.
\end{equation}

Equation \eqref{K1_equiv} replaces the standard integral  definition of the stress intensity factor $K_I$ \citep{Rice_1968}.

\subsection{Problem normalization}
\label{basic}

\subsubsection{Normalized variables}

Let us normalize the problem by introducing the following dimensionless
variables:
\[
\tilde x=\frac{x}{l(t)}, \quad \tilde t = \frac{t}{t_n},\quad \tilde w(\tilde t,\tilde x)=\frac{w(t,x)}{l_*}, \quad L(\tilde
t)=\frac{l(t)}{l_*},\quad\tilde q_l(\tilde t,\tilde x)= \frac{t_n}{l_*} q_l( t, x),\quad \tilde q(\tilde t,\tilde x)=
\frac{t_n}{l_*^2}q(t,x)
\]
\begin{equation}\label{norm_V}
\quad \tilde p(\tilde t, \tilde x)=\frac{t_n}{M}p\,(t,x),\quad \tilde v(\tilde t,\tilde x)=\frac{t_n}{l_*}v(t,x),
\quad \tilde \tau (\tilde t, \tilde x)=\frac{t_n}{M} \tau (t,x), \quad t_n=\frac{M}{k_2},\quad \tilde q_0(\tilde t)=
\frac{t_n}{l_*^2}q_0(t)
\end{equation}
\[
\tilde K_{\{Ic,I,f\}}=\frac{8}{(2\pi)^{3/2}}\frac{1}{k_2\sqrt{l_*}}K_{\{Ic,I,f\}},\quad
\tilde {\cal E}=\frac{32}{E^*l_*\pi}{\cal E},\quad \tilde {\cal E}_c=\frac{32}{E^*l_*\pi}{\cal E}_c.
\]
where $\tilde x\in[0,1]$. The scaling parameter $l_*$ can be taken as convenient.
A similar scaling was already utilized in \cite{wr_mish_2015} and \cite{Perkowska_2016}, where it constituted a basis for the analysis of the time dependent variant of the problem. In this paper we do not investigate directly the transient regime, and as such the following reformulation of the system of governing equations should be considered as an intermediate step in the transition to the self-similar model. However, even without conducting quantitative analysis, some general conclusions on the revised HF formulation can be drawn from the presented equations.

In the normalized variables, the continuity equation \eqref{continuos_1} takes the form:
\begin{equation}
\label{cont_norm}
\frac{\partial \tilde w}{\partial \tilde t}-\frac{L'}{L}\tilde x \frac{\partial \tilde w}{\partial \tilde x}+\frac{1}{L}\frac{\partial \tilde q}{\partial \tilde x}+\tilde q_l=0,
\end{equation}
where the normalized fluid flow rate is:
\begin{equation}\label{Poiseulle_norm}
\tilde q=-\frac{1}{L}\tilde w^3\frac{\partial \tilde p}{\partial \tilde x},\quad \tilde t > 0,\quad 0< \tilde x < 1.
\end{equation}

The normalized particle velocity yields:
\begin{equation}
\label{v_norm}
\tilde v=\frac{\tilde q}{\tilde w}=-\frac{1}{L}\tilde w^2 \frac{\partial \tilde p}{\partial \tilde x},
\end{equation}
while normalized tangential stress gives:
\begin{equation}
\label{tau_norm}
\tilde \tau (\tilde t, \tilde x)=-\frac{1}{2L}\tilde w \frac{\partial \tilde p}{\partial\tilde x}.
\end{equation}

The equivalent of relation \eqref{elasticity_1a} has the form:
\begin{equation}
\label{elasticity_2}
\tilde p=\frac{1}{L} \int_0^1\left(\frac{k_1}{2}\tilde w \frac{\partial \tilde p}{\partial s}+\frac{\partial \tilde w}{\partial s}\right)
\frac{s}{\tilde x^2 -s^2}ds.
\end{equation}

Equation \eqref{tau_w_S_new} is transformed to:
\begin{equation}
\label{tau_w_S1}
-\frac{k_1}{2}\int_{\tilde x}^1 \tilde w \frac{\partial \tilde p}{\partial s}ds+\tilde w=\frac{2}{\pi^2}L\int_0^1 \tilde p \ln \left|\frac{\sqrt{1-\tilde x^2}+\sqrt{1-s^2}}{\sqrt{1-\tilde x^2}-\sqrt{1-s^2}}\right|ds.
\end{equation}

Relation \eqref{K1_equiv} after normalization assumes the form:
\begin{equation}
\label{K1_equiv_norm}
k_1 \tilde w_0 \tilde p_0+\tilde w_0= \frac{8}{\sqrt{2}\pi^2}L(\tilde t)\int_0^{1}\tilde p(s,\tilde t)\frac{ds}{\sqrt{1-s^2}}.
\end{equation}

Observe that, as shown in \cite{wr_mish_2015}:
\begin{equation}
\label{Log_rel}
\int_0^1 \tilde p \ln \left|\frac{\sqrt{1-\tilde x^2}+\sqrt{1-s^2}}{\sqrt{1-\tilde x^2}-\sqrt{1-s^2}}\right|ds=2\sqrt{1-\tilde x^2}\int_0^{1}\tilde p(s,\tilde t)\frac{ds}{\sqrt{1-s^2}}-\int_0^1 \frac{\partial \tilde p}{\partial s} K(s,\tilde x)ds,
\end{equation}
where the kernel $K(s,\tilde x)$ is:
\begin{equation}
\label{K_def}
K(s,\tilde x)=(s-\tilde x)\ln \left|\frac{\sqrt{1-\tilde x^2}+\sqrt{1-s^2}}{\sqrt{1-\tilde x^2}-\sqrt{1-s^2}}\right|+\tilde x \ln \left|\frac{1+s\tilde x+\sqrt{1-\tilde x^2}\sqrt{1-s^2}}{1+s\tilde x-\sqrt{1-\tilde x^2}\sqrt{1-s^2}}\right|.
\end{equation}

When applying \eqref{K1_equiv_norm} and \eqref{Log_rel} to \eqref{tau_w_S1}, we arrive at an alternative form of the latter, which shall be used later on in computations:
\begin{equation}
\label{tau_w_S2}
-\frac{k_1}{2}\int_{\tilde x}^1 \tilde w \frac{\partial \tilde p}{\partial s}ds+\tilde w=\frac{1}{\sqrt{2}}\tilde w_0(k_1\tilde p_0+1)\sqrt{1-\tilde x^2}-\frac{2}{\pi^2}L\int_0^1 \frac{\partial \tilde p}{\partial s} K(s,\tilde x)ds.
\end{equation}

The normalized influx condition now reads:
\begin{equation}
\label{q_0_norm}
\tilde q(\tilde t,0)=\tilde w(\tilde t,0) \tilde v(\tilde t,0) =\tilde q_0(\tilde t), \quad \tilde t > 0,
\end{equation}
while the respective tip conditions produce:
\begin{equation}
\label{w_q_tip}
\tilde w(\tilde t,1)=0 , \quad \tilde q(\tilde t,1)=0, \quad \tilde t > 0.
\end{equation}

Consequently, the asymptotics of the normalized crack opening, the normalized net fluid pressure and the normalized tangential traction now become:
\begin{equation}
\label{w_norm_asymp}
\tilde w(\tilde t, \tilde x)=\tilde w_0(\tilde t)\sqrt{1-\tilde x}+\tilde w_1(\tilde t)(1-\tilde x)+\tilde w_2(\tilde t)(1-\tilde x)^{3/2}\ln(1-\tilde x)+...,\quad \tilde x \to 1,
\end{equation}
\begin{equation}
\label{p_norm_asymp}
\tilde p(\tilde t, \tilde x)=\tilde p_0(\tilde t)\ln(1-\tilde x)+\tilde p_1(\tilde t)+\tilde p_2(\tilde t)\sqrt{1-\tilde x}+\tilde p_3(\tilde t)(1-\tilde x)\ln(1-\tilde x)+..., \quad \tilde x \to 1.
\end{equation}
\begin{equation}
\label{tau_asymp_norm}
\tilde \tau(\tilde t,\tilde x)=\frac{\tilde \tau_0(\tilde t)}{\sqrt{1-\tilde x}}+\tilde \tau_1(\tilde t)+\tilde \tau_2(\tilde t)\sqrt{1-\tilde x}\ln(1-\tilde x)+\tilde \tau_3\sqrt{1-\tilde x}+...,\quad \tilde x\to 1.
\end{equation}

\subsubsection{Normalized fracture criterion. Evaluation of the crack length}

The fracture propagation criterion \eqref{ERR_def_1} is converted, in light of \eqref{ERR_def}, to its dimensionless form:
\begin{equation}
\label{E_dim}
\tilde K_{IC}^2=\tilde K_{I}^2+a\tilde K_{I}\tilde K_f,
\end{equation}
or equivalently
\begin{equation}
\label{frac_cr}
\tilde w_0=\sqrt{L}\frac{1+\tilde \varpi}{\sqrt{1+a\tilde \varpi}}\tilde K_{IC}=\sqrt{L}(\tilde K_{I}+\tilde K_f),
\end{equation}
where
\begin{equation}
\label{om_1}
\tilde \varpi=\frac{\tilde p_0}{\pi(1-\nu)-\tilde p_0},\quad 0<\tilde p_0<\pi(1-\nu).
\end{equation}

As a result, equations \eqref{criterion_1} transform to their normalised equivalents
\begin{equation}
\label{KK_dim}
\tilde K_I=\frac{\tilde K_{IC}}{\sqrt{1+a\tilde \varpi}},\quad \tilde K_f=\frac{\tilde K_{IC}\tilde \varpi}{\sqrt{1+a\tilde \varpi}}.
\end{equation}

By using asymptotic expressions \eqref{w_norm_asymp} -- \eqref{p_norm_asymp}, in combination with definition of the particle velocity, we can convert the Stefan condition \eqref{SE} to:
\begin{equation}
\label{SE_2}
L'\equiv\tilde v_0=-\frac{1}{L}\lim_{\tilde x\to1}\tilde w^2 \frac{\partial \tilde p}{\partial \tilde x}=\frac{\tilde w_0^2 \tilde p_0}{L}.
\end{equation}

When combining \eqref{frac_cr} with \eqref{SE_2}, one obtains a relation for the crack propagation speed in the form:
\begin{equation}
\label{v_0}
\frac{1}{\tilde K^2_{IC}}\tilde v_0(\tilde t)=\tilde p_0 F(\tilde p_0),
\end{equation}
where:
\begin{equation}
\label{F_p}
\quad F(\tilde p_0)=\frac{\pi^2(1-\nu)^2}{\big[\pi(1-\nu)+(3-4\nu)\tilde p_0\big]\big[\pi(1-\nu)-\tilde p_0\big]}.
\end{equation}
The right-hand side of \eqref{v_0}  increases monotonically from zero to infinity over the interval $\tilde p_0\in\big(0,\pi(1-\nu)\big)$, which is a computationally important property.
Taking \eqref{SE_2} into account, one can also rewrite \eqref{v_0} in the form:
\begin{equation}
\label{w_0_L}
\tilde w_0(\tilde t)=\tilde K_{IC}\sqrt{L(\tilde t) F(\tilde p_0)}.
\end{equation}

Equation \eqref{SE_2} can be integrated with respect to time to give a universal formula for the computation of the crack length:
\begin{equation}
\label{speed_2}
L(\tilde t)=\sqrt{L^2(0)+2\int_0^{\tilde t}\tilde w^2_0(\tau)\tilde p_0(\tau)d\tau}.
\end{equation}
Indeed, such an approach has  been effectively used in numerous papers of the authors (see e.g. \cite{M_W_L,Kusmierczyk,wr_mish_2015}), where its advantages have been thoroughly discussed. If required, the crack length can alternatively be determined from \eqref{v_0} or \eqref{frac_cr} as:
\begin{equation}
\label{speed_0}
L(\tilde t)=L(0)+K_{IC}^2\int_0^{\tilde t} \tilde p_0(\tau)F(\tilde p_0(\tau))d\tau,
\end{equation}
\begin{equation}
\label{speed_1}
L(\tilde t)=\frac{\tilde w^2_0 [1+a\tilde \varpi(\tilde p_0)]}{\tilde K^2_{Ic}[1+\tilde \varpi(\tilde p_0)]^2}.
\end{equation}
The first formula is suitable for small $K_{IC}$, while the second one is more convenient for larger values of $K_{IC}$.

\subsubsection{Viscosity dominated regime ($K_{IC}=0$)}

We will now examine in more detail one of the limiting modes of crack propagation - the so called viscosity dominated regime.  In such a case it is assumed that: $\tilde K_{IC}\to0$.
This implies that  $\tilde p_0\to\pi(1-\nu)$ (compare to \eqref{v_0} - \eqref{F_p}). As a result, the following estimates hold:
\begin{equation}
\label{p0_visc}
\tilde p_0=\pi(1-\nu)\left(1-\frac{1}{4(1-\nu)}\frac{\tilde K_{IC}^2L(\tilde t)}{\tilde w_0^2}\right)+O\left(\frac{\tilde K_{IC}^4L^2(\tilde t)}{\tilde w_0^4}\right),
\end{equation}
\begin{equation}
\label{om_visc}
\varpi=\frac{a\tilde w_0^2}{\tilde K^2_{IC}L}+O(1),\quad \tilde K_{IC}\to0,
\end{equation}
where $a$ is defined in (\ref{a_def}). When substituting \eqref{om_visc} into (\ref{KK_dim}), we have
\[
\tilde K_{I}\sim \tilde K^2_{IC}\frac{\sqrt{L}}{a\tilde w_0},\quad  \tilde K_{f}\sim  \frac{\tilde w_0}{\sqrt{L}},\quad \tilde K_{IC}\to0.
\]
Thus, all relationships following from the ERR fracture propagation criterion reach their limit values as $\tilde K_{IC}\to0$, while equation (\ref{w_0_L}) degenerates to identity. As a result, one has in this case:
\begin{equation}
\label{fluid_1}
\tilde p_0=\pi(1-\nu),\quad
\tilde K_{I}=0,\quad \tilde K_{f}=\frac{\tilde w_0}{\sqrt{L}},\quad \tilde v_0(\tilde t)=\pi(1-\nu)\tilde K_f^2,
\end{equation}
\begin{equation}
\label{fluid_2}
\tilde v_0(\tilde t)=\frac{\pi(1-\nu)\tilde w_0^2(\tilde t)}{L(\tilde t)},\quad
L(\tilde t)=\sqrt{L^2(0)+2\pi(1-\nu)\int_0^{\tilde t}\tilde w^2_0(\tau)d\tau}.
\end{equation}

All the other equations that constitute the revised HF model as well as the near-tip asymptotic
behaviour of the solution remain the same. The latter feature creates a fundamental difference as compared to the classic KGD model, where the transition to the viscosity dominated regime entails change in the solution tip asymptotics.

In the following we omit the $\hspace{1mm}\tilde{} \hspace{1mm}$ symbol for convenience. All quantities refer henceforth to the normalized ones.

\section{Self-similar formulation}

In this section, we introduce a self-similar formulation of the revised HF model being considered. The self-similar version of the problem will be solved by means of the particle velocity based universal algorithm for numerical simulation of hydraulic fractures as originally proposed in \cite{wr_mish_2015}. The accuracy of computations will be investigated via comparison to a specifically developed analytical benchmark.

When considering the problem of HFs, there are usually two self-similar formulations available, where the time-dependent component is described either by the power law or exponential function (see e.g. \cite{Spence&Sharp}). In the revised HF model, incorporating the tangential traction on the fracture walls, it is generally not possible (except in the case of incompressible material: $\nu=0.5$, $k_1=0$) to derive the equivalent of the former variant. This is due to the form of the elasticity operator \eqref{elasticity_2}.
For this reason, we consider only the second available option and assume that the imposed influx magnitude is defined as:
\begin{equation}
\label{q0_exp}
q_0(t)=\bar q_0e^{2\alpha t},
\end{equation}
where $q_0$ and $\alpha$ are some positive constants.

Conventionally, we search for a solution in the form:
\begin{equation}
\label{sol_ss_1}
w(t,x)=\sqrt{L_0\bar q_0}e^{\alpha t}\hat w(x), \quad L(t)=L_0^{3/2}e^{\alpha t}\sqrt{\bar q_0},\quad
p(t,x)=\hat p(x),\quad q_l(t,x)=\alpha e^{\alpha t}\sqrt{L_0 \bar q_0}\hat q_l(x),
\end{equation}
\begin{equation}
\label{sol_ss_2}
\tilde {\cal E}=e^{\alpha t}\sqrt{L_0\bar q_0}\hat {\cal E},\quad v(t,x)=\sqrt{\frac{\bar q_0}{L_0}}e^{\alpha t} \hat v(x),\quad  q(t,x)=\bar q_0e^{2\alpha t}\hat q(x),\quad
\tau(t,x)=\frac{1}{L_0}\hat \tau(x),
\end{equation}
\[
\tilde K_{\{Ic,I,f\}}=(L_0 \bar q_0)^{1/4}e^{\frac{\alpha t}{2}}\hat K_{\{Ic,I,f\}}.
\]

We have introduced the notations:
\begin{equation}
\label{sol_ss_3}
\hat v_0=\hat v(1),\quad L_0=\sqrt{\frac{\hat v_0}{\alpha}},
\end{equation}
where the self-similar particle velocity is defined in the following way:
\begin{equation}
\label{v_ss}
\hat v(x)=-\hat w^2(x)\hat p'(x).
\end{equation}

\noindent
Similarly, we define the self-similar fluid flow rate, $\hat q(x)$, and the self-similar tangential stress $\hat \tau(x)$:
\begin{equation}
\hat q(x)=-\hat w^3(x) \hat p'(x),\quad
\label{tau_ss}
\hat \tau(x)=-\frac{1}{2}\hat w(x) \hat p'(x).
\end{equation}

\noindent
The dependent variables are interrelated via the continuity equation:
%The continuity equation \eqref{cont_norm} reduces to the following ODE:
\begin{equation}
\label{cont_ssim}
\hat w -x\frac{d \hat w}{dx}+\frac{1}{\hat v_0}\frac{d \hat q}{dx}+\hat q_l=0,
\end{equation}
and the self-similar analogue of the elasticity operator:
\begin{equation}
\label{elasticity_3}
\hat p=\frac{1}{L_0} \int_0^1\left(\frac{k_1}{2}\hat w \frac{d \hat p}{d s}+\frac{d \hat w}{d s}\right)
\frac{s}{ x^2 -s^2}ds.
\end{equation}

\noindent
The following boundary conditions hold:
\begin{equation}
\label{q_0ss}
\hat q(0)=\hat q_0=1, \quad \hat w(1)=0.
\end{equation}

The self-similar crack opening, the net fluid pressure and the shear stress exhibit the following asymptotic behaviour near the crack tip:
\begin{equation}
\label{w_ss_asymp}
\hat w(x)=\hat w_0\sqrt{1-x}+\hat w_1(1-x)+\hat w_2(1-x)^{3/2}\ln(1-x)+...,\quad x\to 1,
\end{equation}
\begin{equation}
\label{p_ss_asymp}
\hat p(x)=\hat p_0\ln(1-x)+\hat p_1+\hat p_2\sqrt{1-x}+\hat p_3(1-x)\ln(1-x)+...,\quad x\to 1.
\end{equation}
\begin{equation}
\label{tau_asymp}
\hat \tau(x)=\frac{\hat \tau_0}{\sqrt{1-x}}+\hat \tau_1+\hat \tau_2\sqrt{1-x}\ln(1-x)+\hat \tau_3\sqrt{1-x}+...,\quad x\to 1.
\end{equation}
As a result, one can compute (compare to \eqref{sol_ss_3})
\begin{equation}
\label{v_0_ss}
\hat v_0=\hat v(1)=\hat w_0^2 \hat p_0, \quad
L_0=\hat w_0\sqrt{\frac{\hat p_0}{\alpha}},\quad \hat \tau_0=\frac{1}{2}\hat p_0 \hat w_0.
\end{equation}

As it was mentioned above, the inverse formula \eqref{tau_w_S2} is more suitable for computation and its self-similar version takes the form:
\begin{equation}
\label{el_inv_ss}
-\frac{k_1}{2}\int_{x}^1 \hat w \hat p'ds+\hat w=\frac{1}{\sqrt{2}}\hat w_0(k_1\hat p_0+1)\sqrt{1-x^2}-\frac{2}{\pi^2}L_0\int_0^1 \hat p' K(s,x)ds,
\end{equation}
while the counterpart of \eqref{K1_equiv_norm} yields:
\begin{equation}
\label{K1_equiv_ss}
k_1 \hat w_0 \hat p_0+\hat w_0= \frac{8}{\sqrt{2}\pi^2}L_0\int_0^{1}\hat p(s)\frac{ds}{\sqrt{1-s^2}}.
\end{equation}

Similarly to as in the classic KGD model (compare \cite{Spence&Sharp}), the self-similar formulation of the type \eqref{sol_ss_1}--\eqref{sol_ss_2} can be obtained only if the material toughness is a function of time proportional to $\sqrt{L}$. The pertinent assumption has already been introduced in \eqref{sol_ss_2}. When accounting for this feature, the criterion \eqref{frac_cr} converts to:
\begin{equation}
\label{frac_cr_ss}
\hat w_0=\sqrt{L_0}\frac{1+ \varpi}{\sqrt{1+a  \varpi}}\hat K_{IC}=\sqrt{L_0}(\hat K_{I}+\hat K_f),
\end{equation}
where $\varpi$ is computed from (\ref{om_1}) with $\tilde p_0$ replaced by $\hat p_0$.
The interrelations (\ref{E_dim}) and (\ref{KK_dim}) between $\hat K_{Ic}$ and $\hat K_f$ remain unchanged if superscript '$\tilde{\textcolor[rgb]{1.00,1.00,1.00}{a}}$' is replaced by '$\hat{\textcolor[rgb]{1.00,1.00,1.00}{a}}$'.

When combining criterion \eqref{frac_cr_ss} with the condition \eqref{v_0_ss}$_2$ one obtains the following formula for $\hat p_0$:
\begin{equation}
\label{p_0_ss}
(4\nu-3)\hat p_0^2+\pi(1-\nu)(2-4\nu)\hat p_0 -\frac{\pi^2 (1-\nu)^2 \hat K_{Ic}^2}{\sqrt{\alpha}\hat w_0}\sqrt{\hat p_0}+\pi^2(1-\nu)^2=0.
\end{equation}
It can be proved that equation \eqref{p_0_ss}  has a unique solution in the interval \eqref{om_1}$_2$.

Note that for $\hat K_{Ic}=0$ we have:
\begin{equation}
\label{p0_KI_0}
\hat p_0=\pi(1-\nu),\quad
\hat v_0=\pi(1-\nu)\hat w_0^2.
\end{equation}

A natural consequence of \eqref{p0_KI_0}$_2$ is that for the propagating fracture ($\hat v_0>0$) one has $\hat w_0>0$, and thus the LEFM asymptote zone is never reduced to zero.

\section{Numerical results}

In the first part of this section, we verify the accuracy of computations provided by the employed algorithm against an analytical benchmark example. The computational scheme is based on the algorithm proposed in \cite{wr_mish_2015}. Then, having checked the credibility of our numerical results, we shall investigate the consequences of using the modified HF formulation. To this end the following variants of the problem will be compared: i) the modified HF model, ii) the classic KGD model, iii) the KGD model with a new fracture propagation criterion \eqref{frac_cr_ss}. We will analyse some aspects of the modified HF formulation.

\subsection{Accuracy analysis}

In \cite{wr_mish_2015} a set of analytical benchmark solutions was proposed for the KGD model. Unfortunately, when replacing the original KGD elasticity operator with its modified form \eqref{elasticity_3}, one can no longer obtain an analytical benchmark solution using the method given in \cite{wr_mish_2015}. That is why, in order to establish the accuracy of our computations, we now slightly amend  the basic system of equations in a manner which enables us to adopt the benchmark for the toughness dominated regime from \cite{wr_mish_2015}. Specifically, we modify the inverse elasticity equation \eqref{el_inv_ss} in the following way:
\begin{equation}
\label{el_inv_bench}
-\frac{k_1}{2}\int_{x}^1 \hat w \hat p'ds+\hat w=\frac{1}{\sqrt{2}}\hat w_0(k_1\hat p_0+1)\sqrt{1-x^2}-\frac{2}{\pi^2}L_0\int_0^1 \hat p' K(s,x)ds +\Psi(x),
\end{equation}
where $\Psi(x)$ is a known predefined function given by:
\begin{equation}
\label{psi_def}
\Psi(x)=-\frac{k_1}{2}\int_x^1 \hat w \hat p'ds -\frac{k_1}{\sqrt{2}}\hat w_0 \hat p_0.
\end{equation}
Provided that \eqref{el_inv_bench}-\eqref{psi_def} are satisfied, the analytical benchmark solution delivered in \cite{wr_mish_2015} (pp. 56-57 therein) holds. The new fracture propagation condition \eqref{frac_cr_ss} can be imitated by an appropriate combination of the base functions.

The computations are executed according to the algorithm developed in \cite{wr_mish_2015} which is based on two dependent variables: the crack opening and the so-called reduced particle velocity ($\hat v- x \hat v_0$). The only modification of the numerical scheme results from the need to compute an additional integral of the product $\hat w \hat p'$ in \eqref{el_inv_ss} and the inclusion of the predefined function $\Psi (x)$.

\begin{figure}[h!]
%M/N=1/300

    %\hspace{-2mm}
    \includegraphics [scale=0.40]{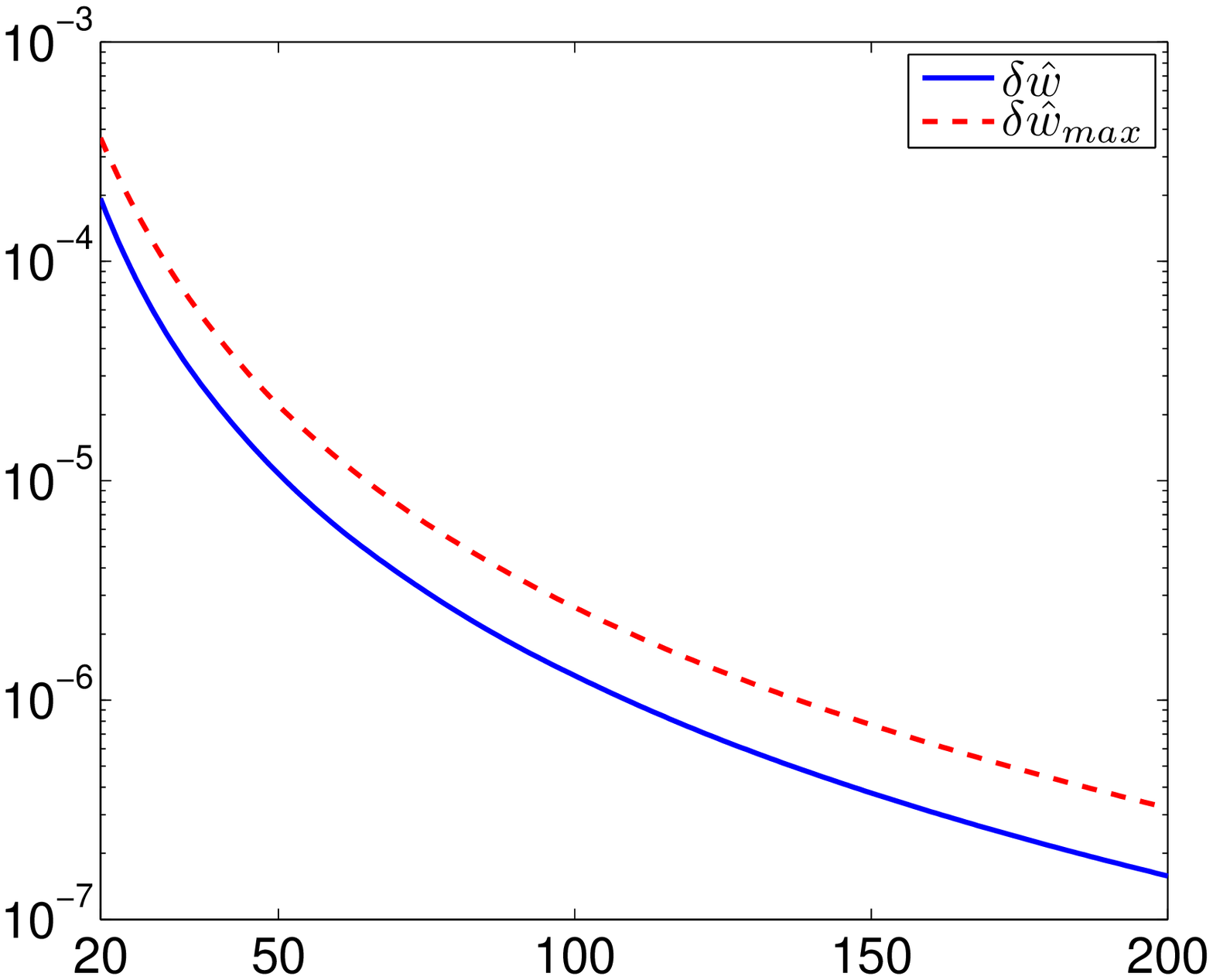}
    \put(-105,0){$N$}
    \put(-230,90){$\delta \hat w$}
    \put(-230,160){$\textbf{a)}$}
%M/N=1/30
    \hspace{2mm}
    \includegraphics [scale=0.40]{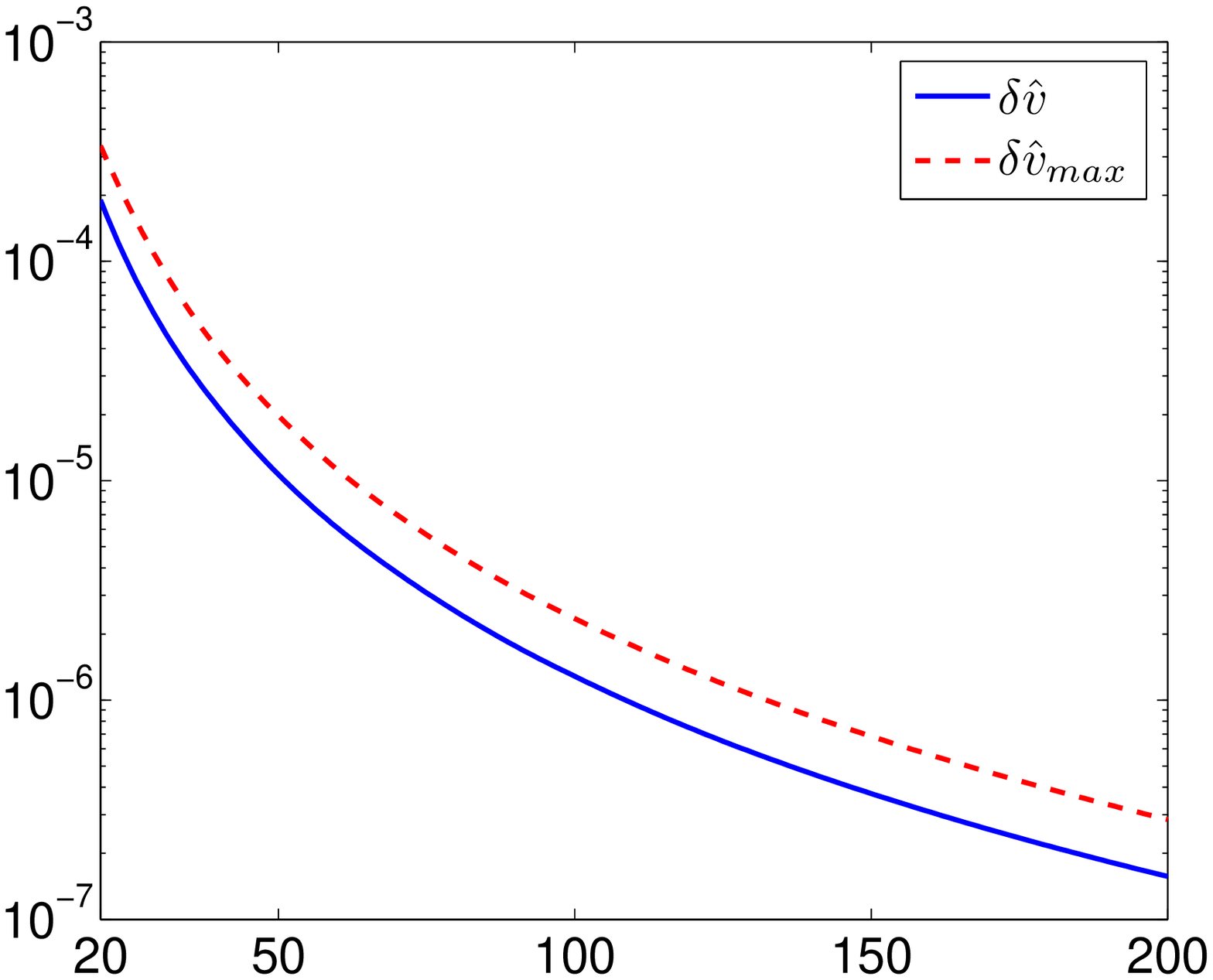}
    \put(-105,0){$N$}
    \put(-230,90){$\delta \hat v$}
    \put(-230,160){$\textbf{b)}$}

    \caption{Average and maximal relative errors of the self-similar solution as a function of the number of nodal points $N$ for: a) the crack opening $\hat w$, b) the particle velocity $\hat v$. }

\label{bledy_od_N}
\end{figure}

\begin{figure}[h!]
 \centering
    \includegraphics[scale=0.40]{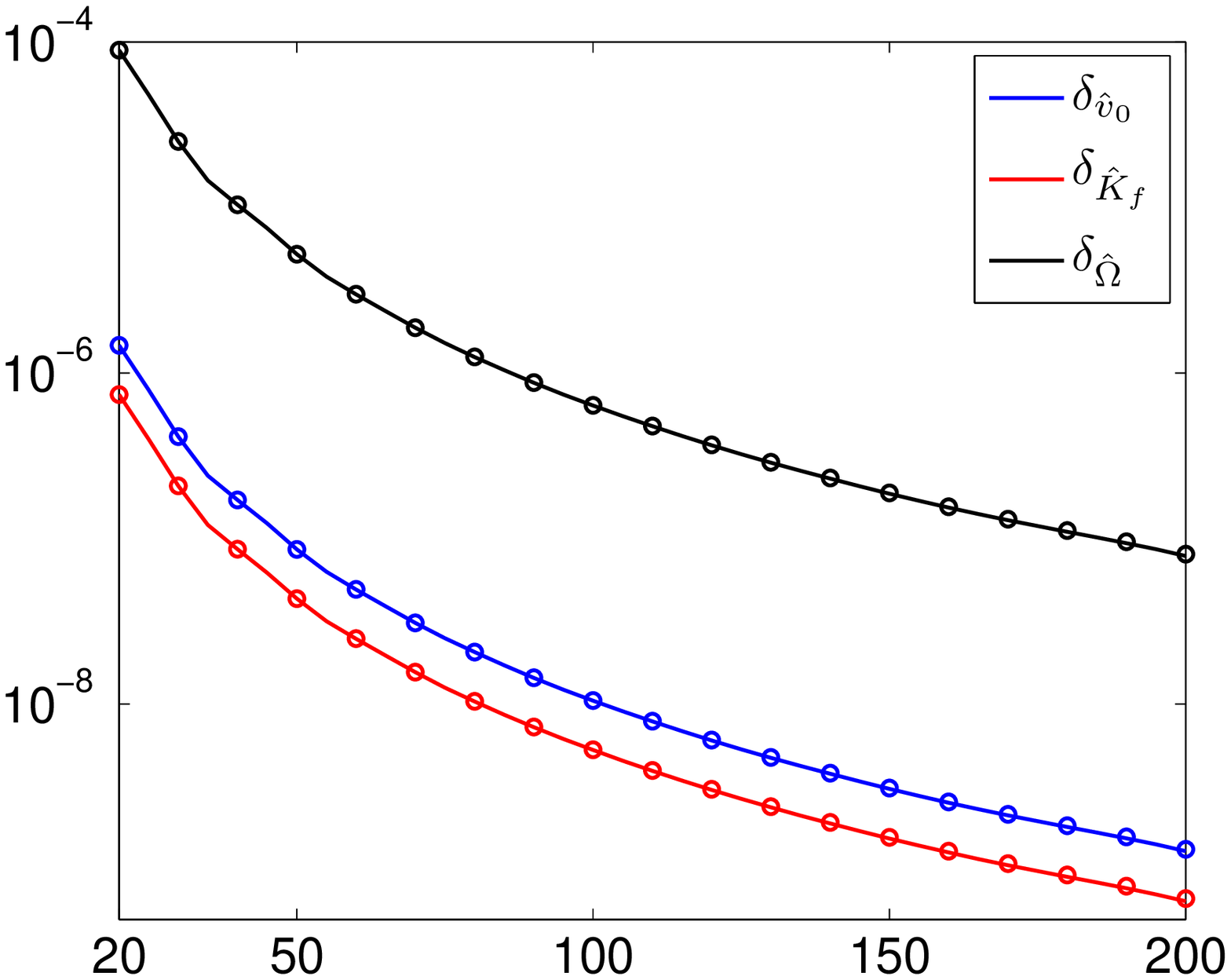}
    \put(-105,0){$\hat N$}
    \put(-230,90){$\delta f$}
 \caption{The convergence of the solution for growing number of nodal points $N$. Solid lines depict the relative errors. Markers correspond to relative deviations of respective parameters from their asymptotic values ($N\to \infty$), obtained numerically by the least-square method.}
 \label{zbieznosc_bench}
\end{figure}

\begin{figure}[h!]
%M/N=1/300

    %\hspace{-2mm}
    \includegraphics [scale=0.40]{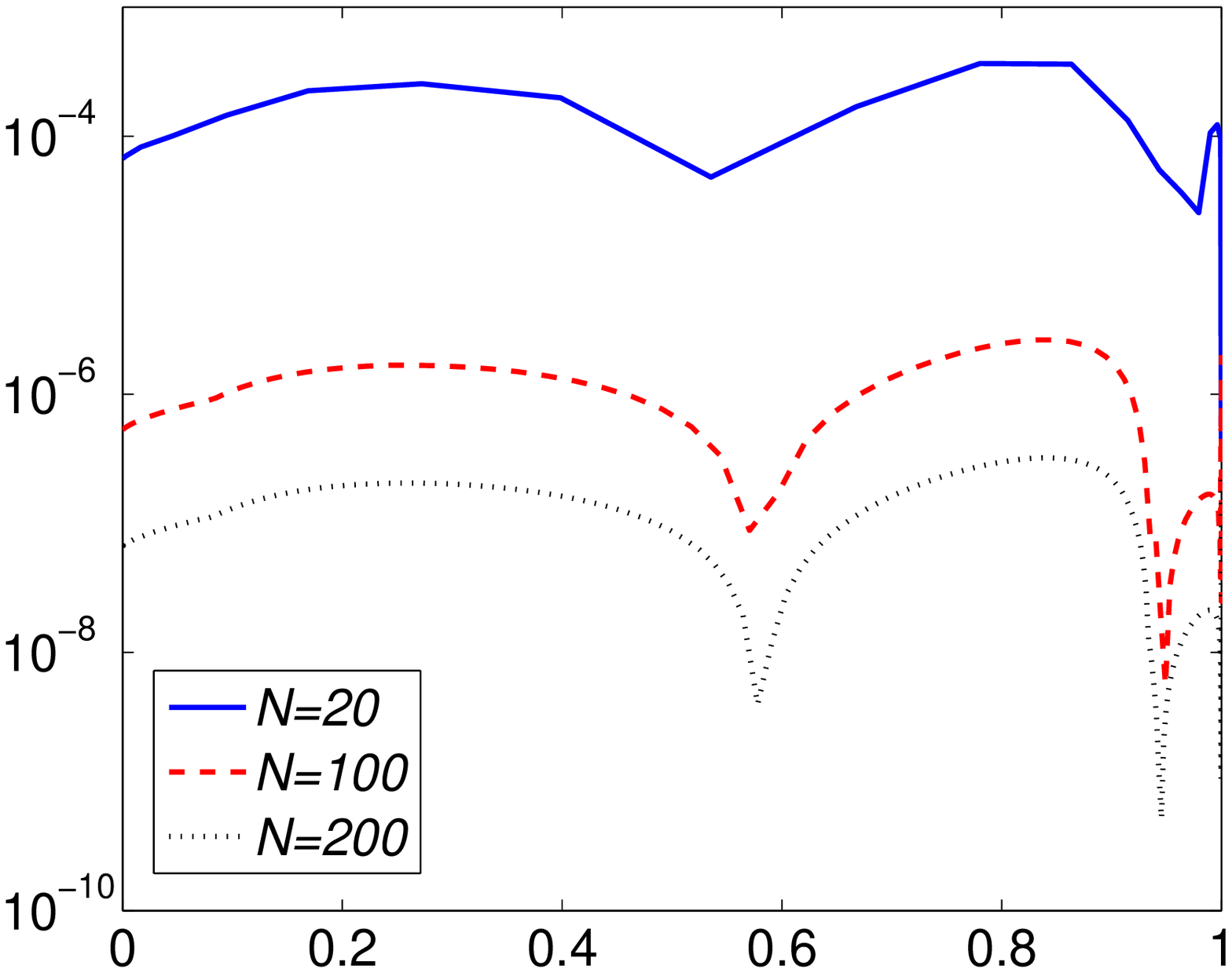}
    \put(-105,0){$x$}
    \put(-230,90){$\delta \hat w$}
    \put(-230,160){$\textbf{a)}$}
%M/N=1/30
    \hspace{2mm}
    \includegraphics [scale=0.40]{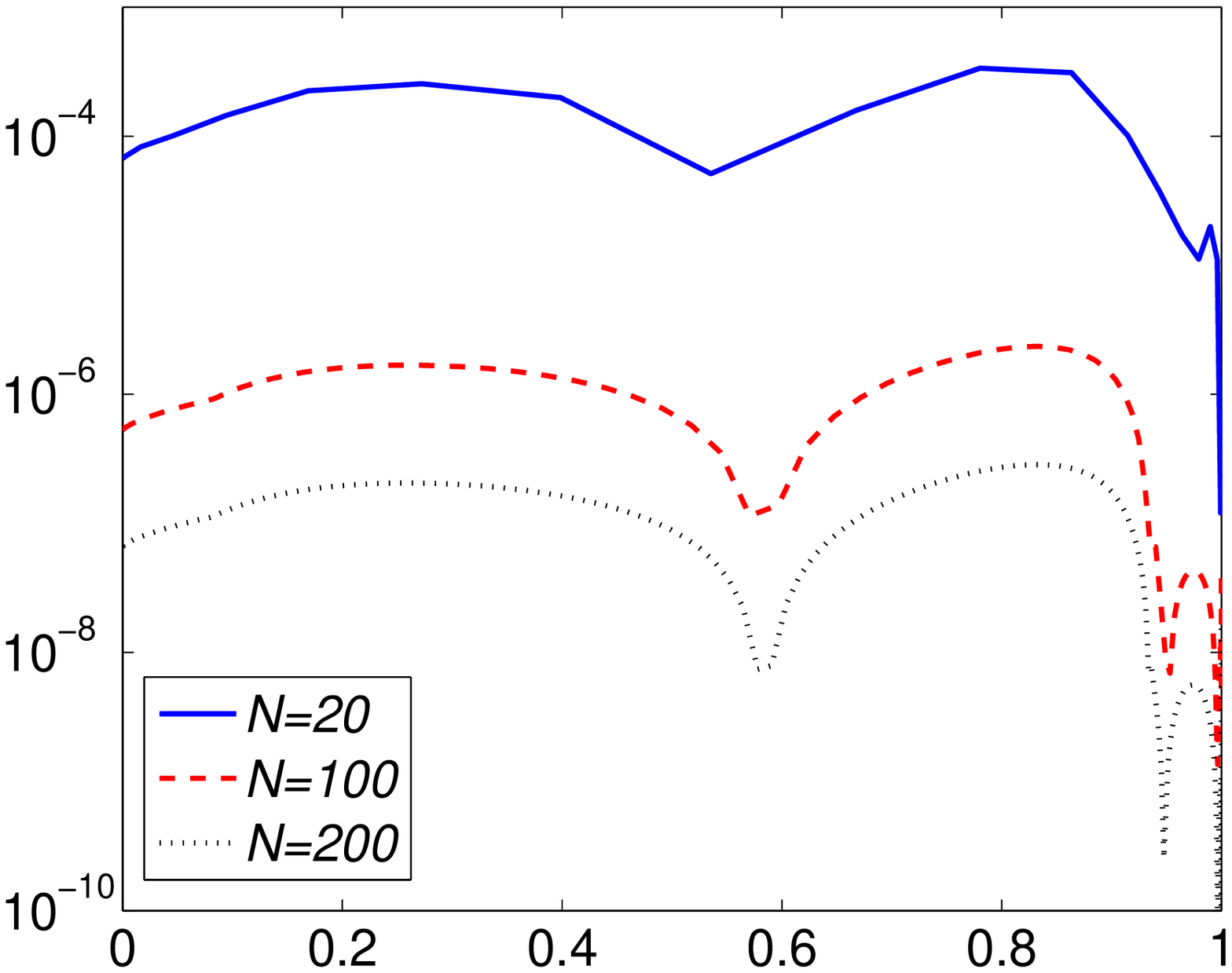}
    \put(-105,0){$x$}
    \put(-230,90){$\delta \hat v$}
    \put(-230,160){$\textbf{b)}$}

    \caption{The spatial distribution of the relative errors of the solution for different numbers of nodal points: a) the crack opening $\hat w$, b) the particle velocity $\hat v$. }

\label{bledy_od_x}
\end{figure}

The results of computations, illustrated by the errors of the crack opening and the particle velocity, are depicted in Figs.\ref{bledy_od_N}-\ref{bledy_od_x}.

In Fig.\ref{bledy_od_N} we show the average (over $x$) and maximal relative errors of solution for varying numbers of nodal points ranging from $N=20$ to $N=200$. The average errors are defined as:
\begin{equation}
\label{bl_av}
\delta g=\sqrt{\int_0^1\left(g-g_n\right)^2dx \left(\int_0^1 g^2dx\right)^{-1}}, \quad g=\{\hat w, \hat v\},
\end{equation}
where the subscript $n$ refers to the numerical solution.

As can be seen, the values of both errors ($\delta \hat w$ and $\delta \hat v$) are very similar, and even for $N=20$ nodal points, the solution accuracy is extremely good (errors much below $0.1\%$). Over the analysed range of $N$, an increase in the number of nodal points entails a monotonic reduction of the errors. It has been established numerically that the convergence of errors with growing $N$ can be described by the relation:
\begin{equation}
\label{err_conv}
\delta g(N)=\frac{d_g}{N^3}, \quad g=\{\hat w, \hat v\},
\end{equation}
where $d_g$ are some constants pertaining to the analysed accuracy parameters.

We also analyse here the convergence of other solution parameters. We consider: i) the self-similar crack propagation speed, $\hat v_0$, ii) the self-similar shear stress intensity factor, $\hat K_f$, iii) the self-similar fracture volume, $\Omega$.  The latter is defined as:
\begin{equation}
\label{om_def}
\Omega =\int_0^1 \hat w(x)dx.
\end{equation}
This shows that the rate of convergence is the same as that of the solution error (see \eqref{err_conv}). In general, the bahaviour of each of the considered parameters can be described by:
\begin{equation}
\label{bench_conv}
f_n=\frac{d_f}{N^3}+f_e, \quad f=\{\hat v_0, \hat K_f,  \Omega\}
 \end{equation}
 where $d_f$ is some constant and the subscripts respectively refer to: '$n$' - numerical value, '$e$' - exact (analytical) value. The results are illustrated in Fig. \ref{zbieznosc_bench}. Solid lines refer to the relative errors of corresponding parameters $\delta f=|f_n-f_e|/f_e$. In the same figure we have used markers to show the relative deviations of $f_n$ from their asymptotic values, obtained numerically by the least-square method when assuming representation \eqref{bench_conv} (in such a case $f_e$ is considered an asymptotic value for $N\to \infty$). As can be seen, the markers coincide perfectly with the respective lines. Thus, we can conclude that in the non-benchmark cases, when the exact solution is unknown, the convergence test can be successfully applied to estimate the errors of computations.

To complement this subsection, we present in Fig.\ref{bledy_od_x} the spatial distribution of the relative errors of $\hat w $ and $\hat v$ for three values of $N$ ($N=20,100,200$). This shows that the solution is very accurate even at the crack tip, which has been proved in  \cite{wr_mish_2015} to be crucial for the stability and accuracy of computations in the transient case.

In the following computations we set $N=100$ which, according to the above analysis, guarantees the accuracy of solution at the level $10^{-6}$ for both the crack opening, $\hat w$, and the particle velocity, $\hat v$.

\subsection{Modified HF formulation - comparison with the classic KGD and the amended KGD models }

Having presented the accuracy of computations using the applied numerical scheme we can now employ it to assess the computational ramifications of the introduction of the modified HF model. To this end we will compare the numerical results for three variants of the problem:
\begin{itemize}
\item{Variant 1: the modified HF formulation described by equations \eqref{q0_exp}--\eqref{p_0_ss}.

This is the most complete variant which includes the hydraulically induced tangential traction in the elasticity operator and appropriate modification of the ERR fracture propagation criterion.}

\item{Variant 2: the classic KGD model.}

In this case equations \eqref{elasticity_3}--\eqref{K1_equiv_ss} hold, provided that $k_1$ is set to zero. The classic crack propagation condition is adopted, which
 assumes equality of the stress intensity factor and material toughness. In the self-similar formulation this yields:
\begin{equation}
\label{farc_cond_KGD}
\hat K_I=\hat K_{Ic}.
\end{equation}
As a result, we have a direct relationship between the crack opening and the stress intensity factor (instead of the general relationship \eqref{p_0_ss}):
\begin{equation}
\label{w0_KGD}
\hat w_0=\sqrt{L_0}\hat K_I.
\end{equation}
\item{Variant 3: the amended KGD formulation.}

Here the elasticity operator is taken directly from the classic KGD model, while the ERR fracture propagation condition is adopted from the modified HF formulation.
This variant of the problem is considered as the intermediate version between Variant 1 and Variant 2. Note that when Poisson's ratio is set to $\nu=0.5$ ($k_1=0$), Variant 1 reduces to Variant 3.
\end{itemize}

\begin{figure}[h!]
 \centering
    \includegraphics[scale=0.40]{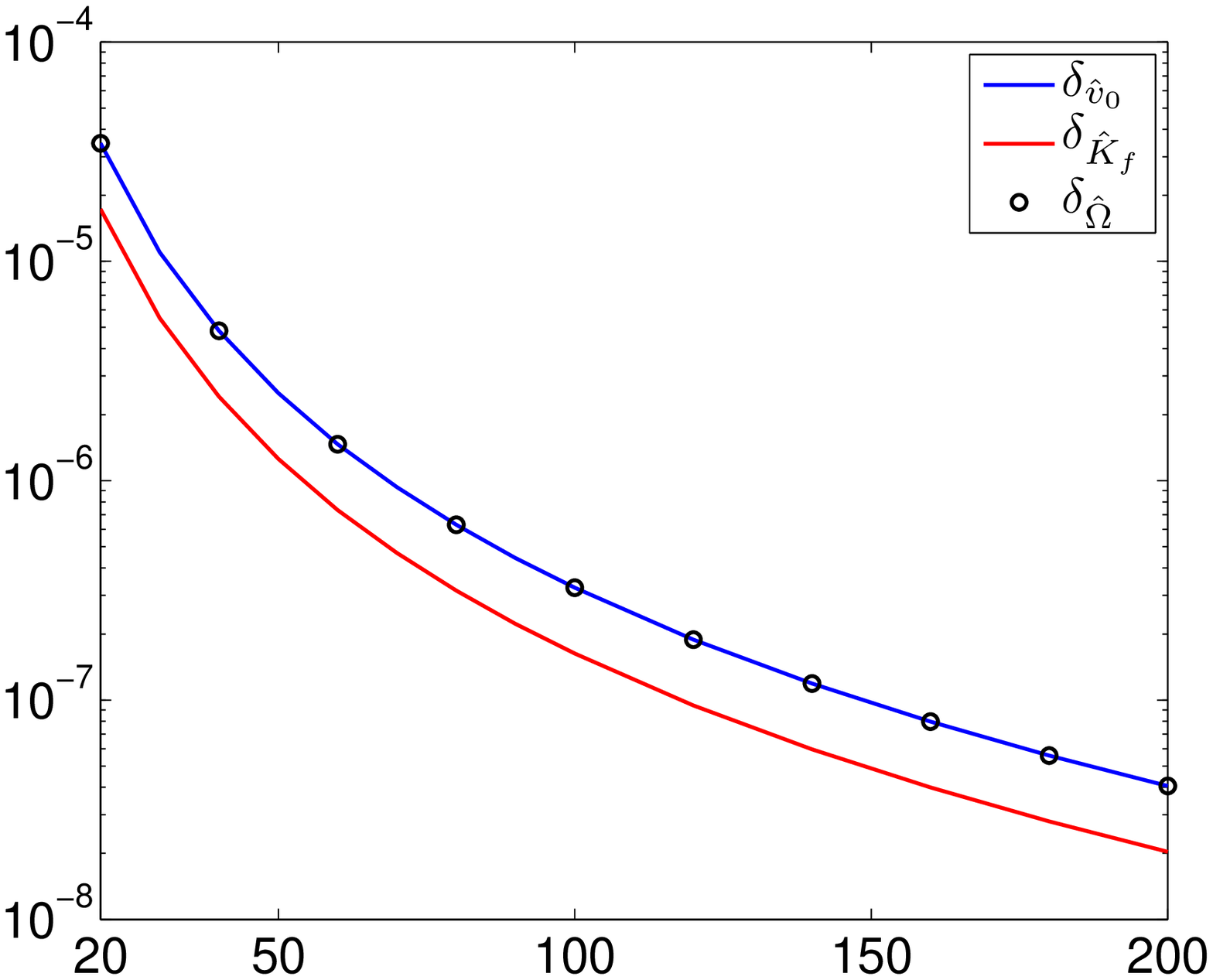}
    \put(-105,0){$\hat N$}
    \put(-230,90){$\delta f$}
 \caption{Modified HF formulation. The convergence of the solution for growing number of nodal points $N$. Presented data corresponds to relative deviations of respective solution parameters from their asymptotic values ($N\to \infty$), obtained numerically by the least-square method.}
 \label{zbieznosc_nb}
\end{figure}

In the following we shall solve the self-similar problem defined in the previous section for different values of Poisson's ratio (and thus for different values of coefficient $k_1$) and the self-similar material toughness $\hat K_{Ic}$. Impermeability of the solid material is assumed ($\hat q_l=0$). The self-similar parameter $\alpha$ is set to $1/3$. Solutions for the three variants mentioned above will be delivered and compared. The solution of  Variant 1 will be treated as a reference solution.
In order to confirm its credibility we present the results of a convergence test similar to that shown in the previous subsection. Obviously, on this occasion, only the relative deviations from the numerically obtained asymptotic values are depicted. The data given in Fig.\ref{zbieznosc_nb} corresponds to $\delta \hat v_0$, $\delta \hat K_f$ and $\delta \Omega$, respectively. It shows that the rate of convergence remains as in \eqref{bench_conv}. Surprisingly, unlike the benchmark case, this time  the level of $\delta \Omega$ does not differ appreciably from the two remaining parameters, being practically identical to $\delta \hat v_0$. The presented data provides an estimate of the accuracy of the numerical solution.

\begin{figure}[h!]
%M/N=1/300

    %\hspace{-2mm}
    \includegraphics [scale=0.40]{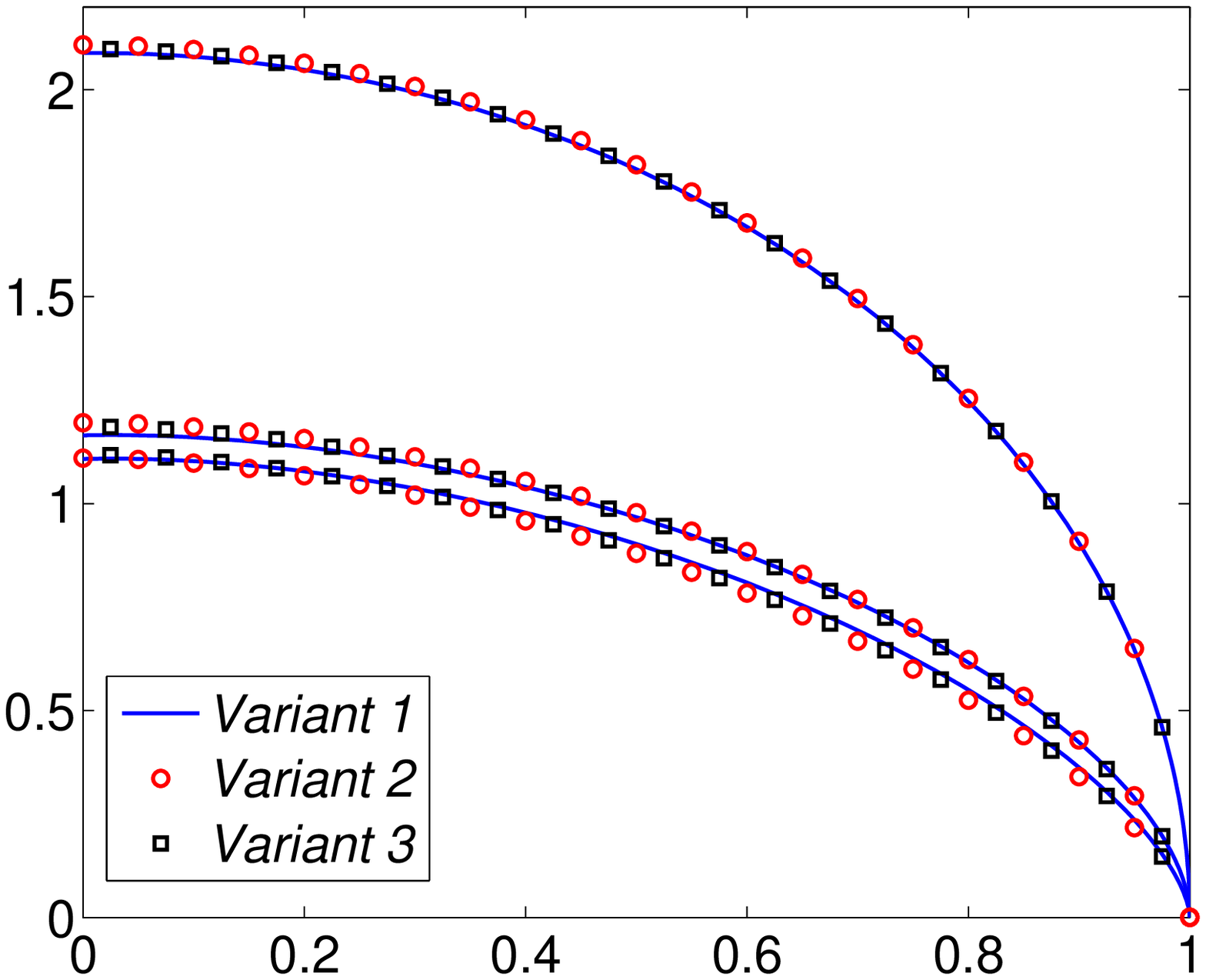}
    \put(-105,0){$x$}
    \put(-220,90){$\hat w$}
    \put(-90,130){$\hat K_{Ic}=3$}
    \put(-120,90){$\hat K_{Ic}=1$}
    \put(-140,65){$\hat K_{Ic}=0$}
    \put(-230,160){$\textbf{a)}$}
%M/N=1/30
    \hspace{2mm}
    \includegraphics [scale=0.40]{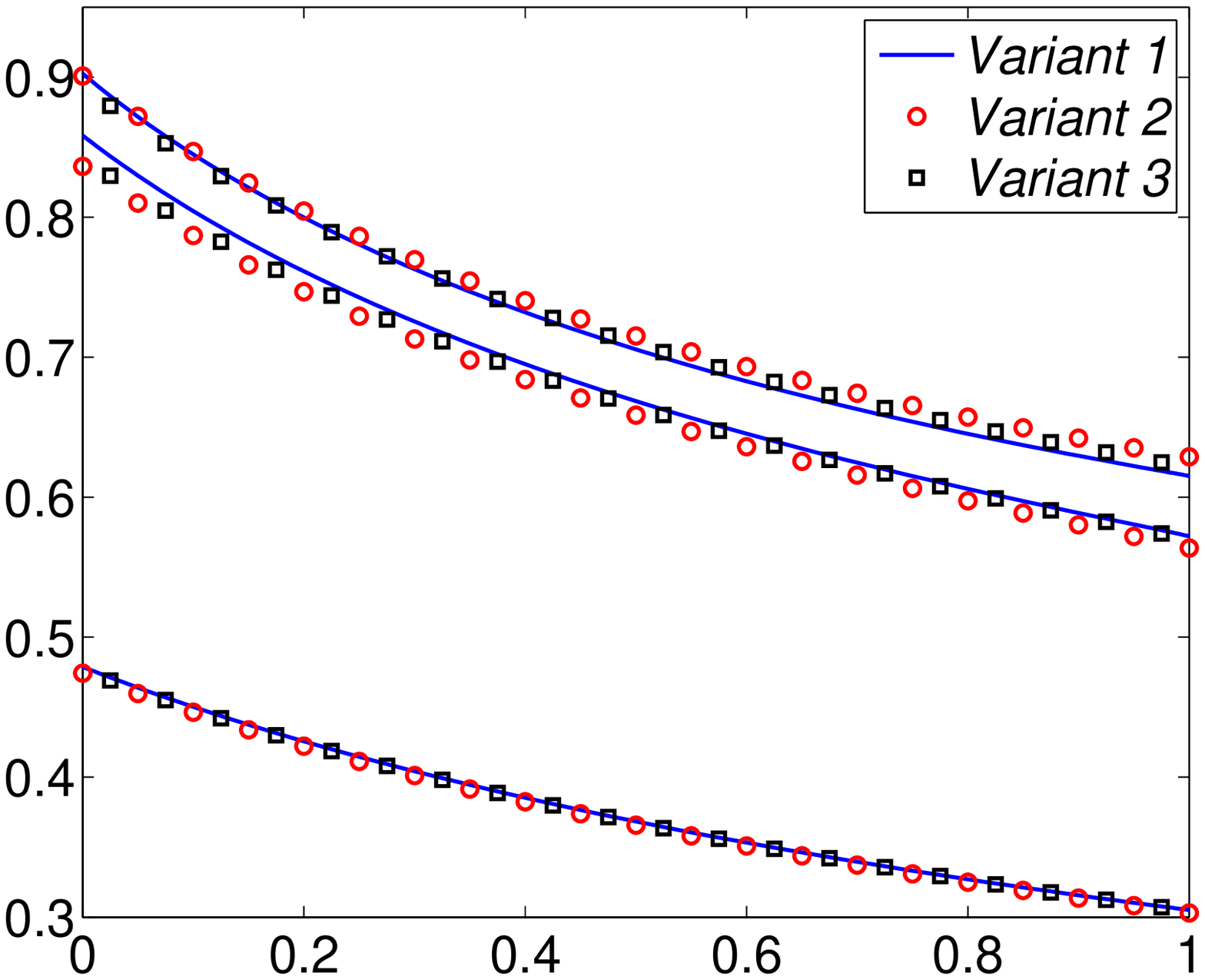}
    \put(-105,0){$x$}
    \put(-220,90){$\hat v$}
    \put(-90,110){$\hat K_{Ic}=0$}
    \put(-120,80){$\hat K_{Ic}=1$}
    \put(-150,48){$\hat K_{Ic}=3$}
    \put(-230,160){$\textbf{b)}$}

    \caption{Self-similar solutions for the three analysed variants of problem: a) the crack opening $\hat w$, b) the particle velocity $\hat v$.
    Respective data corresponds to $\nu=0.3$ and three different values of $\hat K_{Ic}=\{0,1,3\}$. Solid lines refer to the new HF formulation.}

\label{s_sim_fig1}
\end{figure}

\begin{figure}[h!]
%M/N=1/300

    %\hspace{-2mm}
    \includegraphics [scale=0.40]{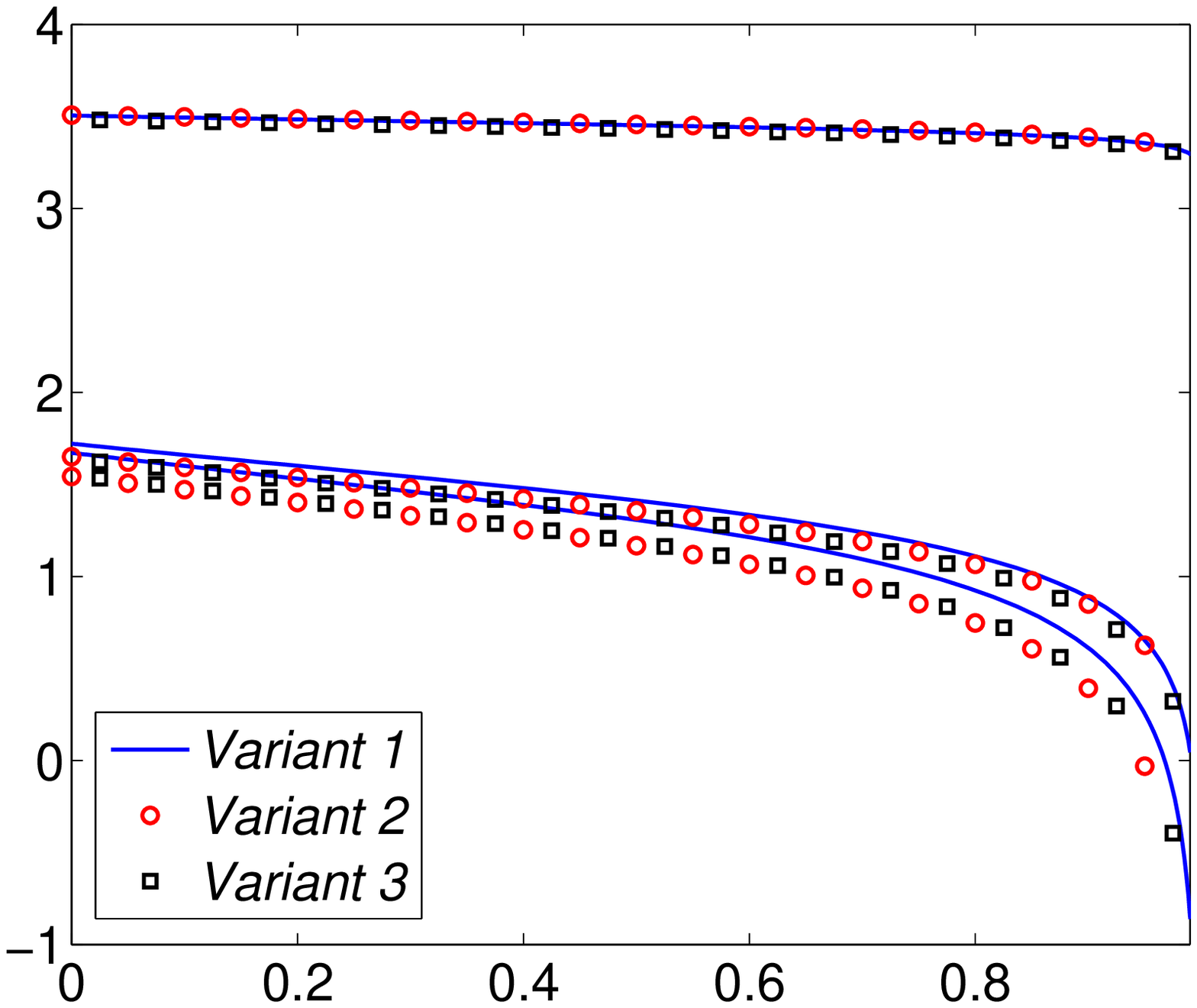}
    \put(-105,0){$x$}
    \put(-220,90){$ \hat p$}
    \put(-90,130){$\hat K_{Ic}=3$}
    \put(-120,92){$\hat K_{Ic}=1$}
    \put(-140,69){$\hat K_{Ic}=0$}
    \put(-230,160){$\textbf{a)}$}
%M/N=1/30
    \hspace{2mm}
    \includegraphics [scale=0.40]{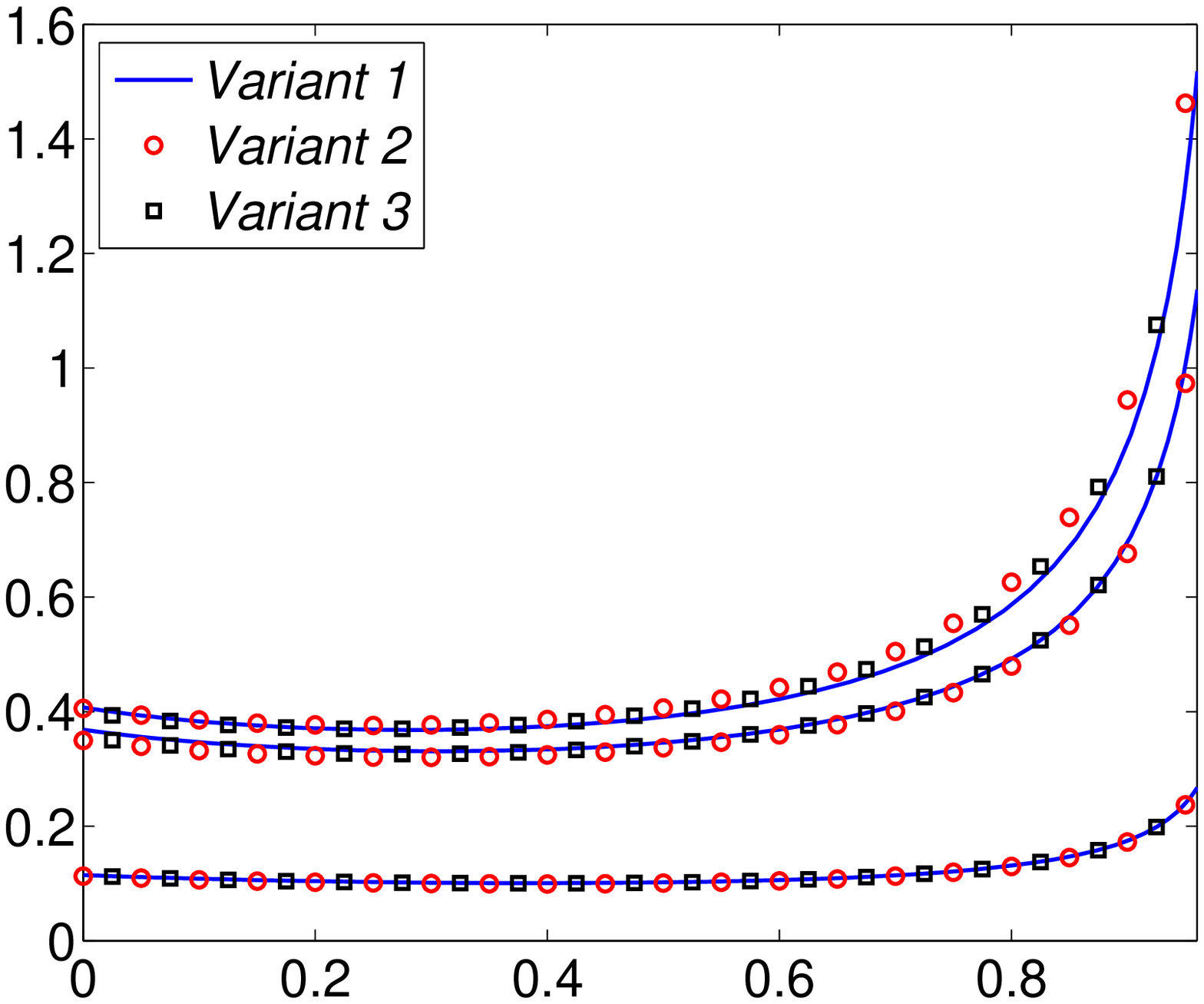}
    \put(-105,0){$x$}
    \put(-220,90){$ \hat \tau$}
    \put(-90,75){$\hat K_{Ic}=0$}
    \put(-70,42){$\hat K_{Ic}=1$}
    \put(-140,32){$\hat K_{Ic}=3$}
    \put(-230,160){$\textbf{b)}$}

    \caption{Self-similar solutions for the three analysed variants of the problem: a) the net fluid pressure $\hat p$, b) the shear stress $\hat \tau$.
    The respective data corresponds to $\nu=0.3$ and three different values of $\hat K_{Ic}=\{0,1,3\}$. Solid lines refer to the new HF formulation. }

\label{s-sim_fig2}
\end{figure}

\begin{figure}[h!]
%M/N=1/300

    %\hspace{-2mm}
    \includegraphics [scale=0.40]{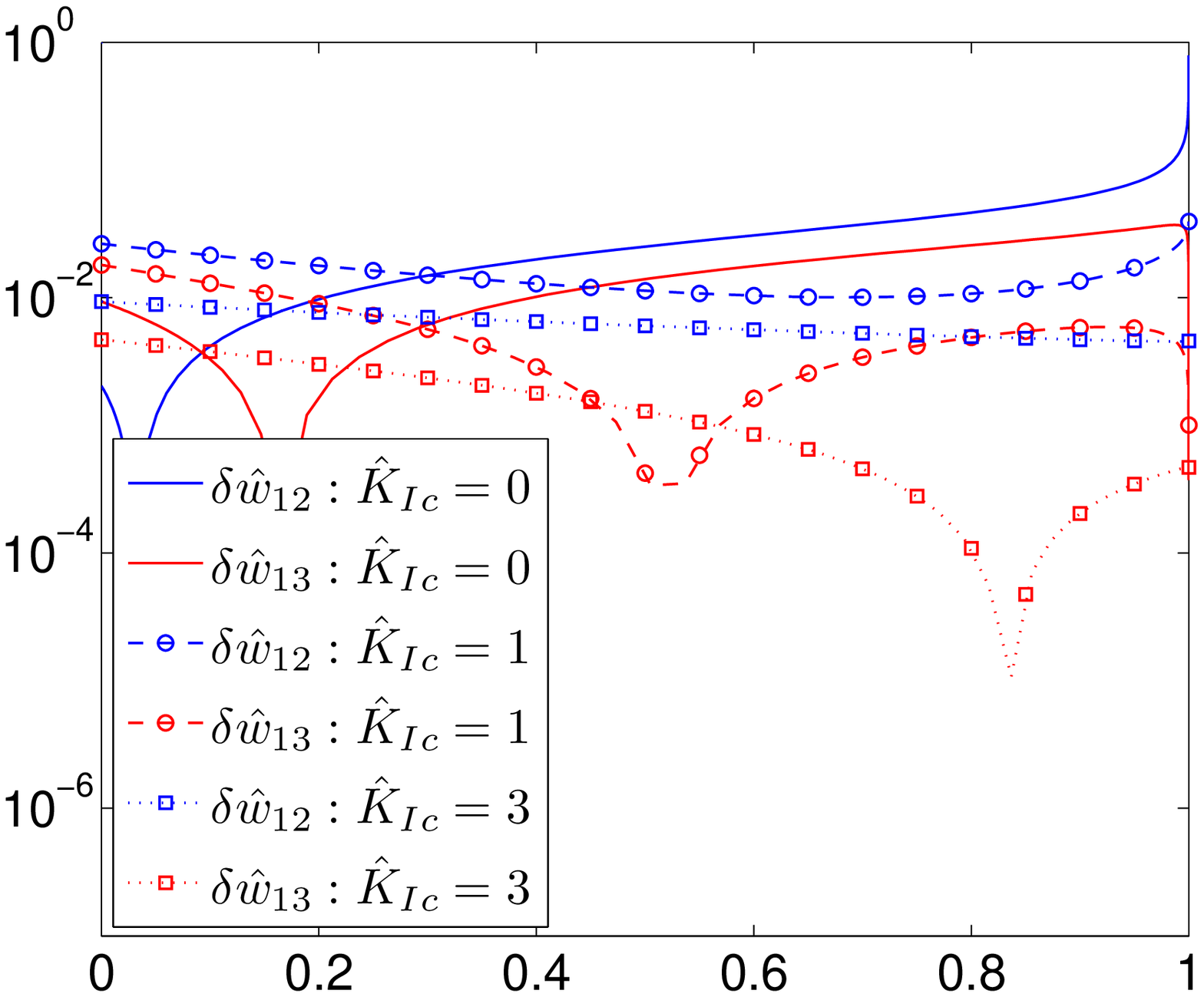}
    \put(-105,0){$x$}
    \put(-230,90){$\delta \hat w$}
    \put(-230,160){$\textbf{a)}$}
%M/N=1/30
    \hspace{2mm}
    \includegraphics [scale=0.40]{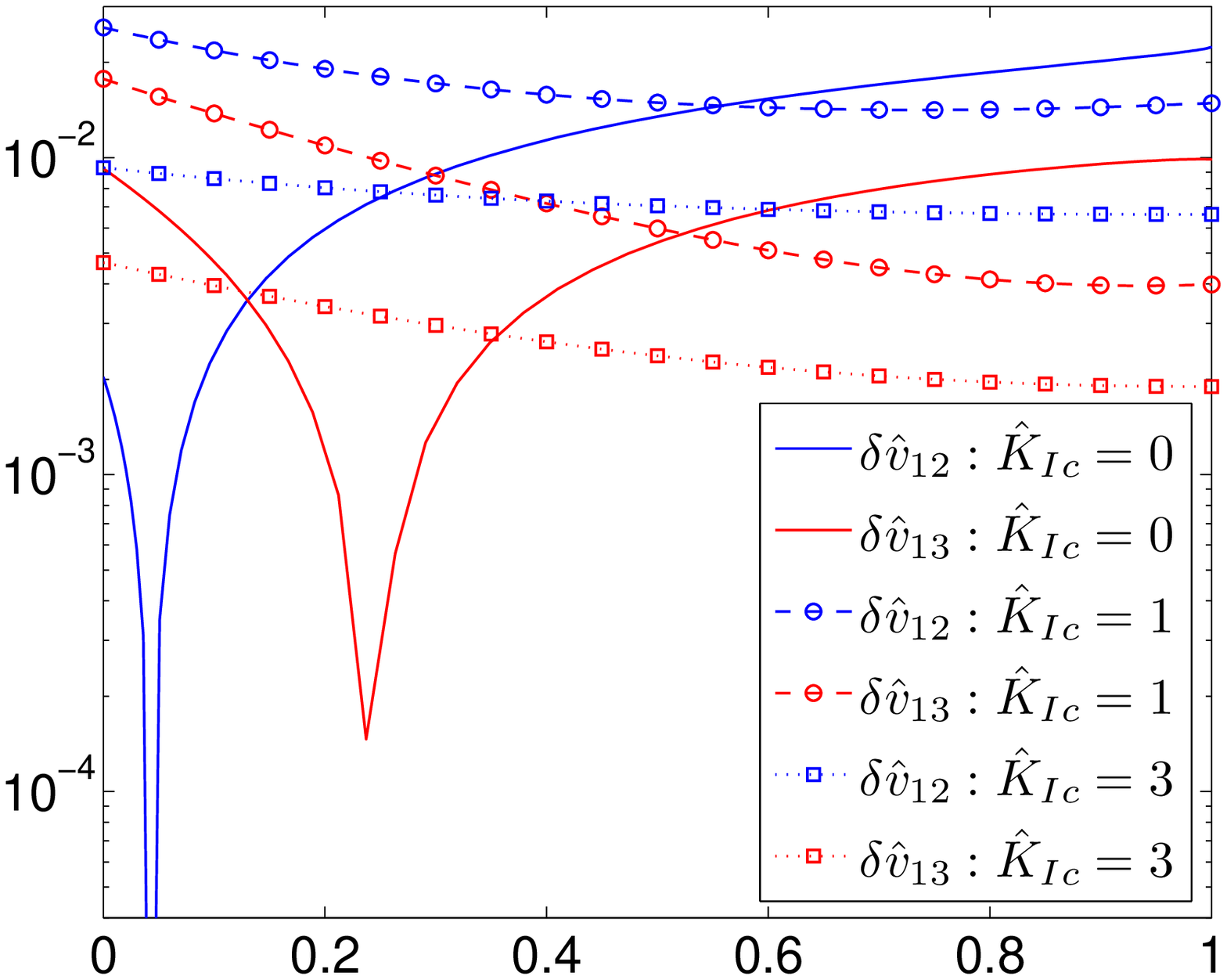}
    \put(-105,0){$x$}
    \put(-230,90){$\delta \hat v$}
    \put(-230,160){$\textbf{b)}$}

    \caption{Relative deviations of the solutions for the classic KGD and amended KGD models from the reference solution obtained in the modified HF formulation: a) the crack opening $\hat w$, b) the particle velocity $\hat v$. }

\label{odchylki_1}
\end{figure}

\begin{figure}[h!]
%M/N=1/300

    %\hspace{-2mm}
    \includegraphics [scale=0.40]{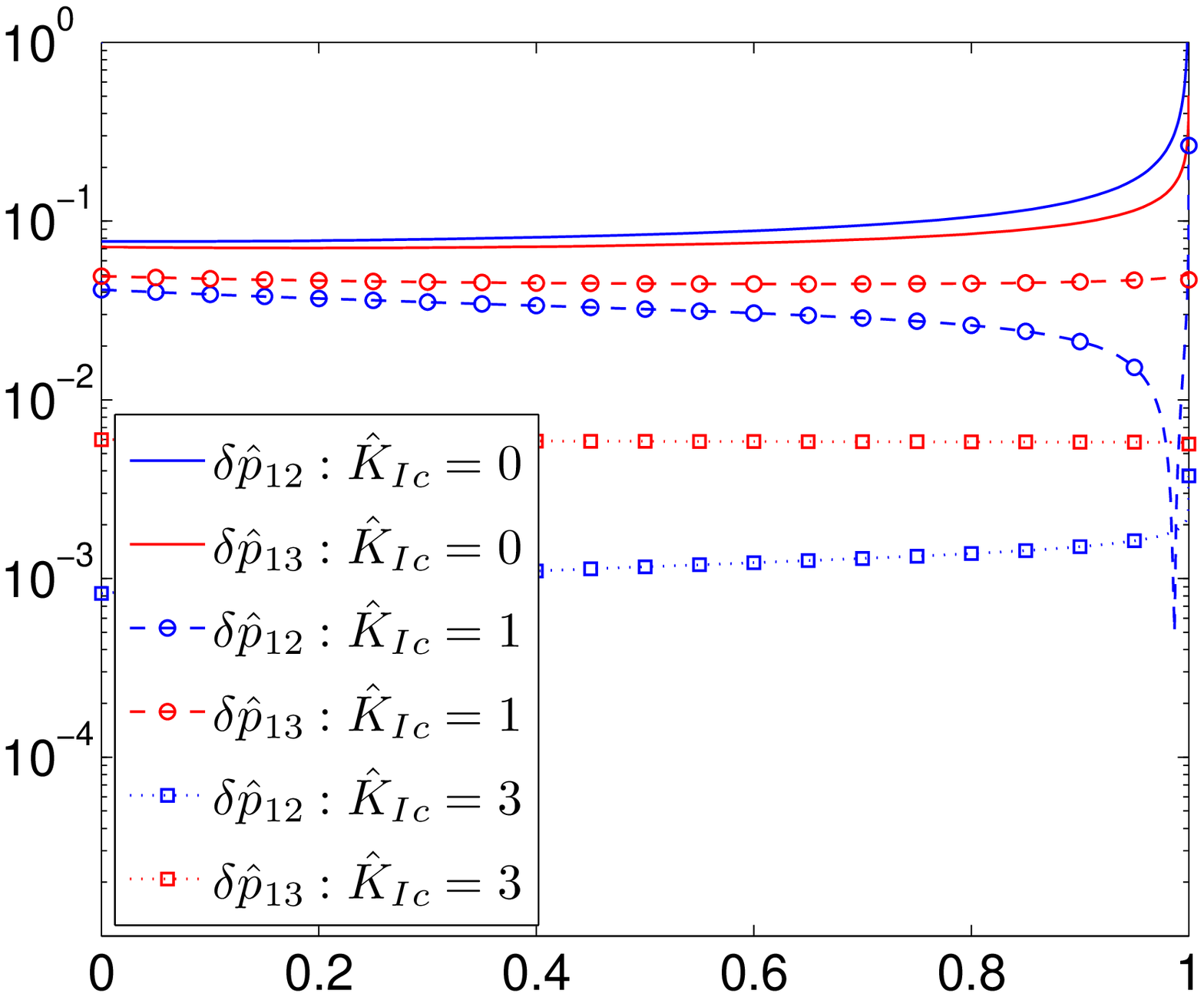}
    \put(-105,0){$x$}
    \put(-230,90){$\delta \hat p$}
    \put(-230,160){$\textbf{a)}$}
%M/N=1/30
    \hspace{2mm}
    \includegraphics [scale=0.40]{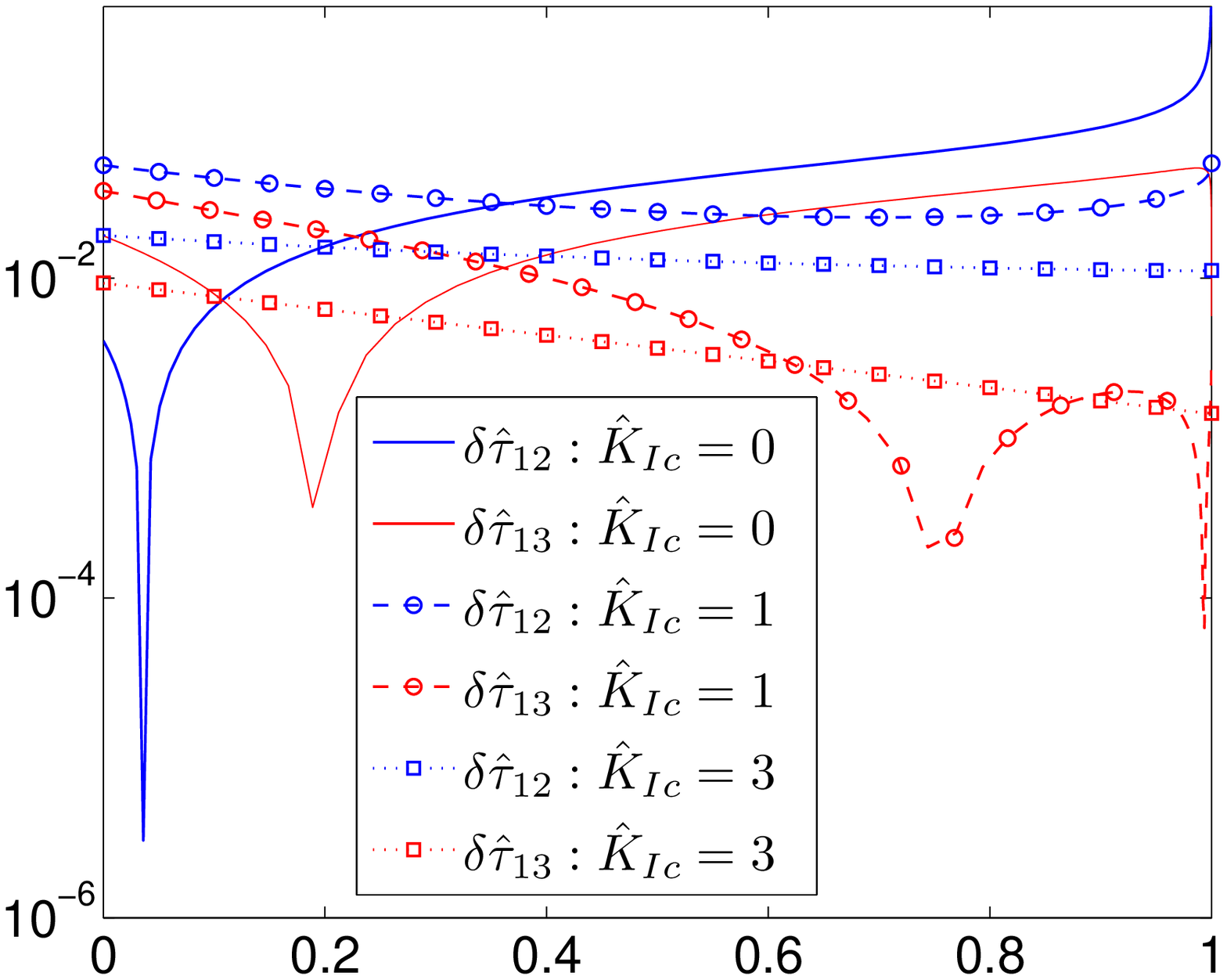}
    \put(-105,0){$x$}
    \put(-230,90){$\delta \hat \tau$}
    \put(-230,160){$\textbf{b)}$}

    \caption{Relative deviations of the solutions for the classic KGD and amended KGD models from the reference solution obtained in the modified HF formulation: a) the net fluid pressure $\hat p$, b) the shear stress $\hat \tau$. }

\label{odchylki_2}
\end{figure}

Having identified the accuracy level, we now consider the self-similar problem for $\nu=0.3$ and three different values of the self-similar material toughness $\hat K_{Ic}=\{0,1,3\}$, where the first corresponds to the viscosity dominated regime of crack propagation.
The value of Poisson's ratio was set to a magnitude which is characteristic for a number of mineral materials and, for example, ice \citep{Gercek_2006}. The influence of $\nu$ itself shall be discussed later in this subsection.
In Figs.\ref{s_sim_fig1}--\ref{s-sim_fig2} the graphs for the crack opening, $\hat w$, the particle velocity, $\hat v$, the net fluid pressure, $\hat p$, and the shear stress, $\hat \tau$, are presented. As can be seen, the curves for the respective variants of the problem are rather close to each other. The greatest discrepancies can be observed in the net fluid pressure, where Variants 2 and 3 slightly underestimate the values of $\hat p$. In general, the larger the value of $\hat K_{Ic}$, the less pronounced the differences between the relevant variants of the problem. To illustrate this trend more clearly, in Figs.\ref{odchylki_1}-\ref{odchylki_2} we show  the relative deviations of the solutions for Variants 2 and 3 from the reference solution (Variant 1).  For the net fluid pressure the following measure was accepted in order to avoid dividing by zero inside the interval:
\begin{equation}
\label{p_dev}
\delta \hat p_{1i}=\left |\frac{\hat p^{(i)}-\hat p^{(1)}}{\hat p^{(1)}(0)}\right |, \quad i=2,3,
\end{equation}
where superscripts refer to the variant of the problem. The notation accepted in the graphs is: $\delta g_{12}$ - deviation of the solution for Variant 2 from the reference data, $\delta g_{13}$ - deviation of the solution for Variant 3 from the reference data ($g=\{\hat w,\hat v, \hat p, \hat \tau\}$).

As can be seen in the figures, these are not only values but also distributions of the relative deviations that change with $\hat K_{Ic}$. In particular, for the viscosity dominated regime, pronounced deviations at the crack tip are observed, while for growth in $K_{Ic}$ the maxima of $\delta \hat w$ decrease and move towards the fracture inlet. Similar trends hold for the particle velocity and the shear stress (because of definition of $\delta  \hat p$ we do not extend this conclusion to the net fluid pressure). For $\hat K_{Ic}=3$, all deviations, except for those of the shear stress, are significantly below 1$\%$.

To complement the analysis, let us now investigate the influence of the value of Poisson's ratio on the mutual relations between respective solutions. This time we analyse the relative deviations of the self-similar fracture volume, $\Omega$, and the relative deviations of the self-similar crack propagation speed, $\hat v_0$. Observe that the latter can be used to compute the deviation of the self-similar crack length by relation \eqref{sol_ss_3}.

\begin{figure}[h!]
%M/N=1/300

    %\hspace{-2mm}
    \includegraphics [scale=0.40]{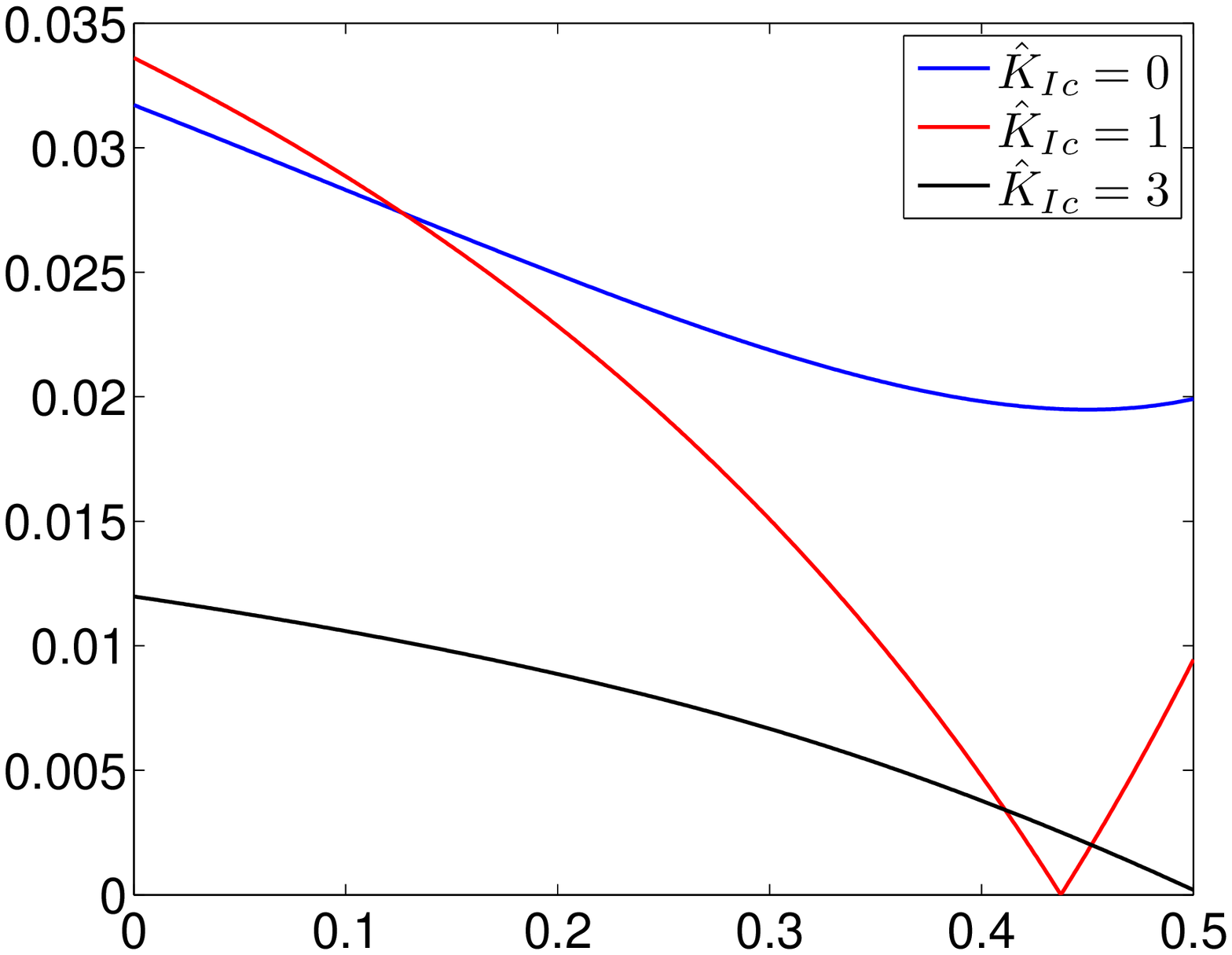}
    \put(-105,0){$\nu$}
    \put(-230,90){$\delta \Omega$}
    \put(-230,160){$\textbf{a)}$}
%M/N=1/30
    \hspace{2mm}
    \includegraphics [scale=0.40]{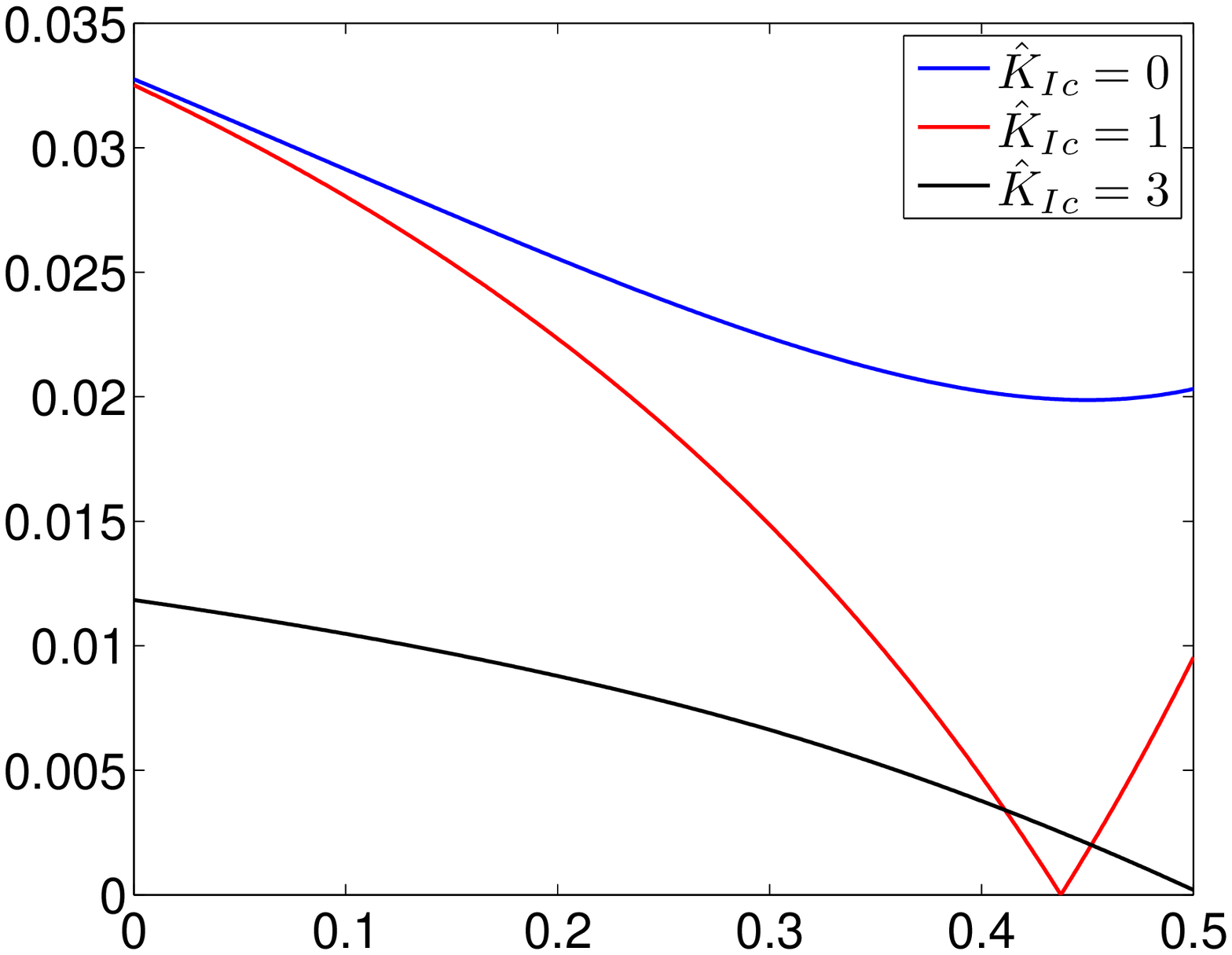}
    \put(-105,0){$\nu$}
    \put(-230,90){$\delta \hat v_0$}
    \put(-230,160){$\textbf{b)}$}

    \caption{The relative deviations of the solution for variant 2 (the classic KGD model) from the one for variant 1 (modified HF formulation): a) the fracture volume $\Omega$, b) the crack propagation speed $\hat v_0$. }

\label{odchylki_KGD_ni}
\end{figure}

\begin{figure}[h!]
%M/N=1/300

    %\hspace{-2mm}
    \includegraphics [scale=0.40]{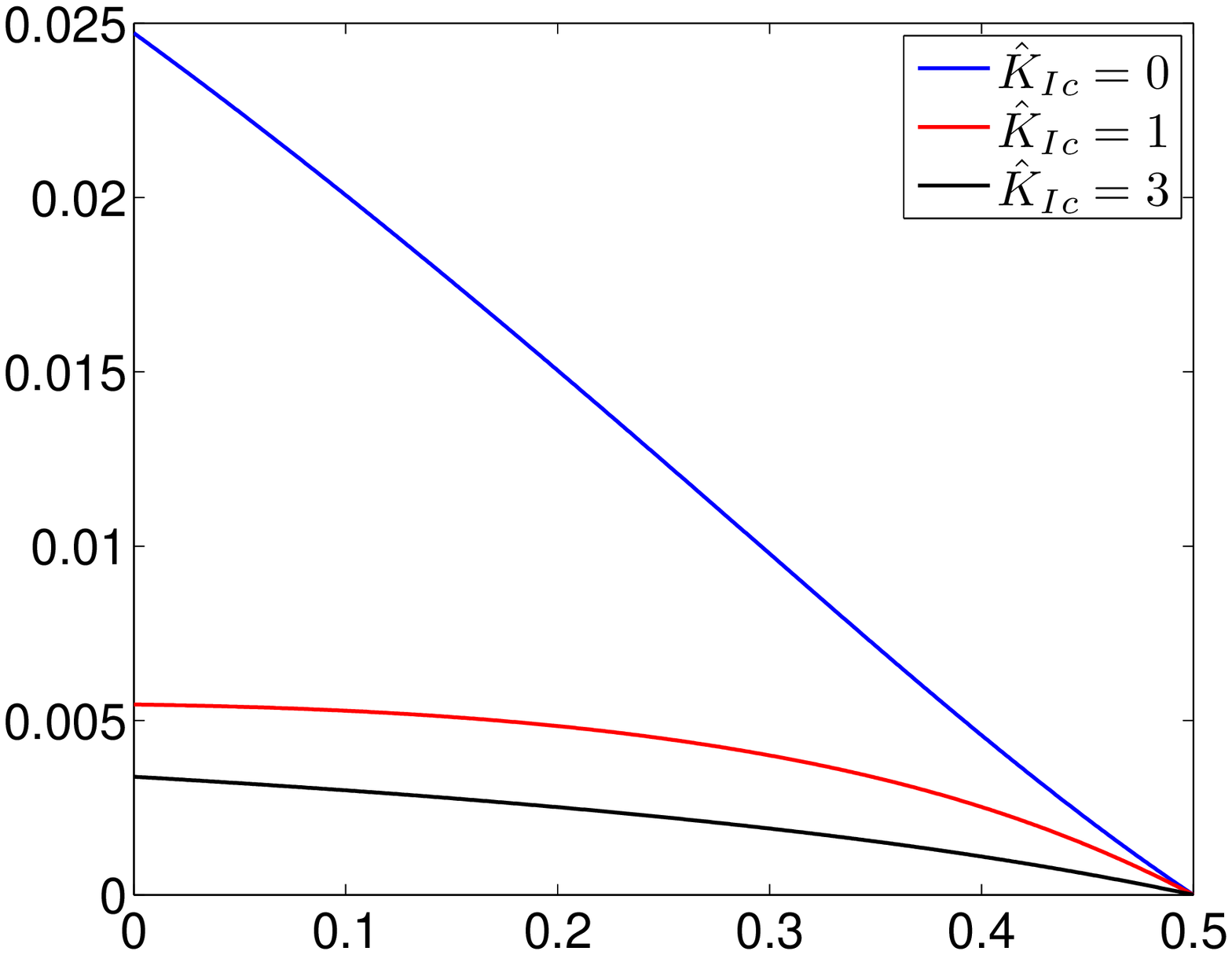}
    \put(-105,0){$\nu$}
    \put(-230,90){$\delta \Omega$}
    \put(-230,160){$\textbf{a)}$}
%M/N=1/30
    \hspace{2mm}
    \includegraphics [scale=0.40]{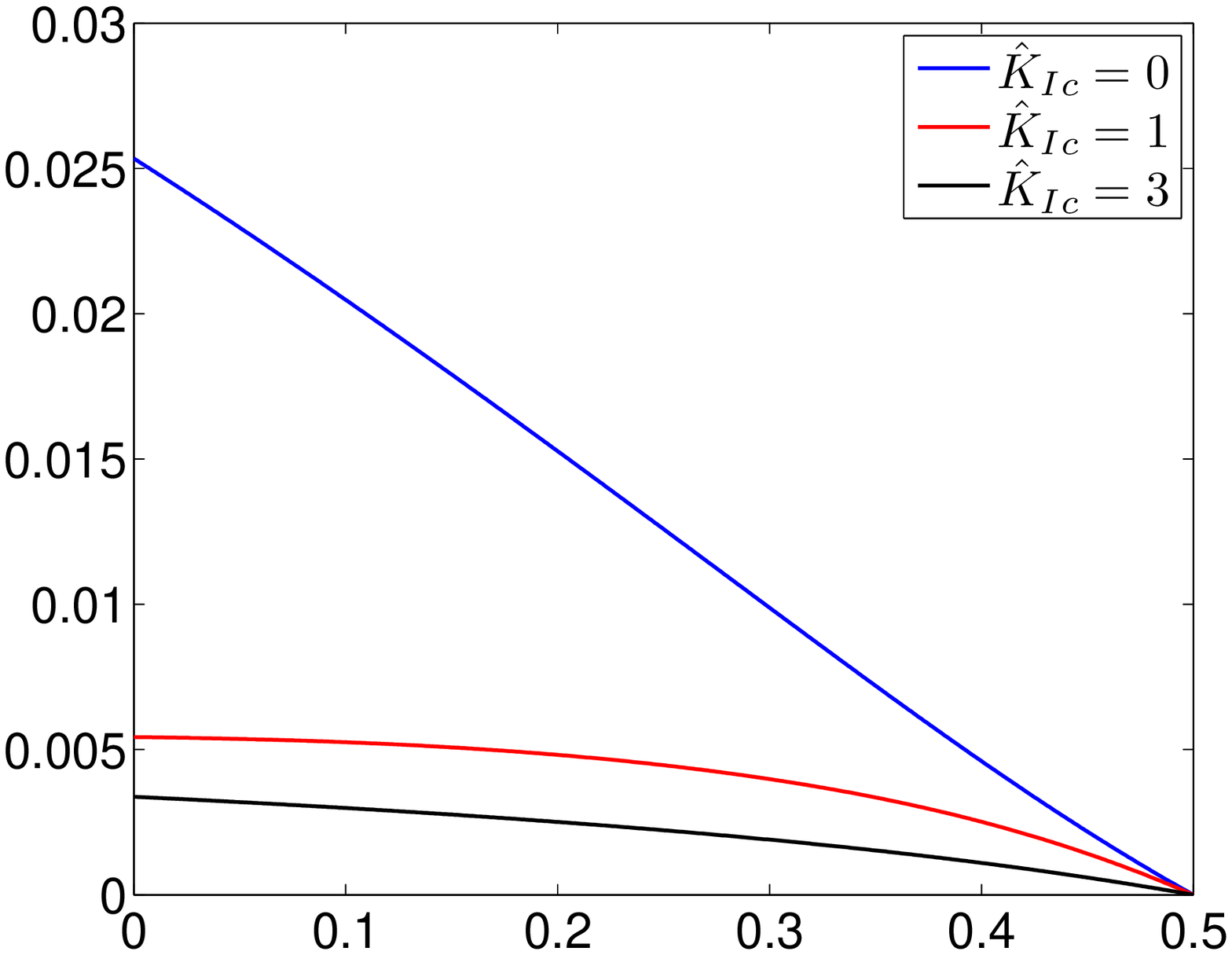}
    \put(-105,0){$\nu$}
    \put(-230,90){$\delta \hat v_0$}
    \put(-230,160){$\textbf{b)}$}

    \caption{The relative deviations of the solution for variant 3 (the amended KGD model) from the one for variant 1 (modified HF formulation): a) the fracture volume $\Omega$, b) the crack propagation speed $\hat v_0$.}

\label{odchylki_amend_KGD_ni}
\end{figure}

The graphs for $\delta \Omega$ and $\delta v_0$ are shown in Fig.\ref{odchylki_KGD_ni} (the deviation of Variant 2 from Variant 1) and Fig.\ref{odchylki_amend_KGD_ni} (the deviation of Variant 3 from Variant 1).

As anticipated, a better agreement between the reference solution and the results for Variant 3 holds.  For both analysed cases, the values of $\delta \Omega$ and $\delta v_0$ decrease with growing $\hat K_{Ic}$. The deviation of the results for Variant 3 decays as $\nu$ approaches 0.5 (incompressible material), which is natural as the multiplier $k_1$ of the shear stress term in the elasticity equation \eqref{elasticity_3} goes to zero (for $\nu=0.5$, Variant 1 and Variant 3 are equivalent). A similar relation for Variant 2 is observed only for $\hat K_{Ic}=3$. For smaller values of the self-similar material toughness, the minima of deviations are located in the range $0.4<\nu<0.5$.

\vspace{3mm}
Some general conclusions can be drawn from the above analysis:
\begin{itemize}
\item{Under certain conditions (e.g. large values of the self-similar material toughness) the classic KGD model and the amended KGD model can be considered to be good approximations of the modified HF formulation. }
\item{The amended KGD formulation (Variant 3) produces results which are more consistent with those given by the revised model (i.e. the most complete one, which takes into account the effects introduced by hydraulically induced tangential traction)}.
\item{The principal computational advantage of the modified and amended HF formulations over the classic KGD model is that the former do not change their qualitative asymptotic behaviour for $\hat K_{Ic} \to 0$. Indeed, as will be shown in the next subsection, the length of the process zone for the solution tip asymptote \eqref{w_ss_asymp}, \eqref{p_ss_asymp}, \eqref{tau_asymp} is not reduced to zero. Thus, no special measures are required to enforce the different asymptotic behavior of the solution. Consequently, in the small toughness case, the efficiency of computations for the modified and amended HF formulations is an improvement on that for calculations using the classic KGD model.   }
\end{itemize}

\subsection{Modified HF formulation - analysis}

Let us now have some insight into the peculiarities of the solution obtained for the modified HF formulation which takes into account the effects related to the hydraulically induced shear stress.

\begin{figure}[h!]
%M/N=1/300

    %\hspace{-2mm}
    \includegraphics[scale=0.40]{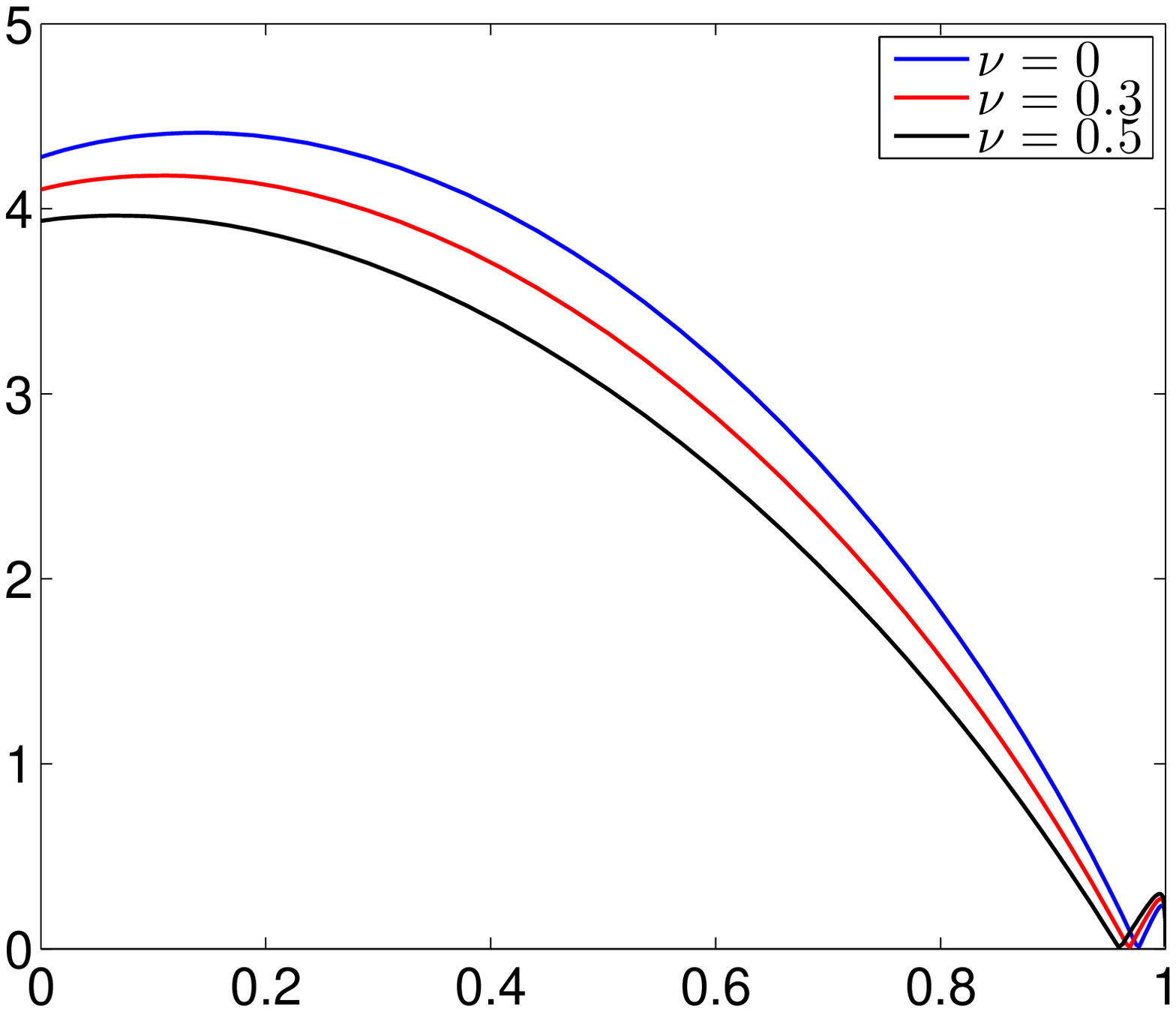}
    \put(-105,0){$x$}
    \put(-230,90){$\left |\frac{\hat p}{\hat \tau} \right |$}
    \put(-140,92){$\hat K_{Ic}=0$}
    \put(-230,160){$\textbf{a)}$}
%M/N=1/30
    \hspace{2mm}
    \includegraphics[scale=0.40]{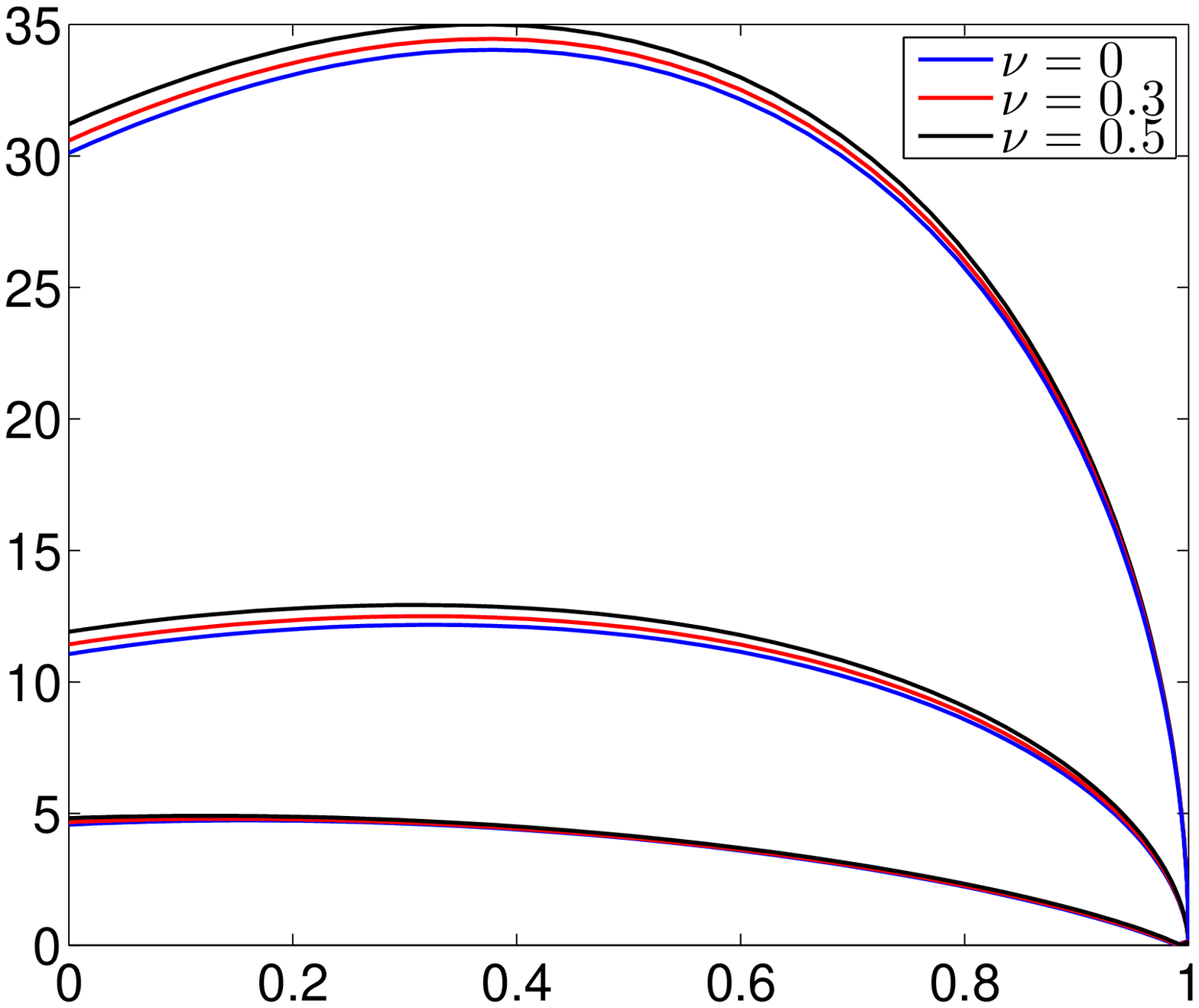}
    \put(-105,0){$x$}
    \put(-230,90){$\left |\frac{\hat p}{\hat \tau} \right |$}
    \put(-230,160){$\textbf{b)}$}
    \put(-140,78){$\hat K_{Ic}=2$}
    \put(-140,42){$\hat K_{Ic}=1$}
    \put(-140,145){$\hat K_{Ic}=3$}

    \caption{Absolute value of the ratio fluid pressure/tangential traction ($|\hat p/\hat \tau |$) for: a) viscosity dominated regime ($\hat K_{Ic}=0$), b) toughness dominated regime for three values of the self-similar material toughness ($\hat K_{Ic}=\{1,2,3\}$). }

 \label{tau_p_fig}
\end{figure}

First, we quantitatively analyse  the relationship between the magnitudes of the fluid pressure, $\hat p$, and the shear stress, $\hat \tau$. As already mentioned in subsection 2.1, the tip singularity of $\hat \tau$ is stronger than that of $\hat p$. For this reason, the common justification for the omission of the tangential traction in the analysis ($\hat p\gg\hat \tau$) is rather questionable. In Fig.\ref{tau_p_fig}, we depict the ratio $|\hat p/\hat \tau|$ for four values of the self-similar material toughness $\hat K_{Ic}=\{0,1,2,3\}$. It shows that away from the fracture tip the fluid pressure indeed has  greater values than the shear stress, however the extent of this dominance depends essentially on the material toughness (a weaker dependence on Poisson's ratio is also observed). In general, the larger the value of $\hat K_{Ic}$ the greater the value of the ratio $|\hat p/\hat \tau|$ outside the near-tip zone. In the viscosity dominated regime, $|\hat p/\hat \tau|$ achieves the maximal value of about four, and it requires   $\hat K_{Ic}=2$ to take this ratio above ten. We should, however, still remember that when approaching the crack tip, the magnitude of $\hat \tau$ is much greater than that of $\hat p$. Thus, a straightforward conclusion from this part of the analysis can be drawn: unless the fracture propagates in the high toughness mode, the value of the ratio $|\hat p/\hat \tau|$ cannot be treated as a justification for neglecting the hydraulically induced tangential traction.

\begin{figure}[h!]
%M/N=1/300

    %\hspace{-2mm}
    \includegraphics[scale=0.40]{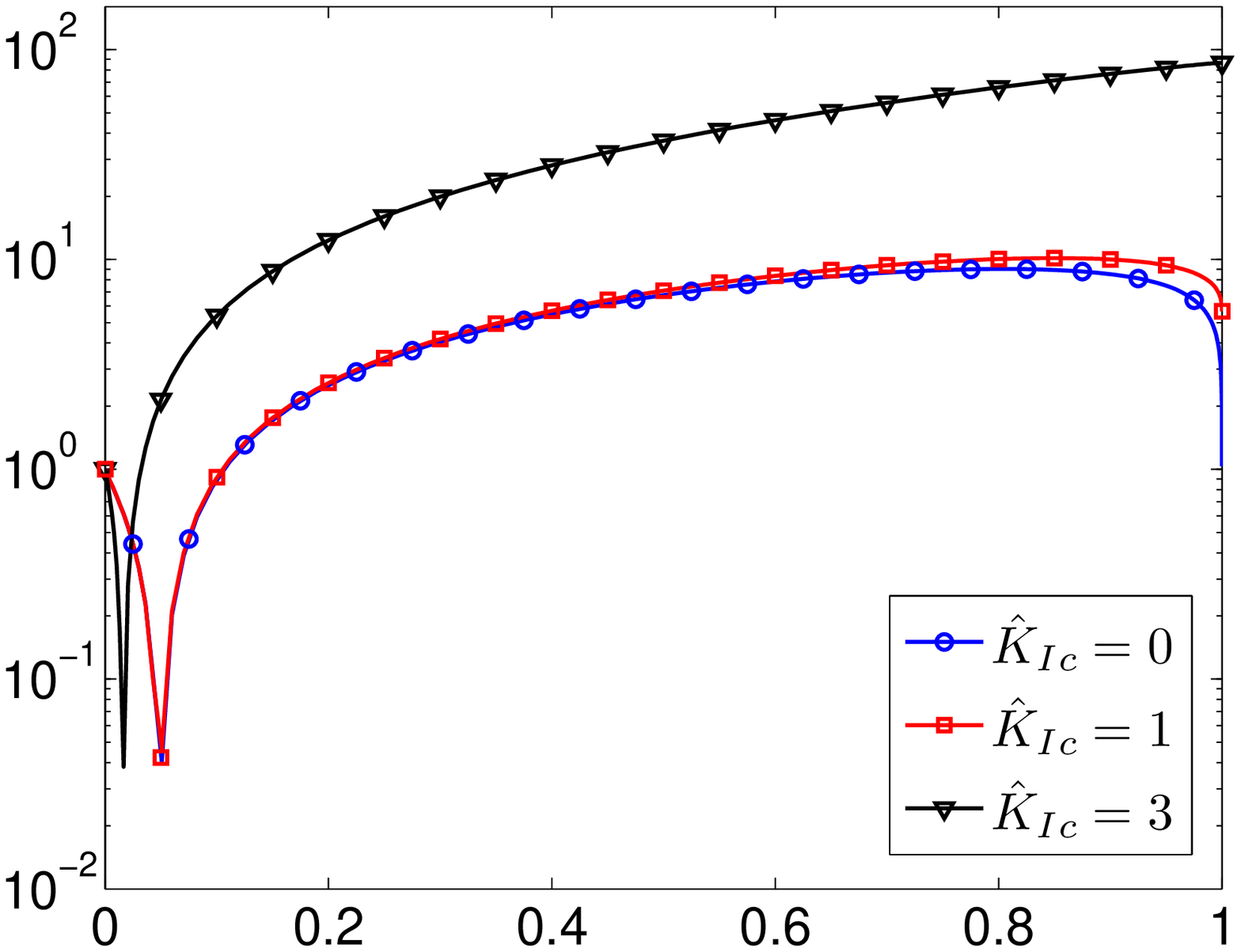}
    \put(-105,0){$x$}
    \put(-230,90){$\left |\frac{\hat w'}{k_1 \hat \tau} \right |$}
    \put(-230,160){$\textbf{a)}$}
%M/N=1/30
    \hspace{2mm}
    \includegraphics[scale=0.40]{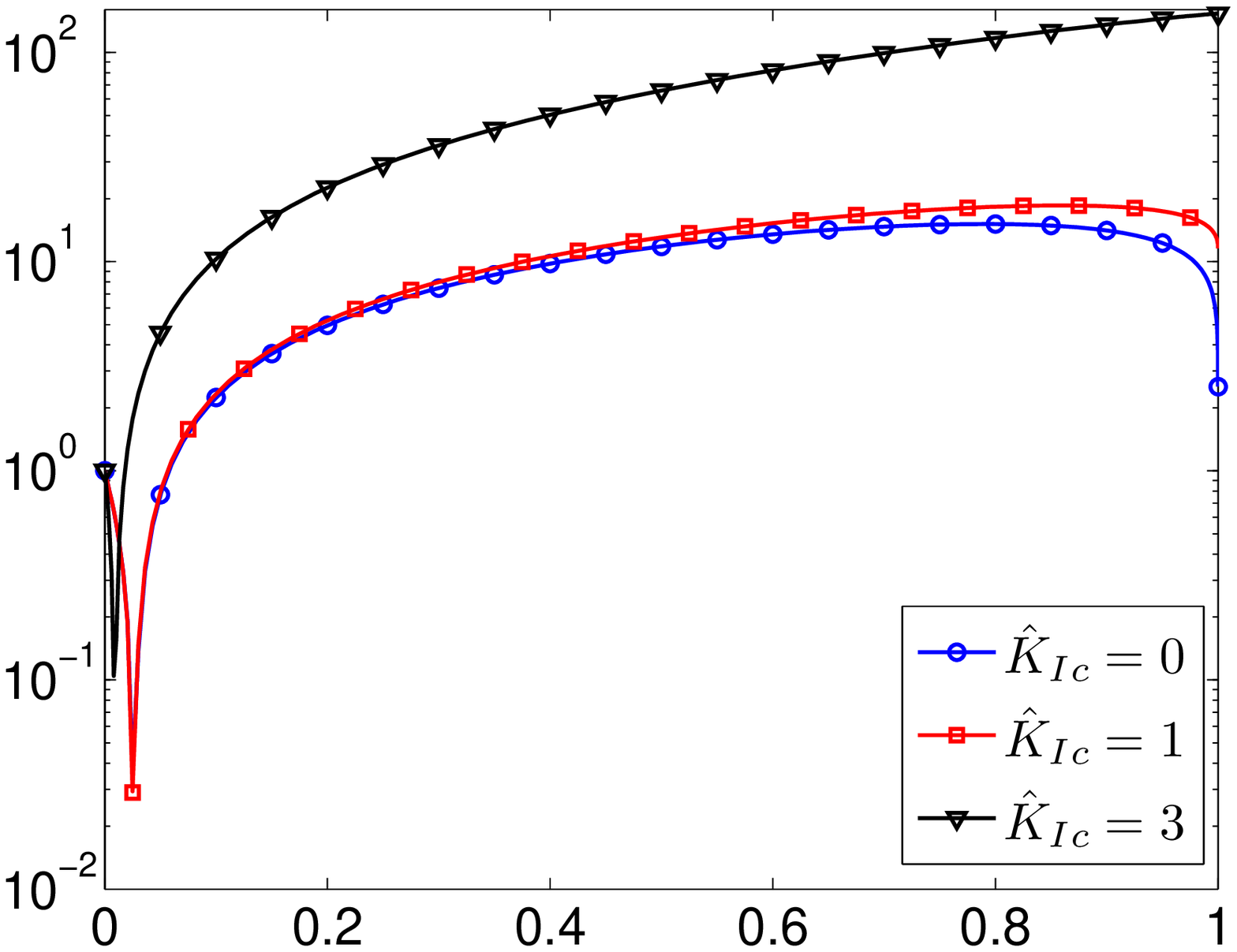}
    \put(-105,0){$x$}
    \put(-230,90){$\left |\frac{\hat w'}{k_1 \hat \tau} \right |$}
    \put(-230,160){$\textbf{b)}$}

    \caption{Relations between the entries of the elasticity operator \eqref{elasticity_3}: $\frac{d \hat w}{dx}$ and $k_1 \hat \tau$, for a) $\nu=0$ ($k_1=\pi^{-1}$), b) $\nu=0.3$ ($k_1=4/7\pi^{-1}$). The respective curves were obtained for three values of the self-similar material toughness $\hat K_{Ic}=\{0,1,3\}$. }

 \label{wp_tau}
\end{figure}

On the other hand, as shown in the previous subsection, the differences between the results for the various HF formulations are relatively small. In order to explain this apparent paradox, let us analyse the relationships between respective entries in the elasticity operator \eqref{elasticity_3}. In Fig.\ref{wp_tau} the ratio $\left|\frac{\hat w'}{k_1 \hat \tau}\right|$ is depicted for three values of the self-similar material toughness, $\hat K_{Ic}=\{0,1,3\}$. The data in Fig.\ref{wp_tau}a) corresponds to the case $\nu =0$ which gives the maximal value of $k_1$ ($k_1$ changes continuously with $\nu$ from $\pi^{-1}$ to 0 - see \eqref{k_1_2}). In this case we observe the largest influence of the shear stress component on the elasticity equation, and consequently the greatest deviations between the results of the respective HF formulations analysed in the previous subsection. As can be seen in the figure, it is only in the immediate neighborhood of the crack inlet that the absolute values of $k_1 \hat \tau$ are comparable to the those of the spatial derivative of the crack opening, $\hat w'$. Over almost the entire length of the spatial interval the latter component is a few times greater than the former, and this trend magnifies with growing $\hat K_{Ic}$. This fact contributes to an explanation of why the results of the respective HF formulations (Variant 1 - Variant 3) are relatively close to each other. For comparison, we show similar analysis for $\nu =0.3$ in Fig.\ref{wp_tau}b). As can be seen, the effect described above is magnified to produce even smaller contribution of the tangential traction term, $k_1 \hat \tau$, in the elasticity operator.

The basic feature of the introduced modified HF formulation is the presence of the hydraulically induced shear stresses at the crack faces. As a result, the energy release rate depends on two components: the classic stress intensity factor, and the newly introduced shear stress intensity factor. When the material toughness tends to zero, the first one (as well es the entire ERR) decays. However, the shear stress intensity factor assumes a non-zero value. As a consequence, the standard asymptote of LEFM is not superseded by qualitatively different asymptotic behaviour as it is in the classic KGD model. In other words, in the revised formulation the process zone for the governing asymptotics is never reduced to zero. Referring to the self-similar problem, it has already been shown that in the viscosity dominated regime the multiplier of the leading term of the fluid pressure asymptote assumes a non-zero value - see \eqref{p0_KI_0}. Consequently, for the multiplier of the leading term of the crack aperture asymptote one has from \eqref{frac_cr_ss} and \eqref{p0_KI_0} the following relation:
\begin{equation}
\label{w_0_visc}
\hat w_0=\sqrt{\frac{\pi(1-\nu)}{\alpha}}\hat K_f^2>0.
\end{equation}

\begin{figure}[h!]
%M/N=1/300

    %\hspace{-2mm}
    \includegraphics [scale=0.40]{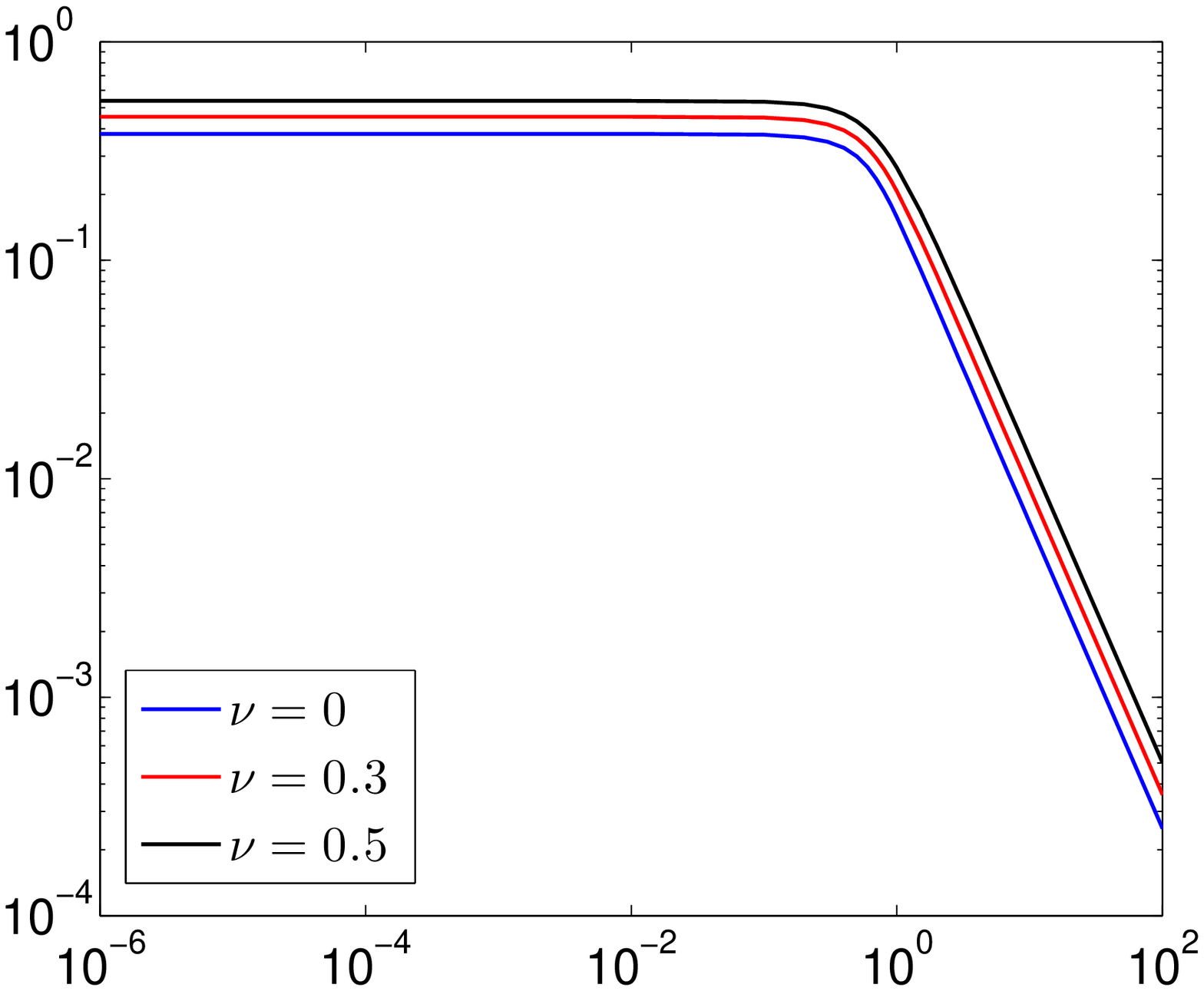}
    \put(-105,0){$\hat K_{Ic}$}
    \put(-230,90){$\hat K_f$}
    \put(-230,160){$\textbf{a)}$}
%M/N=1/30
    \hspace{2mm}
    \includegraphics [scale=0.40]{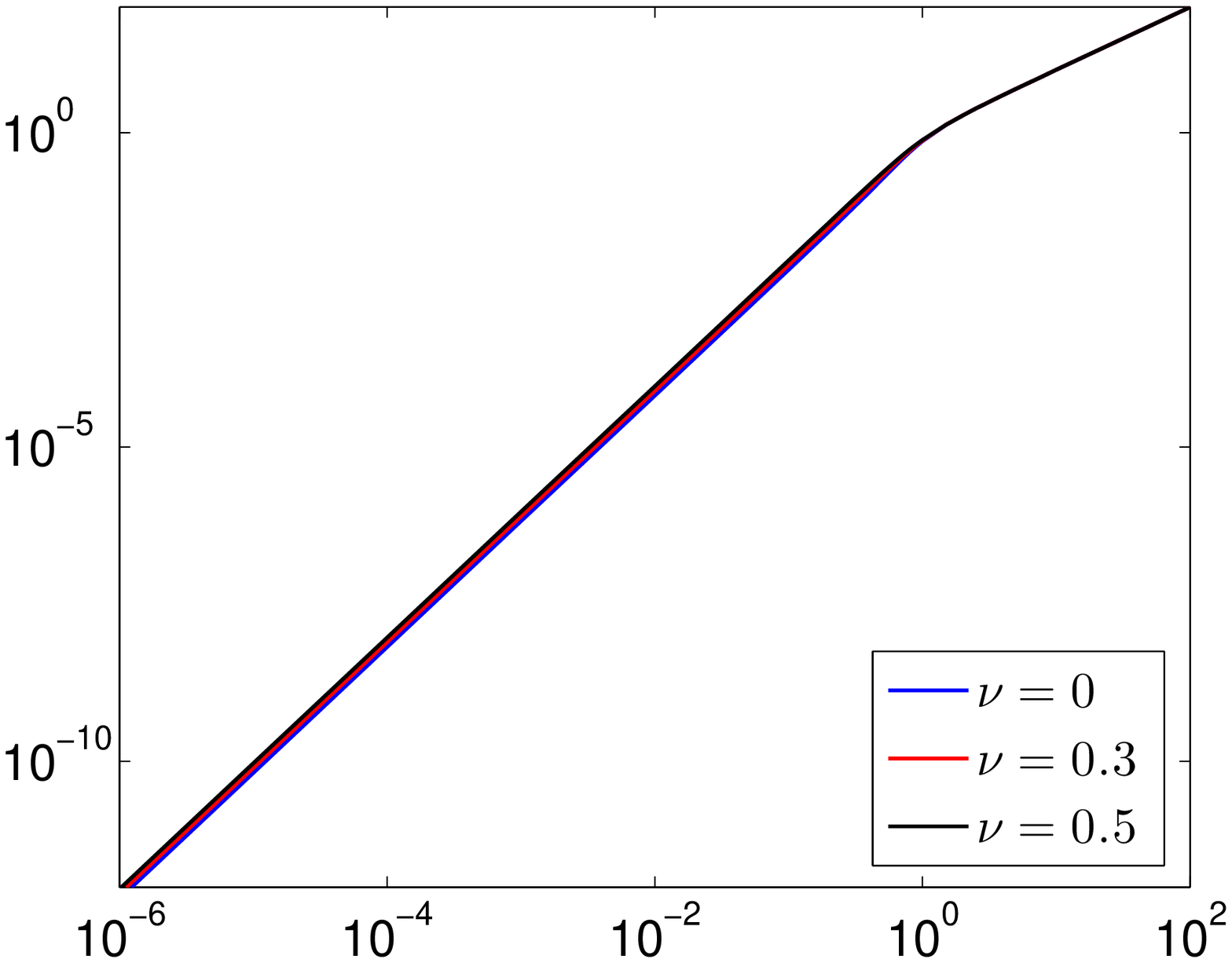}
    \put(-105,0){$\hat K_{Ic}$}
    \put(-230,90){$\hat K_I$}
    \put(-230,160){$\textbf{b)}$}

    \caption{Relationship between the self-similar material toughness, $\hat K_{Ic}$, and: a) the shear stress intensity factor, $\hat K_f$, b) the stress intensity factor, $\hat K_I$.}

\label{KI_Kf_w0}
\end{figure}

\begin{figure}[h!]
%M/N=1/300

    %\hspace{-2mm}
    \includegraphics [scale=0.40]{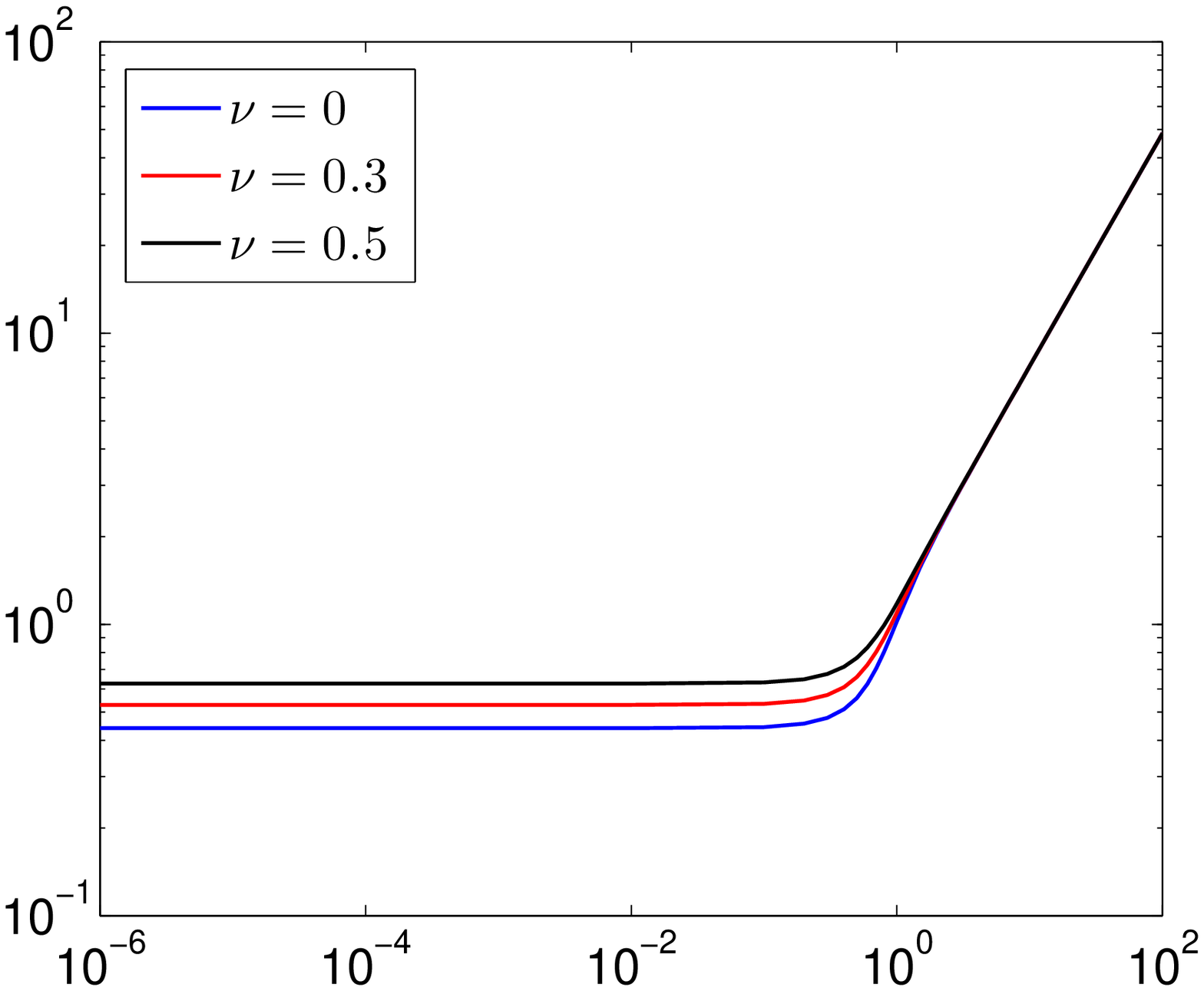}
    \put(-105,0){$\hat K_{Ic}$}
    \put(-230,90){$\hat w_0$}
    \put(-230,160){$\textbf{a)}$}
%M/N=1/30
    \hspace{2mm}
    \includegraphics [scale=0.40]{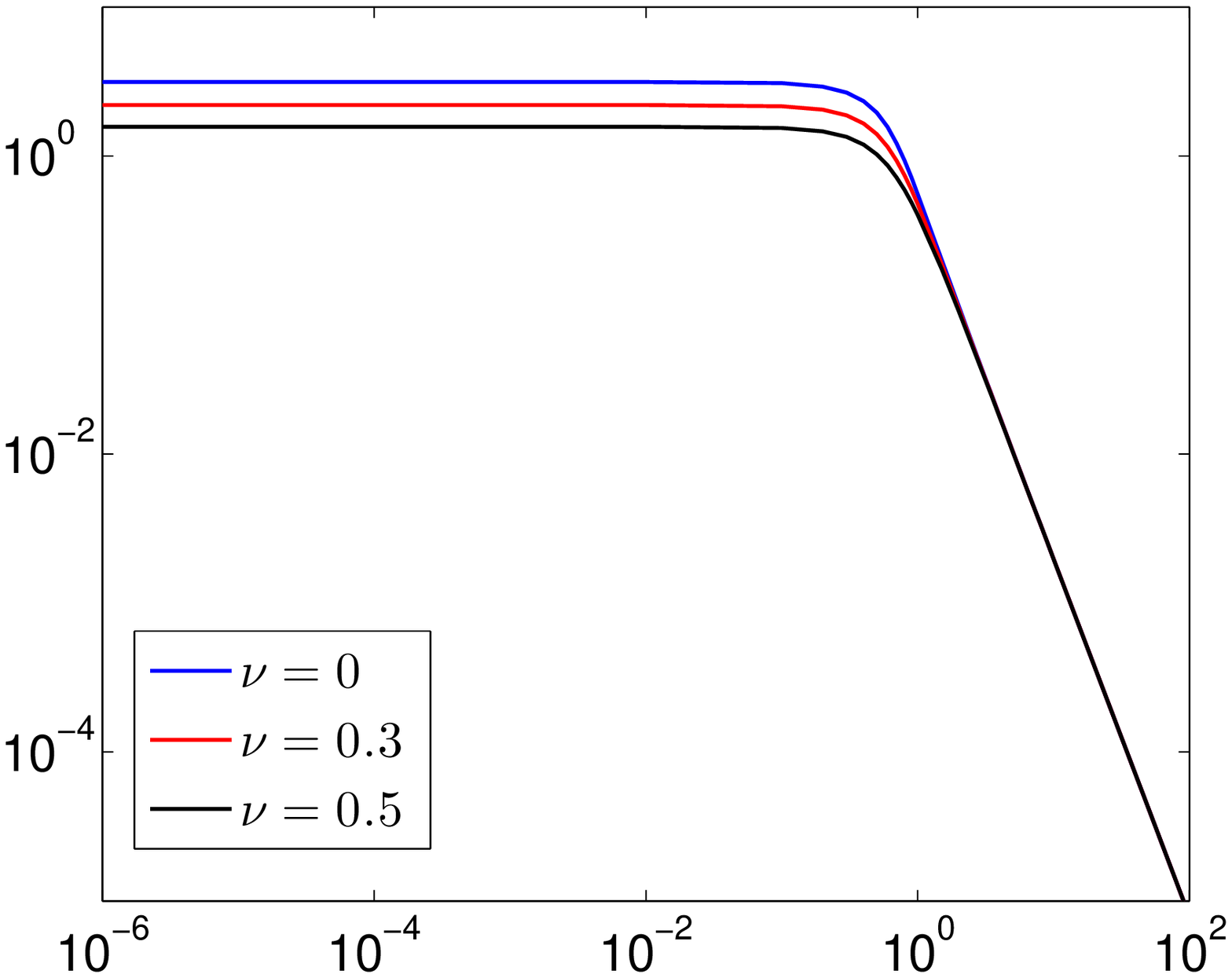}
    \put(-105,0){$\hat K_{Ic}$}
    \put(-230,90){$\hat p_0$}
    \put(-230,160){$\textbf{b)}$}

    \caption{Relationship between the self-similar material toughness, $\hat K_{Ic}$, and: a) the multiplier of the leading (square root) term of the crack opening tip asymptote, $\hat w_0$, b) the multiplier of the leading (logarithmic) term of the fluid pressure tip asymptote, $\hat p_0$.}

\label{KI_p0_v0}
\end{figure}

To illustrate this, and describe some general trends for different values of $\hat K_{Ic}$, in Fig.\ref{KI_Kf_w0}-Fig.\ref{KI_p0_v0} we depict  the following characteristics obtained by numerical simulations: $\hat K_f\left(\hat K_{Ic}\right)$, $\hat K_I\left(\hat K_{Ic}\right)$, $\hat w_0\left(\hat K_{Ic}\right)$, $\hat p_0\left(\hat K_{Ic}\right)$, $\hat v_0\left(\hat K_{Ic}\right)$, $\hat L_0\left(\hat K_{Ic}\right)$ and $\hat \tau_0\left(\hat K_{Ic}\right)$. The value of parameter $\alpha$ was set to $1/3$.

\begin{figure}[h!]
%M/N=1/300

    %\hspace{-2mm}
    \includegraphics [scale=0.40]{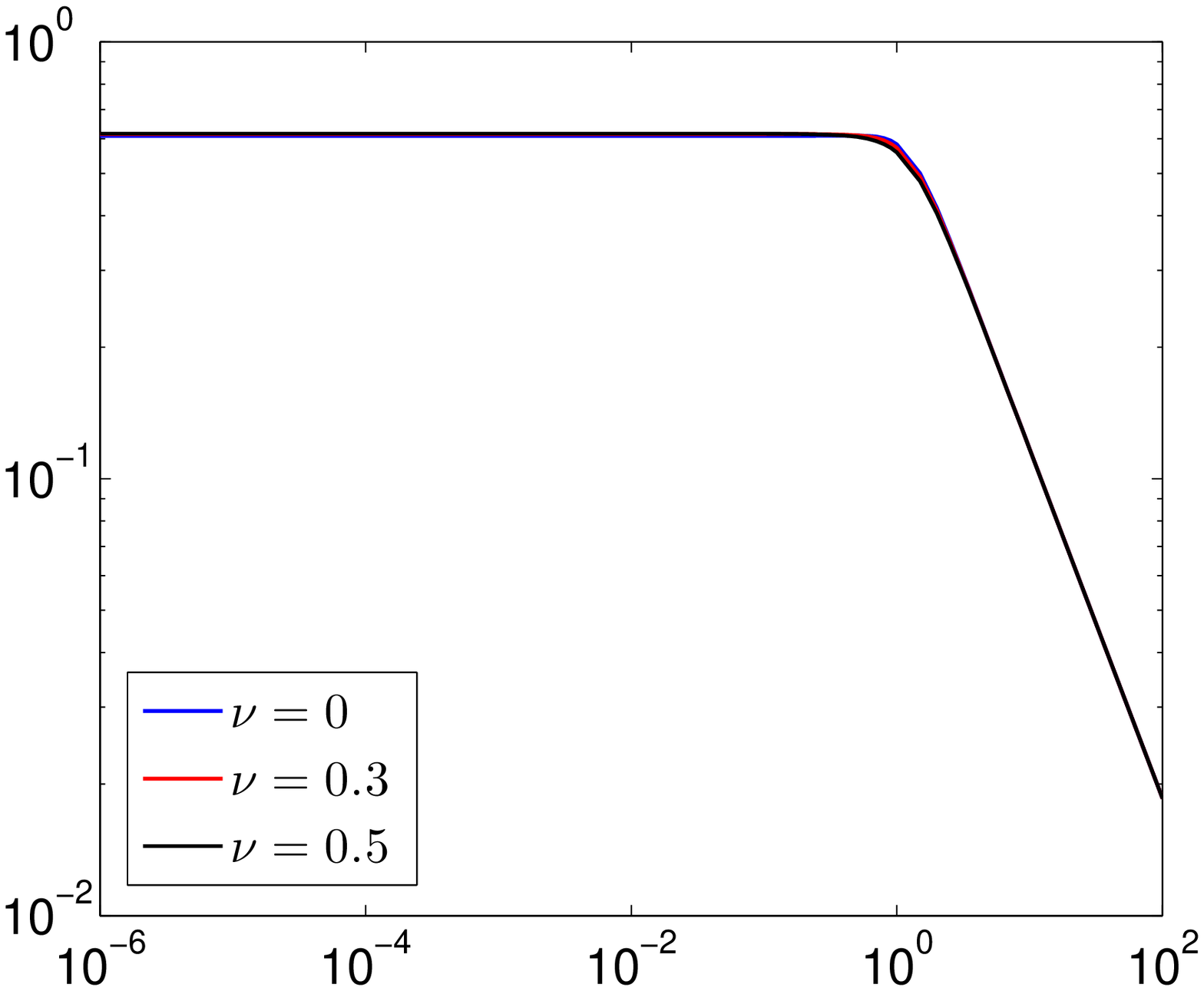}
    \put(-105,0){$\hat K_{Ic}$}
    \put(-230,90){$\hat v_0$}
    \put(-230,160){$\textbf{a)}$}
%M/N=1/30
    \hspace{2mm}
    \includegraphics [scale=0.40]{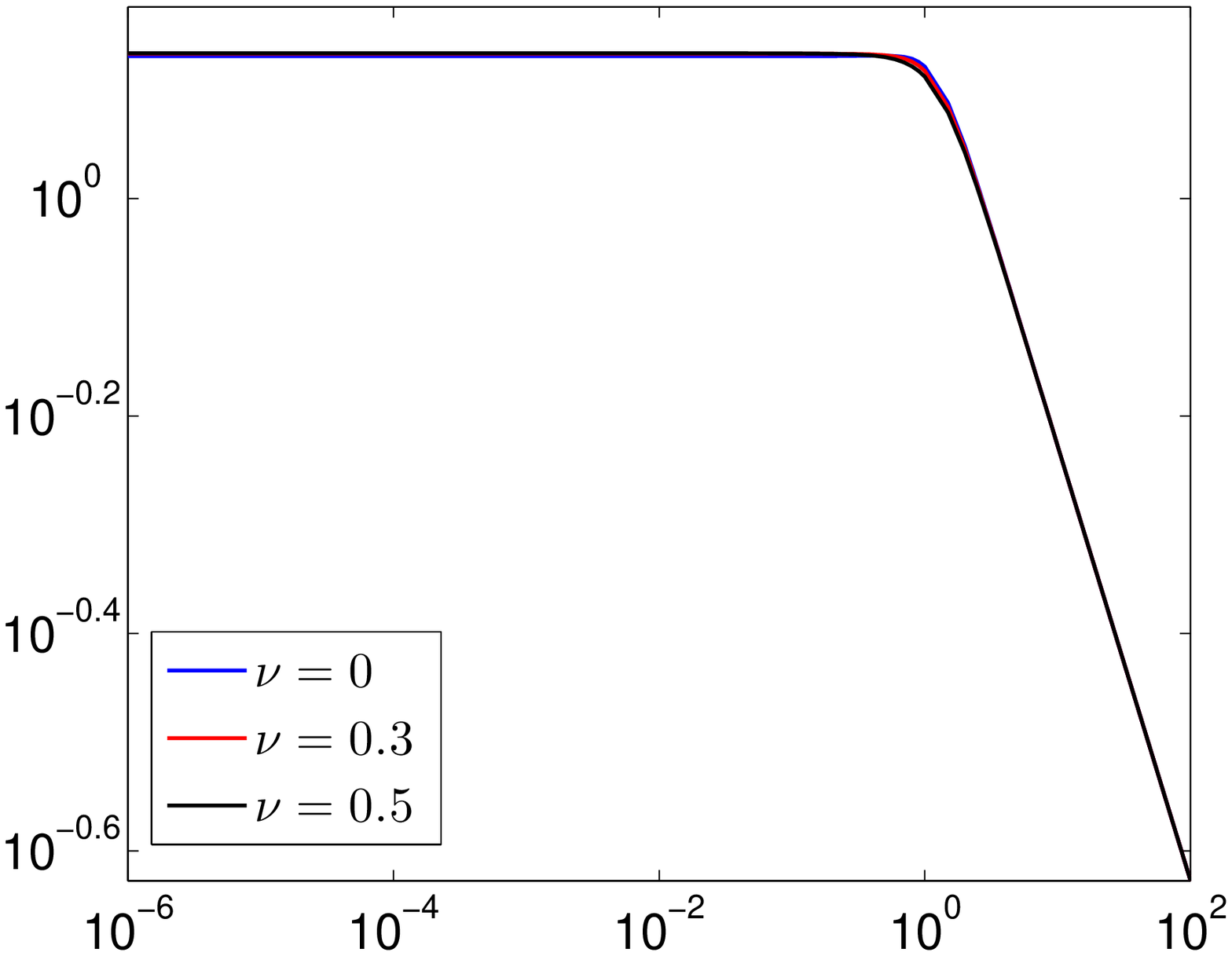}
    \put(-105,0){$\hat K_{Ic}$}
    \put(-230,90){$\hat L_0$}
    \put(-230,160){$\textbf{b)}$}

    \caption{Relationship between the self-similar material toughness, $\hat K_{Ic}$, and: a) the self-similar crack propagation speed, $\hat v_0$ ,   b) the self-similar crack length, $\hat L_0$,}

\label{KI_L0_T0}
\end{figure}

\begin{figure}[h!]
\includegraphics[scale=0.40]{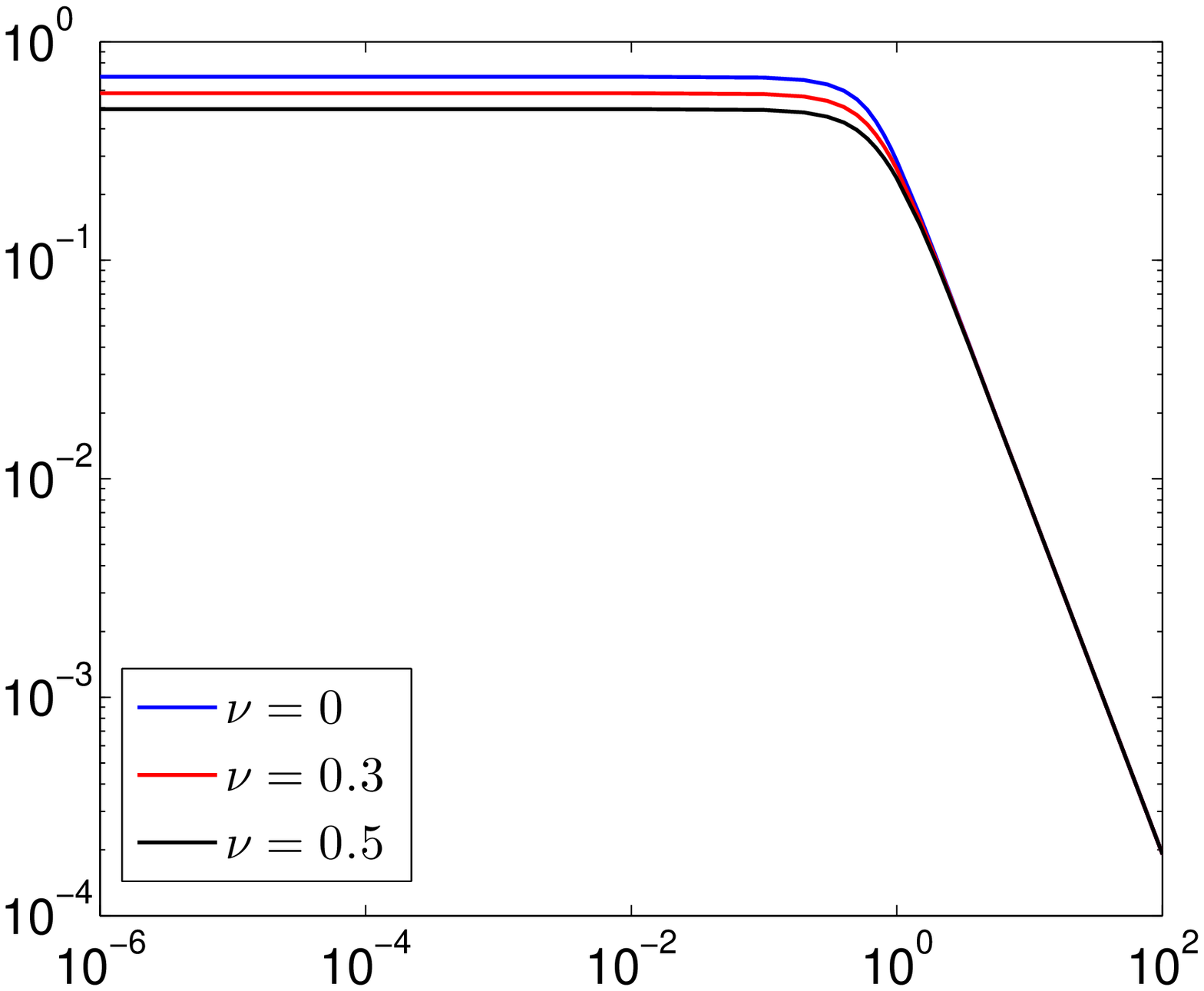}
    \put(-105,0){$\hat K_{Ic}$}
    \put(-230,90){$\hat \tau_0$}
    \put(-230,160){$\textbf{a)}$}
%M/N=1/30
    \hspace{2mm}
    \includegraphics[scale=0.40]{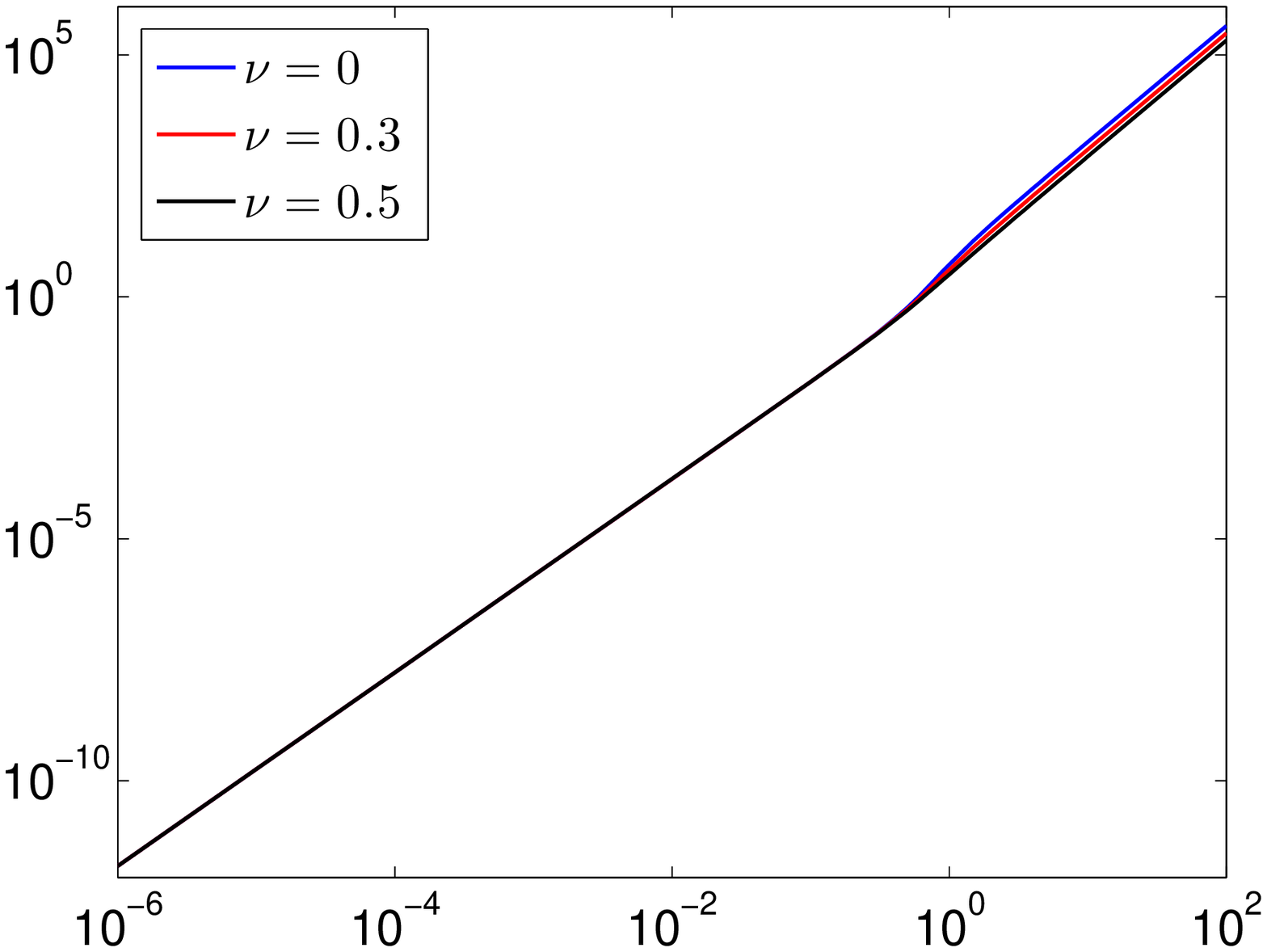}
    \put(-105,0){$\hat K_{Ic}$}
    \put(-235,90){$\frac{\hat K_I}{\hat K_f}$}
    \put(-230,160){$\textbf{b)}$}

 \caption{Relationship between the self-similar material toughness, $\hat K_{Ic}$, and: a) the multiplier of the leading  term of the shear stress tip asymptote, $\hat \tau_0$, b) the ratio of the self-similar stress intensity factor, $\hat K_I$, and the self-similar shear stress intensity factor, $\hat K_f$.  }
 \label{KI_T0}
\end{figure}

As can be seen in the figures, the viscosity dominated regime is reached for approximately $\hat K_{Ic}=0.1$. For $\hat K_{Ic}<0.1$, the shear stress intensity factor stabilizes at a constant level depending on Poisson's ratio. A similar situation is observed for the other analysed parameters (except for $\hat K_I$), however, the self-similar crack propagation speed and the crack length exhibit low sensitivity to the value of $\nu$. Note that, as mentioned above, $\hat w_0$ is greater than zero (see \eqref{w_0_visc}) for the whole range of $\hat K_{Ic}$, which means that the LEFM asymptote holds for any value of the material toughness.

The numerical results and the analysis of the governing equations enabled us to deliver the following asymptotic estimations for $\hat K_{Ic} \to 0$. The multiplier of the leading term of the fluid pressure reaches its limiting value as:
\begin{equation}
\label{p_0_low_KI}
\hat p_0=\pi (1-\nu)-\zeta \hat K_{Ic}^2+O(\hat K_{Ic}^4),\quad \hat K_{Ic}\to0,
\end{equation}
where parameter $\zeta$ depends on $\alpha$ (the parameter determining the temporal evolution of the self-similar solution) and Poisson's ratio $\nu$. It exhibits an almost linear distribution with respect to $\nu$ and can be approximated by the following empirical relation:
\begin{equation}
\label{zeta_low_KI}
\zeta(\alpha,\nu) \approx \frac{4\sqrt{5}}{3}\alpha^{-0.55}(1-\nu).
\end{equation}
Approximation \eqref{zeta_low_KI} exhibits the relative accuracy of the order $10^{-3}$ for  a broad range of $\alpha$ ($0.05<\alpha<10$).

The remaining parameters depicted in Figs.\ref{KI_Kf_w0}-\ref{KI_T0} can be asymptotically described as:
\begin{equation}
\label{KF_low_KI}
\hat K_f=\sqrt{\frac{\pi}{4\zeta}}+O(\hat K_{Ic}^2),
\quad
\hat K_I=\sqrt{\frac{\zeta}{\pi}}\frac{1}{2(1-\nu)}\hat K_{Ic}^2+O(\hat K_{Ic}^4), \quad \hat K_{Ic}\to0,
\end{equation}
\begin{equation}
\label{w0_low_KI}
\hat w_0=\frac{\pi^{3/2}}{4\sqrt{\alpha}\zeta}\sqrt{1-\nu}+O(\hat K_{Ic}^2),
\quad
\hat v_0=\frac{\pi^4}{16\alpha\zeta^2}(1-\nu)^2+O(\hat K_{Ic}^2), \quad \hat K_{Ic}\to0,
\end{equation}
\begin{equation}
\label{L0_low_KI}
\hat L_0=\frac{\pi^2}{4\alpha\zeta}(1-\nu)+O(\hat K_{Ic}^2),
\quad
\hat \tau_0=\frac{\pi^{5/2}}{4\sqrt{\alpha}\zeta}(1-\nu)^{3/2}+O(\hat K_{Ic}^2),\quad \hat K_{Ic}\to0.
\end{equation}

For large values of the self-similar material toughness, all the parameters exhibit power law dependence on $\hat K_{Ic}$. The transition area between the viscosity dominated and large toughness asymptotes is relatively small. For the latter regime ($\hat K_{Ic}\gg1$) the following estimates hold for any $\alpha$ and $\nu$:
\[
\hat p_0=s^{2/5}\alpha^{3/5}\hat K_{Ic}^{-12/5}+O\left(\hat K_{Ic}^{-24/5}\right), \quad \hat K_{Ic}\to\infty,
\]
\begin{equation}
\label{KF_large_KI}
\hat K_f=\frac{\alpha^{3/5}s^{2/5}}{\pi(1-\nu)}\hat K_{Ic}^{-7/5}+O\left(\hat K_{Ic}^{-19/5}\right),
\quad
\hat K_I=\hat K_{Ic}+O\left(\hat K_{Ic}^{-7/5}\right), \quad \hat K_{Ic}\to\infty,
\end{equation}
\begin{equation}
\label{w0_large_KI}
\hat w_0=\left(\frac{s}{\alpha}\right)^{1/5}\hat K_{Ic}^{4/5}+O\left(\hat K_{Ic}^{-8/5}\right),
 \quad \hat v_0=s^{4/5}\alpha^{1/5}\hat K_{Ic}^{-4/5}+O\left(\hat K_{Ic}^{-16/5}\right),
\quad \quad \hat K_{Ic}\to\infty,
\end{equation}
\begin{equation}
\label{v0_large_KI}
\hat L_0=\left(\frac{s}{\alpha}\right)^{2/5}\hat K_{Ic}^{-2/5}+O\left(\hat K_{Ic}^{-14/5}\right),
\quad \hat \tau_0=\frac{s^{3/5}\alpha^{2/5}}{2}\hat K_{Ic}^{-8/5}+O\left(\hat K_{Ic}^{-4}\right), \quad \hat K_{Ic}\to\infty,
\end{equation}
where $s=2^{3/2}\pi^{-1}$. As can be seen from the above formulae (and is confirmed by the graphs in Figs.\ref{KI_Kf_w0}-\ref{KI_T0}), for the large toughness asymptotics it is only the self-similar shear stress intensity factor, $\hat K_f$, that exhibits dependence on Poisson's ratio $\nu$.

\section{Discussion and conclusions}

A novel approach to HF that better describes the underlying physics of the process, by accounting for the substantial effect of hydraulically induced shear stress at the crack faces, has been developed. A new form of the boundary integral equation of elasticity has been utilized alongside the modified fracture propagation condition. The latter is based on the energy release rate and depends on the stress intensity factor and newly introduced parameter, {\it the shear stress intensity factor}. In this formulation,  the linear elastic fracture mechanics (LEFM) asymptotics holds regardless of the material toughness. A rescaling and transformation of the problem to the self-similar form has been given. A number of numerical simulations by means of the universal HF algorithm developed by \cite{wr_mish_2015} have been conducted to compare the new formulation with its classic counterpart.
The asymptotics of small and large toughness have been considered in detail.

%The asymptotic estimations for various parameters have been provided for the small and large toughness %cases.

 A notable feature of our approach is that the paradox of the viscosity dominated regime of HF, where the LEFM asymptotics does not hold, has been solved. A computational consequence is that the difficult problem of the so-called small toughness regime has been resolved.

Additionally the following conclusions can be drawn:
\begin{itemize}
\item{ The deviation of the results for the classic KGD model from those of the new formulation decreases with increasing material toughness and Poisson's ratio, due to the decreasing influence of the additional shear stress term in \eqref{elasticity_1}.  The greatest discrepancies are observed for the fluid pressure and the particle  velocity.}
\item{In the new HF model, even in the viscosity dominated regime of fracture propagation, the standard asymptotics of LEFM holds. This fact has serious computational ramifications. Unlike the classic HF model, no dedicated mechanisms are needed in the computational scheme to account for the transition between different crack tip singularities as the length of the LEFM process zone is never reduced to infinitesimal value. No special measures are required to retain the proper tip behaviour and stabilize the algorithm in the so called small toughness case. This substantially increases the reliability of the numerical results. In the new formulation, we have not observed any differences in the computational performance of the algorithm between the viscosity dominated, small and moderate toughness regimes.}
    \item{The transition interval between the viscosity dominated and large toughness regimes is relatively small. Note that, in the latter regime the asymptotic behaviour of all quantities, except for the shear stress intensity factor, do not depend on Poisson's ratio.
\item{The asymptotic analysis of the modified HF formulation shows that the temporal evolution of the fracture is not associated with qualitative changes in the tip behaviour. This property of the solution has a crucial consequence for modelling 2D and 3D hydraulic fracture problems in inhomogeneous media when the material properties may differ along the fracture contour. New model allows to consider such cases within the same computational scheme capable of treating each point of the crack front in a unified way}.   }
 \item{Yet another important observation is that for a constant injection flux rate, the pressure drop singularity becomes less pronounced as the crack grows. As a result, both parameters $p_0$ and $\varpi$ decay in time leading to a monotonic temporal decrease of the ratio $K_f/K_I$, provided that $K_{IC}>0$.  In other words, the crack propagation regime is changing from the viscosity dominated ($K_f>0$, $K_I \to 0$) to toughness dominated ($K_f \to 0$, $K_I> 0$).}

\end{itemize}

Further analysis is need to examine the proposed model and respective consequences for the numerical simulation. As an example, one can mention the phenomenon of a lag between the fracture tip and fluid front, which has been understood so far to have a negligible influence on the process under typical high confinement conditions encountered in reservoir stimulation \citep{garagash_large_toughenss,Gar_Det_Ad}. In the modified HF formulation such a problem can be considered in the framework of a regular perturbation, where the standard LEFM crack propagation criterion \eqref{K_IC} is used in conjunction with the new elasticity operator \eqref{elasticity_1}.

\vspace{2mm}

\noindent
{\bf Acknowledgements} MW and GM are grateful to the FP7 EU funded projects: PIAP-GA-2009-251475-HYDROFRAC and IRSES-GA-2013-610547-TAMER. AP acknowledges the support of the project: PCIG13-GA-2013-618375-MeMic.

\vspace{10mm}

% \clearpage
%%%%%%%%%%%%%%%%%%%%%%%%%%%%%%%%%%%%%%%%%%%%%%%%%%%%%%%%%%%%%%%%%%%%%%%
\noindent
\section*{Appendix}
\appendix
\renewcommand{\theequation}{\thesection.\arabic{equation}}

\section{Definition of functions $\bPhi_j(\theta)$}
\setcounter{equation}{0}
\label{app1}

Below we collect the radial vector-functions $\bPhi_j(\theta)$
involved in the asymptotic representation of the displacement field near the crack tip. They are evaluated in a standard manner by substitution of the representation (\ref{ass_U}) into the Lame's equations where respective boundary conditions resulting from the known bahaviour of the fluid field along the crack surfaces and symmetry of the solution hold.

\begin{equation}
\Phi_1^r(\theta) =
\frac{1}{E^*} \cos \frac{\theta }{2} \left[3-\nu ^* - \left(1+\nu^*\right) \cos \theta\right],
\end{equation}

\begin{equation}
\Phi_1^\theta(\theta) =
-\frac{1}{E^*} \sin \frac{\theta }{2} \left[3-\nu ^*-\left(1+\nu^*\right) \cos \theta\right],
\end{equation}

\begin{equation}
\Phi_{\tau_0}^r(\theta) =
-\frac{2}{E^*} \left(1+\nu ^*\right) \cos \frac{3 \theta}{2},\quad
\Phi_{\tau_0}^\theta(\theta) =
\frac{2}{E^*} \left(1+\nu ^*\right) \sin \frac{3 \theta}{2},
\end{equation}

\begin{equation}
\tilde\Phi_{p_0}^r(\theta) =
-\frac{1-\nu ^*}{E^*},\quad
\tilde\Phi_{p_0}^\theta(\theta) = 0,
\end{equation}

\begin{equation}
\tilde\Phi_{\tau_1}^r(\theta) =
\frac{2}{\pi E^*} \left(\cos ^2\theta-\nu ^* \sin ^2\theta\right),
\quad
\tilde\Phi_{\tau_1}^\theta(\theta) =
-\frac{\left(1+\nu ^*\right)}{\pi E^*} \sin (2 \theta ),
\end{equation}

\begin{equation}
\Phi_{2}^r(\theta) =
\frac{4}{E^*} \left(\cos ^2\theta-\nu ^* \sin ^2\theta\right),\quad
\Phi_{2}^\theta(\theta) =
-\frac{2 \left(1+\nu ^*\right)}{E^*} \sin (2 \theta),
\end{equation}

\begin{equation}
\Phi_{p_0}^r(\theta) =
\frac{1+\nu ^*}{2 E^*} (1 - 3 \cos 2 \theta),\quad
\Phi_{p_0}^\theta(\theta) =
\frac{1}{E^*} \left[ 3\left(1+\nu ^*\right) \sin (\theta ) \cos(\theta )-2 \theta  \right],
\end{equation}

\begin{equation}
\Phi_{p_1}^r(\theta) =
\frac{\left(1+\nu ^*\right)}{E^*} \cos 2 \theta,
\Phi_{p_1}^\theta(\theta) =
-\frac{\left(1+\nu ^*\right)}{E^*} \sin 2 \theta,
\end{equation}

\begin{equation}
\Phi_{\tau_1}^r(\theta) =
-\frac{\left(1+\nu ^*\right)}{\pi E^*} \sin \theta \left[\sin\theta+2 \theta \cos \theta\right],
\end{equation}

\begin{equation}
\Phi_{\tau_1}^\theta(\theta) =
-\frac{1}{\pi  E^*} \left[ \left(1+\nu ^*\right) \left(\frac{1}{2} \sin (2 \theta )+\theta  \cos (2 \theta )\right)-2\theta \right],
\end{equation}

\begin{equation}
\tilde\Phi_{p_2}^r(\theta) =
\frac{1}{6 \pi  E^*} \left[ \left(3-5 \nu ^*\right) \cos \left(\frac{\theta }{2}\right)+\left(1+\nu^*\right) \cos \left(\frac{5 \theta}{2}\right) \right],
\end{equation}

\begin{equation}
\tilde\Phi_{p_2}^\theta(\theta) =
-\frac{1}{3 \pi E^*} \sin \frac{\theta }{2} \left[ \left(1+\nu ^*\right) \left( \cos (\theta )+  \cos (2 \theta) \right) -4\right],
\end{equation}

\begin{equation}
\tilde\Phi_{\tau_2}^r(\theta) =
\frac{1}{6 E^*} \left[ \left (3 - 5 \nu ^*\right) \cos \frac{\theta }{2} + 5 \left(1+\nu^*\right) \cos \frac{5 \theta}{2} \right],
\end{equation}

\begin{equation}
\tilde\Phi_{\tau_2}^\theta(\theta) =
\frac{1}{6 E^*} \left[ \left(9+\nu ^*\right) \sin \frac{\theta}{2}-5 \left(1+\nu ^*\right) \sin \frac{5 \theta }{2} \right]
\end{equation}

\begin{equation}
\Phi_{3}^r(\theta) =
\frac{1}{2 E^*} \left[ \left(3-5 \nu ^*\right) \cos \frac{\theta}{2}+\left(1+\nu^*\right) \cos \left(\frac{5 \theta}{2}\right) \right],
\end{equation}

\begin{equation}
\Phi_{3}^\theta(\theta) =
-\frac{1}{E^*} \sin \frac{\theta}{2} \left[ \left(1+\nu ^*\right) \left( \cos (\theta)+ \cos (2 \theta) \right) -4 \right],
\end{equation}

\begin{equation}
\Phi_{p_2}^r(\theta) =
-\frac{1}{6 \pi E^*}
\left[
2\left(1+\nu^*\right) \left( \cos \frac{\theta }{2} - \cos \frac{5 \theta}{2} + \frac{\theta}{2} \sin \frac{5 \theta }{2} \right) + \left(3-5 \nu ^*\right)\theta \sin \frac{\theta}{2}
\right],
\end{equation}

\begin{equation}
\Phi_{p_2}^\theta(\theta) =
\frac{1}{6 \pi E^*}
\left[
2 \left(1+\nu ^*\right)
\left(
\sin \frac{\theta}{2} - \sin \frac{5 \theta}{2} - \frac{\theta}{2} \cos \frac{5 \theta}{2}
\right)
+ \left(9+\nu ^*\right) \theta \cos \frac{\theta}{2}
\right],
\end{equation}

\begin{equation}
\Phi_{\tau_2}^r(\theta) =
-\frac{1}{18 E^*}
\left[
\left(1+\nu ^*\right)
\left(
6 \cos \frac{\theta }{2} + 2 \cos \frac{5 \theta}{2}
+ 15 \theta \sin \frac{5 \theta}{2}
\right)
+ 3 \left(3-5 \nu^*\right) \theta \sin \frac{\theta}{2}
\right],
\end{equation}

\begin{equation}
\Phi_{\tau_2}^\theta(\theta) =
\frac{1}{18 E^*}
\left[
\left(1+\nu^*\right)
\left(
6 \sin \frac{\theta}{2} + 2 \sin \frac{5 \theta}{2} - 15 \theta \cos \frac{5 \theta}{2}
\right)
+ 3 \left(9+\nu^*\right) \theta \cos \frac{\theta}{2}
\right].
\end{equation}

\end{document}